\documentclass[
aip,
jcp,%
amsmath,amssymb,
dvipsnames,
twocolumn,
reprint,%
]{revtex4-2}

\usepackage[utf8]{inputenc}
\usepackage{graphicx}
\usepackage{dcolumn}
\usepackage{bm}
\usepackage{tikz}
\usepackage{placeins}

\usepackage{ulem}

\newcommand{\zeff}{Z_{\mathrm{eff}}}

\begin{document}

\title{Towards Quantum Monte Carlo Forces on Heavier Ions: Scaling Properties}
\author{Juha Tiihonen}
\affiliation{ 
Materials Science and Technology Division, Oak Ridge National Laboratory, Oak Ridge, TN 37831, USA
}%
\author{Raymond C. Clay III}
\affiliation{ 
Sandia National Laboratories, Albuquerque, NM 87185, USA
}%
\author{Jaron T. Krogel}
\affiliation{ 
Materials Science and Technology Division, Oak Ridge National Laboratory, Oak Ridge, TN 37831, USA
}

\date{\today}

\begin{abstract}
Quantum Monte Carlo (QMC) forces have been studied extensively in recent decades because of their importance with spectroscopic observables and geometry optimization.  Here we benchmark the accuracy and statistical cost of QMC forces.  The zero-variance zero-bias (ZVZB) force estimator is used in standard variational and diffusion Monte Carlo simulations with mean-field based trial wavefunctions and atomic pseudopotentials. Statistical force uncertainties are obtained with a recently developed regression technique for heavy tailed QMC data [P. Lopez Rios and G. J. Conduit, Phys. Rev. E 99, 063312 (2019)]. By considering selected atoms and dimers with elements ranging from H to Zn ($1\leq \zeff \leq 20$), we assess the accuracy and the computational cost of ZVZB forces as the effective pseudopotential valence charge, $\zeff$, increases.  We find that the cost of QMC energies and forces approximately follow simple power laws in $\zeff$. The force uncertainty grows more rapidly, leading to a best case cost scaling relationship of approximately $\zeff^{6.5(3)}$ for DMC. We find the accessible system size at fixed computational cost decreases as $\zeff^{-2}$, insensitive to model assumptions or the use of ``space warp'' variance reduction. Our results predict the practical cost of obtaining forces for a range of materials, such as transition metal oxides where QMC forces have yet to be applied, and underscore the importance of further developing force variance reduction techniques, particularly for atoms with high $\zeff$.
\end{abstract}

\maketitle

\section{Introduction}

The ability to compute forces accurately and efficiently is a critical part of \textit{ab initio} electronic structure calculations.  Forces are used in geometry optimization in solids and molecules, but beyond this, they are useful for studying vibrational properties of molecules and solids \cite{Vrbik1990}, chemical reaction pathways \cite{Saccani2013}, electron phonon-coupling \cite{Kresse1995}, and general thermodynamic properties via molecular dynamics.  Moreover, the ability to compute forces increases the amount of training data one can use to construct classical force-fields, potentially allowing the construction and evaluation of \textit{ab initio} quality potential energy surfaces (PES) for a fraction of the computational cost \cite{Ercolessi1994,Nakano2021}.  

Quantum Monte Carlo (QMC) is well known to produce highly accurate PES with high efficiency, and the ability to compute forces directly within QMC has long been desired. QMC force estimators have faced challenges in terms of both systematic bias \cite{Assaraf2003} and statistical ill-posedness (i.e. the infinite variance problem \cite{Assaraf2000,Rios2019}), but zero-variance zero-bias (ZVZB) estimators from Assaraf and Caffarel\cite{Assaraf2000,Assaraf2003} have gained significant traction in recent years due to their accuracy and greatly improved statistical efficiency over bare Hellman-Feynman style estimators. This is in large part due to steady advancements in capabilities \cite{Badinski2007,Badinski2008a}, improved statistical properties \cite{Filippi2000,Attaccalite2008}, and algorithmic efficiency \cite{Sorella2010,Filippi2016}. It has also lead to numerous successful applications in the structural optimization of molecules \cite{Barborini2012,Barborini2012a, Coccia2012,Coccia2012a,Coccia2014, Zen2013,Varsano2014,Coccia2014,Barborini2015,Zen2015}, minimum energy pathways \cite{Saccani2013}, and molecular dynamics of molecular \cite{Mouhat2017} and bulk systems \cite{Luo2014,Mazzola2017,Mazzola2018,Nakano2021}. 

Most of the applications to date are limited to first and second row elements, although exciting prospects await in transition metal elements of third row and beyond. Transition metals have innumerable applications in materials research, because of their properties in, \textit{e.g.}, catalysis \cite{Reen2019} and superconductivity \cite{Hardy1954}, and prevalence in semiconductors \cite{Lany2015} and prospective 2D-materials \cite{Kalantar-zadeh2016,Yang2018}. Since transition-metal oxides are famous for their strong electronic correlation\cite{Wagner2007,Shin2017} and also often display sensitivity in optimized lattice structure to the description of correlation, the ability to quickly compute forces in beyond-density functional theory (DFT) methods like QMC is  greatly needed for a fully consistent description of these materials. QMC simulation of transition metal systems is well established \cite{DoblhoffDier2016,Melton2016,Dubecky2016,Santana2016,Kylanpaa2017,Yu2017,Kent2020} and feasible with modern pseudopotentials \cite{Burkatzki2007,Burkatzki2008, Bennett2017,Bennett2018,Annaberdiyev2018,Wang2019}, but without forces the structural effects due to, \textit{e.g.}, lattice defects and phonons can only be considered at the mean-field level. 

In this work, we seek to characterize and understand how the computational cost of QMC forces scales with the effective valence charge $\zeff$ of the employed pseudopotential.  The computational cost is central to practical use of forces in chemical and materials science applications.  It is based on factors such as intrinsic variance of the estimator, wavefunction quality, and numerical implementation, but here we only focus on the estimator properties while using a standard Slater-Jastrow wavefunction most common in large-scale applications. For the estimator, we will use the standard ZVZB force estimator \cite{Assaraf2003} with a recent tail-regression technique \cite{Rios2019} to regularize the infinite variance problem. We will also use the popular space warp transformation technique \cite{Filippi2000,Sorella2010} to show that it significantly reduces variance, but remains subject to an apparent cost-scaling effect with $\zeff$.  By gaining explicit knowledge of how the intrinsic variance and the computational cost of the force scale with $\zeff$ and system size, we can make projections of what systems are affordable with the computing resources of the present or the future.

The remainder of the work is organized as follows: In Sec.~\ref{sec:estimator} we lay out the estimators used in this work to consider VMC and DMC energies and forces, and their statistical properties.  Technical details of the simulations and data sets are given in Sec.~\ref{sec:comp_details}.  In Section \ref{sec:zvzbforces} we validate the accuracy of the VMC and DMC energies and forces by comparing estimated bond lengths and vibration frequencies of selected dimers to experimental data and earlier QMC works.  In Section \ref{sec:efficiency} we analyze the scaling of the uncertainties and computational costs of energies and forces with $\zeff$. Moreover, we project their statistical implications in large-scale applications. We conclude with a summary of results made in this work in Sec.~\ref{sec:conclusions}. Additional results and data are available in supplemental Material and external resources \cite{Tiihonen2021repository}.

\section{QMC Force estimators}
\label{sec:estimator}

QMC force estimators are based on derivatives of the statistically sampled total energy, which can be written as
\begin{align}
\label{eq:qmcenergy}
E = \frac{\int \mathrm{d}R\, E_L(R) \Phi(R)\Psi_T(R)}{\langle \Phi | \Psi_T \rangle} \equiv \langle E_L \rangle_{\Phi\Psi_T},
\end{align}
where $R$ is the set of coordinates of $N$ electrons,
\begin{align}
\label{eq:localenergy}
E_L(R) = \frac{\hat{H} \Psi_T(R)}{\Psi_T(R)}
\end{align}
is the local energy, $\Psi_T$ is a trial wave function and $\Phi$ denotes a complementary sampling wave function. Sampling of $\Phi(R) \Psi_T(R)/\langle \Phi | \Psi_T \rangle$ is done with an appropriate Monte Carlo procedure: in VMC $\Phi=\Psi_T$ and in DMC $\Phi$ is the projected fixed-node ground state. 

Let $\lambda$ refer to a parameter affecting the energy, such as an ionic coordinate. The associated force is given by the negative gradient of $E$ w.r.t. $\lambda$. It is a straightforward exercise to show that
\begin{eqnarray}
\frac{d E}{d \lambda} & = & \left\langle \frac{\partial}{\partial \lambda} E_L\right\rangle_{\Phi\Psi_T} \label{eq:zv-orig} \\
 &+& \left\langle (E_L-E)\left[\Phi^{-1}\frac{\partial \Phi}{\partial \lambda}+\Psi_T^{-1}\frac{\partial \Psi_T}{\partial \lambda}\right]\right\rangle_{\Phi\Psi_T} \label{eq:zb-orig}\\
 &+& \sum_{i=1}^{N_c}\frac{\partial E}{\partial c_i} \frac{\partial c_i}{\partial \lambda} \label{eq:sc-orig}.
\end{eqnarray}
The terms within the expectation values of Eqs.~\eqref{eq:zv-orig} and \eqref{eq:zb-orig} comprise, respectively, the zero-variance (ZV) zero-bias (ZB) force estimators for a given instance of the Hamiltonian and the sampling wavefunction\cite{Assaraf2003}. In the last term, the set of variables $\lbrace c_1,\ldots,c_{N_c}\rbrace$ are parameters of the wavefunction that only implicitly depend on $\lambda$. If the system is at a variational minimum ($\frac{\partial E}{\partial c_i}=0$) or the implicit parameters are not allowed to vary with changing $\lambda$ ($\frac{\partial c_i}{\partial\lambda}=0$), then $\frac{dE}{d\lambda}$ can be calculated exactly by taking an expectation value over the ZVZB estimator in brackets.  If either of these assumptions is false, then there will be an error incurred between the sampled energy derivative and the true energy derivative, the magnitude of which is given by Eq.~\eqref{eq:sc-orig}.  We will refer to this error as ``self-consistency error'' to highlight the inconsistency between the exact and estimated energy derivatives.

\subsection{VMC Forces}

Let $\lambda$ refer to a specific ionic coordinate.  Recalling that the force is simply $F=-\frac{dE}{d\lambda}$ and using Eqs. \ref{eq:zv-orig} and \ref{eq:zb-orig} with $\Phi=\Psi_T$, the ZVZB VMC force estimator is:
\begin{equation}
\hat{F}^{VMC}_{ZVZB}= -\frac{\partial}{\partial \lambda} E_L(R)-2 \frac{(E_L(R)-E^{VMC})\frac{\partial \Psi_T}{\partial \lambda}}{\Psi_T(R)}
\end{equation}
The total VMC force is just:
\begin{equation}
F^{VMC}_{ZVZB} = \langle \hat{F}^{VMC}_{ZVZB} \rangle_{|\Psi_T|^2}
\end{equation}
This will differ from the exact VMC PES derivative by the negative of Eq. \ref{eq:sc-orig}.

\subsection{DMC Forces}

For DMC, the role of $\Phi$ is taken by the fixed-node wave function.  In contrast to VMC, it is not straightforward to construct an exact ZVZB estimator for DMC on account of the $\Phi^{-1}\frac{\partial \Phi}{\partial \lambda}$ term in Eq. \ref{eq:zb-orig}.  There are three known ways to deal with this term:  direct evaluation using forward walking or pure DMC \cite{Assaraf2003}, approximate evaluation using the variational drift-diffusion approximation\cite{Moroni2014}, or the Reynold's approximation\cite{Reynolds1986}.  Due to its simplicity, we will benchmark the Reynold's approximation in this work.  This simply makes the assumption that: 
\begin{equation}
\Phi^{-1}\frac{\partial \Phi}{\partial \lambda} = \Psi_T^{-1}\frac{\partial \Psi_T}{\partial \lambda}
\end{equation}
at the cost of an error that scales like $O(\Phi-\Psi_T)$. 

Under this assumption, the DMC ZVZB force estimator takes the following analogous expression to VMC:
\begin{equation}
F^{DMC}_{ZVZB} = \langle \hat{F}^{DMC}_{ZVZB} \rangle_{\Phi\Psi_T}
\end{equation}
\begin{equation}
\hat{F}^{DMC}_{ZVZB}= -\frac{\partial}{\partial \lambda} E_L(R)-2 \frac{(E_L(R)-E^{DMC})\frac{\partial \Psi_T}{\partial \lambda}}{\Psi_T(R)}
\end{equation}

The systematic error in the above expression comes from two sources.  In addition to the term discussed in Eq. \ref{eq:sc-orig} (except now it should be $\frac{\partial E^{DMC}}{\partial c_i}=0$, which is not what the VMC optimization guarantees), there is an unquantified $O(\Phi-\Psi_T)$ error term coming from the Reynold's approximation.  These two contributions account for all deviations of the DMC force estimator from the true derivative of the DMC PES. 

One proposed method to mitigate this $O(\Phi-\Psi_T)$ error is to use a ``hybrid estimator" \cite{Assaraf2003}:
\begin{align}
\label{eq:hybrid}
    F^{\mathrm{hybrid}}_{ZVZB} = 2 \langle \hat{F}^{DMC}_{ZVZB} \rangle_{\Phi\Psi_T} - \langle \hat{F}^{VMC}_{ZVZB} \rangle_{|\Psi_T|^2}
\end{align}
This is coincidentally identical in form to the extrapolated estimator used to mitigate the DMC mixed-estimator bias, though we stress that the origin of this estimator is different.

\subsection{Space Warp Transformation}

Space warp transformation is a way to reduce variance of the ZVZB estimator by including more physically motivated properties into $\Psi_T$.  Originally developed as a means to perform efficient correlated sampling QMC calculations between systems with slightly different ionic geometries \cite{Filippi2000}, the differential form \cite{Assaraf2003,Sorella2010} includes the physical assumptions of the original model into the ZVZB estimator itself.

Let $\mathbf{r}_i$ and $\mathbf{R}_I$ denote the respective coordinates of electron $i$ and ion $I$, and $\Delta \mathbf{R}_I$ a finite displacement of the latter.  The space warp transformation is defined as:
\begin{equation}
\mathbf{r}'_i = \mathbf{r}_i + \Delta \mathbf{R}_I\omega_I(\mathbf{r_i}).
\end{equation}
where
\begin{equation}
\omega_I(\mathbf{r})=\frac{F(|\mathbf{r}-\mathbf{R}_I|)}{\sum_I F(|\mathbf{r}-\mathbf{R}_I|)}.
\end{equation}

We choose $F(r)=r^{-4}$ as is commonly done in the literature \cite{Filippi2000,Sorella2010}.  In differential form, one first adds the following expression to the $\frac{\partial}{\partial \lambda}E_L$ term:
\begin{equation}
\mathbf{F}^{ZV-SW}_I=-\sum_{i=1}^{N_{elec}} \omega_I(\mathbf{r}_i)\nabla_i E_L    
\end{equation}
Lastly, one adds the following term to the of $\frac{\nabla_I \Psi_T}{\Psi_T}$:
\begin{equation}
\sum_{i=1}^{N_{elec}} \omega_I(\mathbf{r}_i)\nabla_i \log(\Psi_T) + \frac{1}{2}\nabla_i \omega_I(\mathbf{r}_i)
\end{equation}
As we will see later, the improvement to the statistical performance of the ZVZB estimator by including the above two terms can be significant depending on the force component and system.

\subsection{Estimation of statistical error}
\label{sec:err_est}

Statistical uncertainty and variance are key concepts in the integration and error estimation of Monte Carlo observables. Following Ref.~\onlinecite{Rios2019}, let $A$ denote a random variable with a probability distribution $P(A)$. Then, the expectation value of $A$ is given by
\begin{align}
\label{eq:stdmean}
    \langle A \rangle = \int_{-\infty}^\infty P(A) A \, \mathrm{d}A,
\end{align}
and its variance by
\begin{align}
\label{eq:stdvar}
    \sigma^2_A = \int_{-\infty}^\infty P(A) (A - \langle A \rangle )^2 \, \mathrm{d}A.
\end{align}
The central limit theorem states that $N\rightarrow \infty$ uncorrelated random samples from $P(A)$ produce an estimate of $\langle A \rangle$ with an uncertainty (or error bar) of $\sigma_e =  \sqrt{\frac{1}{N-1}} \sigma_A$. This is referred to as the standard estimate of the uncertainty of the mean. The central limit theorem only holds when $\sigma_A$ is well defined, \textit{i.e.}, when the integral in \eqref{eq:stdvar} is non-divergent. This means that $P(A)$ should not have heavy asymptotic tails of the form $P(A) \sim |A-\langle A\rangle|^{-\mu}$, where $\mu \leq 3$, if the standard estimate of the uncertainty of the mean is to be used\cite{Rios2019}. If such heavy tails are present, the estimator of the mean has manifestly infinite variance: the estimates of statistical uncertainty become formally undefined, and practically unstable.

A potential remedy to this problem was put forward by R{\^i}os and Conduit\cite{Rios2019}, in which they introduce a tail regression estimator tailored to the particular distributions encountered in the QMC.  The tail regression estimator idea builds on the foundation of tail index estimation established in the statistics community.  The approach reformulates the estimator of the mean of a heavy tailed distribution in such a way that the statistical variance of the distribution of the estimated means remains finite.  The basic idea is to use regression approaches to model the asymptotics of the tail region in a stable fashion relative to variations in the discrete sample distribution.  In the tail regression approach, the distribution $P(A)$ is estimated as 
\begin{align}
  P(A) = 
  \begin{cases}
    \sum\limits_{n=0}^{M_L} c^L_n \lvert A-A_C\rvert^{-\mu-n\Delta}     , &  A \in (-\infty,A_L) \\
    P_C(A) , & A\in [A_L,A_R] \\
    \sum\limits_{n=0}^{M_R} c^R_n \lvert A-A_C\rvert^{-\mu-n\Delta}     , &  A \in (A_R,\infty)
  \end{cases}
\end{align}
Given a set of $N_C$ samples drawn from $P(A)$ denoted $\{A_i\}_{i=1}^{N_C}$, $P_C(A)$ is approximated by the samples falling in the interval $[A_L,A_R]$, while the asymptotic parameters $\mu, c^L_n, c^R_n$ are fitted based on the samples falling outside the interval (\textit{i.e.} in the tail regions).  The distribution center, $A_C$, is approximated by the sample median.  Following the fitting procedure, the sample mean is estimated with the tail regression estimator as
\begin{align}
    \langle A \rangle \approx &\frac{1}{N_C}\sum_{A_i\in[A_L,A_R]}A_i \nonumber \\
      &+ \sum_{n=0}^{N_L} c^L_n \int_{-\infty}^{A_L}dA A \lvert A-A_C \rvert^{-\mu-n\Delta} \nonumber \\
      &+ \sum_{n=0}^{N_R} c^R_n \int_{A_R}^{\infty}dA A \lvert A-A_C \rvert^{-\mu-n\Delta}
\end{align}
where the continuum integrals are evaluated analytically.  The fitting procedure determining the asymptotic parameters has been shown to result in Gaussian distributed sample means with finite variance\cite{Rios2019} for distributions with polynomial tails.  The statistical uncertainty, or standard deviation, of the sample means is determined in practice through the bootstrap resampling technique.  Practical application of tail regression estimators to QMC energy and force data are available through the tail-regression estimator toolkit (TREAT), which has been made publicly available by the authors of Ref. \onlinecite{Rios2019}.

In this work, we solely use tail regression estimator (TRE) to obtain finite variance estimates of QMC forces and their attendant statistical uncertainties.  We also estimate statistical autocorrelation for energies and forces via reblocking techniques (see supplemental information for details).  We have also applied tail regression estimators to QMC total energies, which are formally expected to have heavy tails ($\mu=4$)\cite{Rios2019}.  However, similar to Ref. \onlinecite{Rios2019} we find no significant difference between the tail regression and standard estimates of energy means or uncertainties, and therefore we simply report the results from the standard estimator of the energy throughout.

\section{Computational details}
\label{sec:comp_details}

We carry out standard VMC and DMC simulations of atomic and molecular systems in open boundary conditions, as implemented in QMCPACK \cite{Kim2018} and documented in Ref.~\onlinecite{Kim2018} and the QMCPACK user manual. A conservative DMC time-step of 0.005 Ha$^{-1}$ is used, which does not lead to any significant bias in any of the systems studied here, according to our tests.  The DMC walker population is set to 1000, which also does not introduce significant bias. We use correlation consistent effective core potentials (ccECP) from Refs.~\onlinecite{Bennett2017,Bennett2018,Annaberdiyev2018,Wang2019}, and sampling of the nonlocal projector is done with Casula T-moves \cite{Casula2006}. The QMC workflows are managed by the Nexus workflow system \cite{Krogel2016a}. The ZVZB force estimators from Sec.~\ref{sec:estimator} have been available since QMCPACK version 3.9 and with space warp transformation since version 3.11. The estimator contains tail-regression analysis carried out with a tail-regression estimator toolbox, TREAT \cite{Rios2019}, as discussed earlier and detailed in the supplemental material. Full documentation and postprocessing features are not yet available in the QMCPACK repository, but raw data and curated scripts can be found from external data repository \cite{Tiihonen2021repository}.

Trial wavefunctions used in this work are standard Slater-Jastrow wavefunctions 
\begin{equation}
    \Psi_T(R) = \mathrm{e}^{J(R)} D^{\uparrow}(R) D^{\downarrow}(R),
\end{equation}
where $D^{\uparrow\downarrow}(R)$ are determinants for electrons with up and down spins. While more sophisticated wavefunctions could be used to better match the experiment, we have chosen the simple form, because it is commonly used in materials science. Spin-unrestricted self-consistent field calculation of the Slater determinant is done with GAMESS \cite{Gordon2005} using DFT with the PBE0 functional \cite{Adamo1999} and aug-VTZ bases from the ccECPs. Jastrow factor contains terms electron--ion (J1) and electron--electron (J2) terms between each combination of ions and the electron spin species. Jastrow optimization is performed via energy minimization with the linear method \cite{Toulouse2007a}. No orbital rotations are done during Jastrow optimization, leading to a bias as discussed in Sec.~\ref{sec:zvzbforces}. These single-reference trial wave functions are typically expected to recover 90\% of the correlation energy at the DMC level \cite{Williams2020}, which corresponds to 75-85\% of correlation energy recovered at the VMC level for trial wave functions used in this work.

Supplemental information contains further details on the basis sets and studies regarding advanced trial wave functions and orbital rotations.

\begin{figure}[b]
    \centering
    \begin{tikzpicture}
    \node at (-4.5, 0) {\includegraphics[width=8cm]{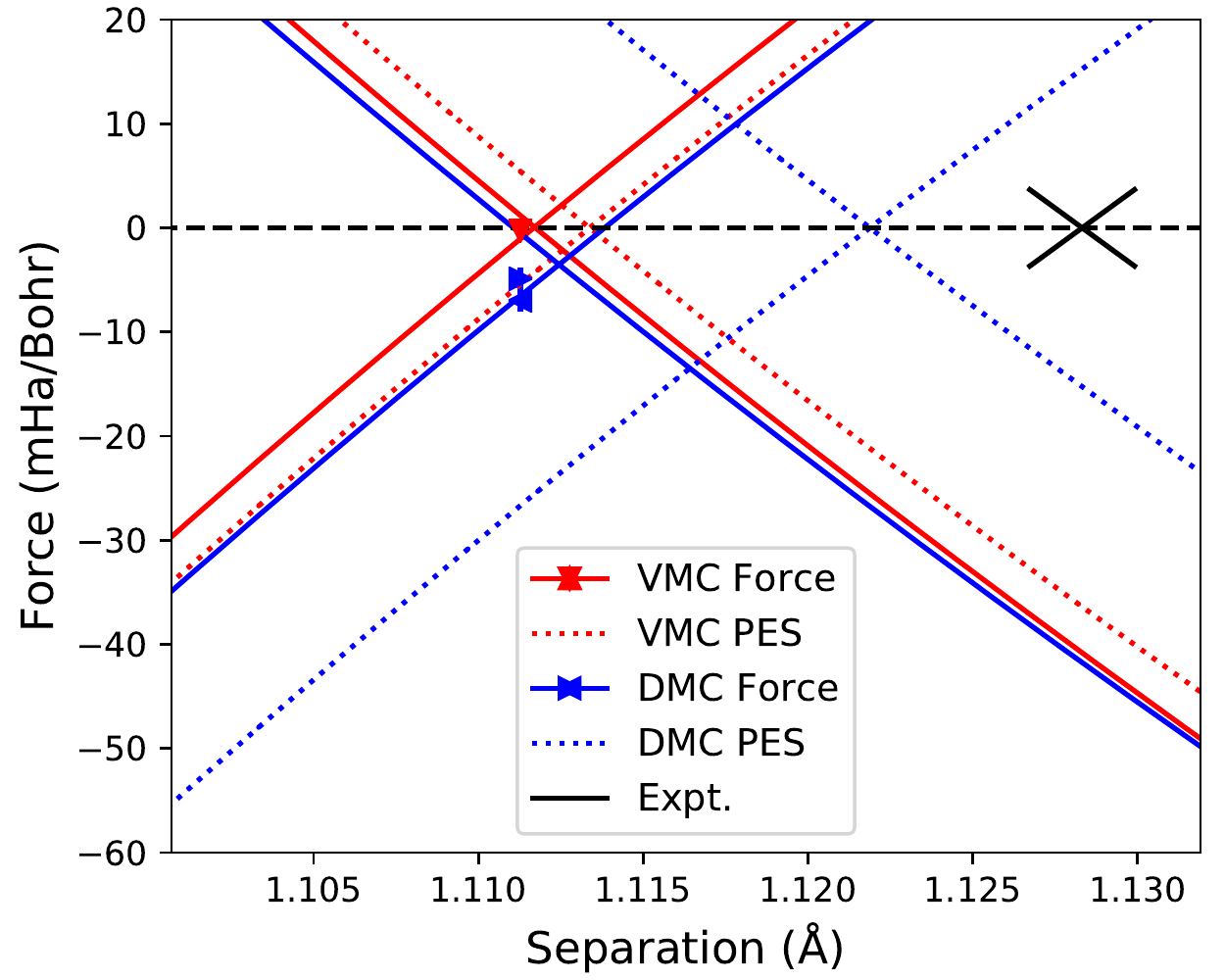}};
    \end{tikzpicture}
    \caption{Comparison of fitted forces (solid lines) and PES gradients (dotted lines) from VMC (red) and DMC (blue) near the equilibrium geometry of the CO dimer. Markers and error bars represent the original data.  The experimental equilibrium geometry and force constant are represented by a black cross.}
    \label{fig:F_CO_curves}
\end{figure}

\begin{figure*}[t]
    \centering
    \begin{tikzpicture}
    \node at (-4.5, 0) {\includegraphics[width=8cm]{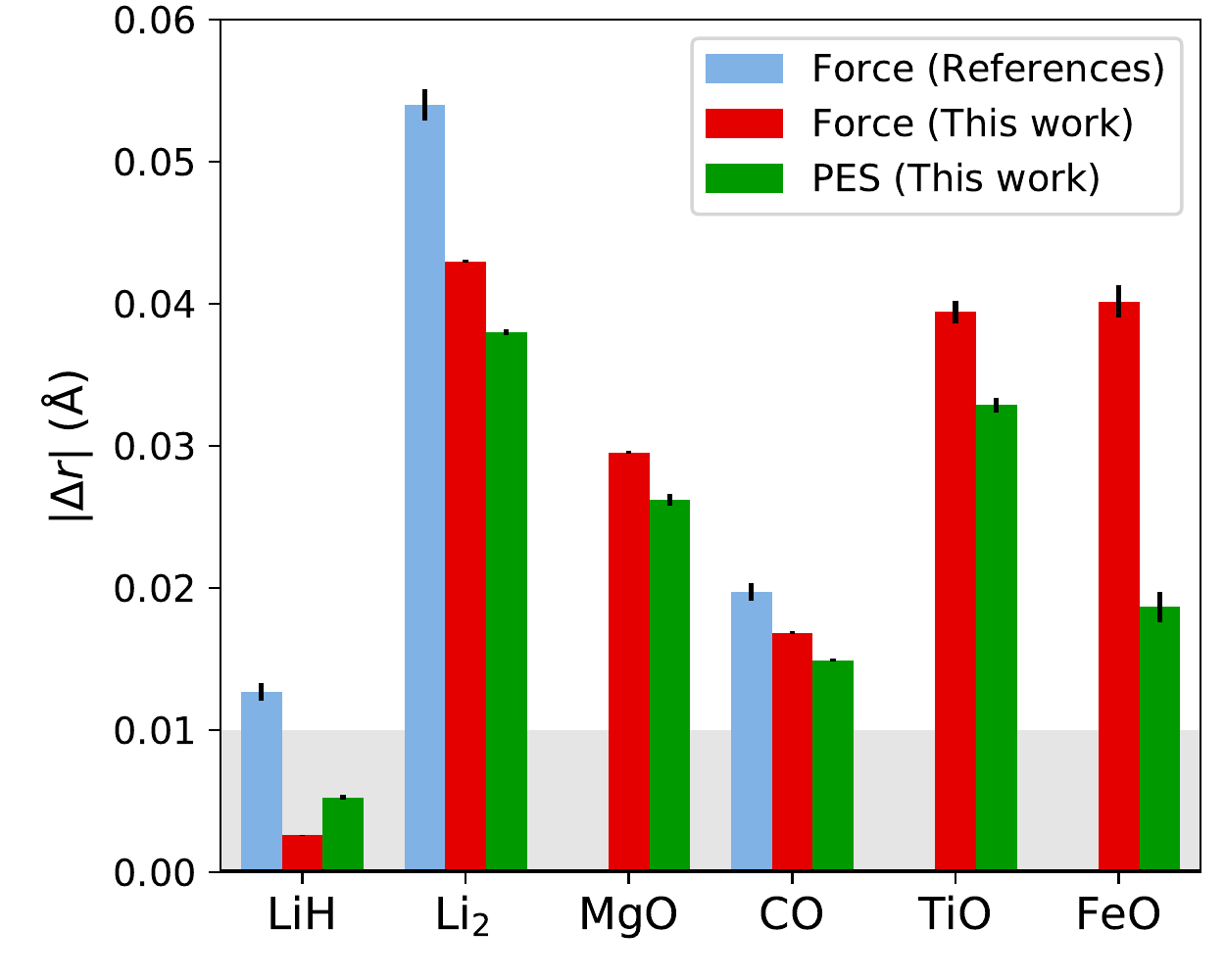}};
    \node at (4.5, 0) {\includegraphics[width=8cm]{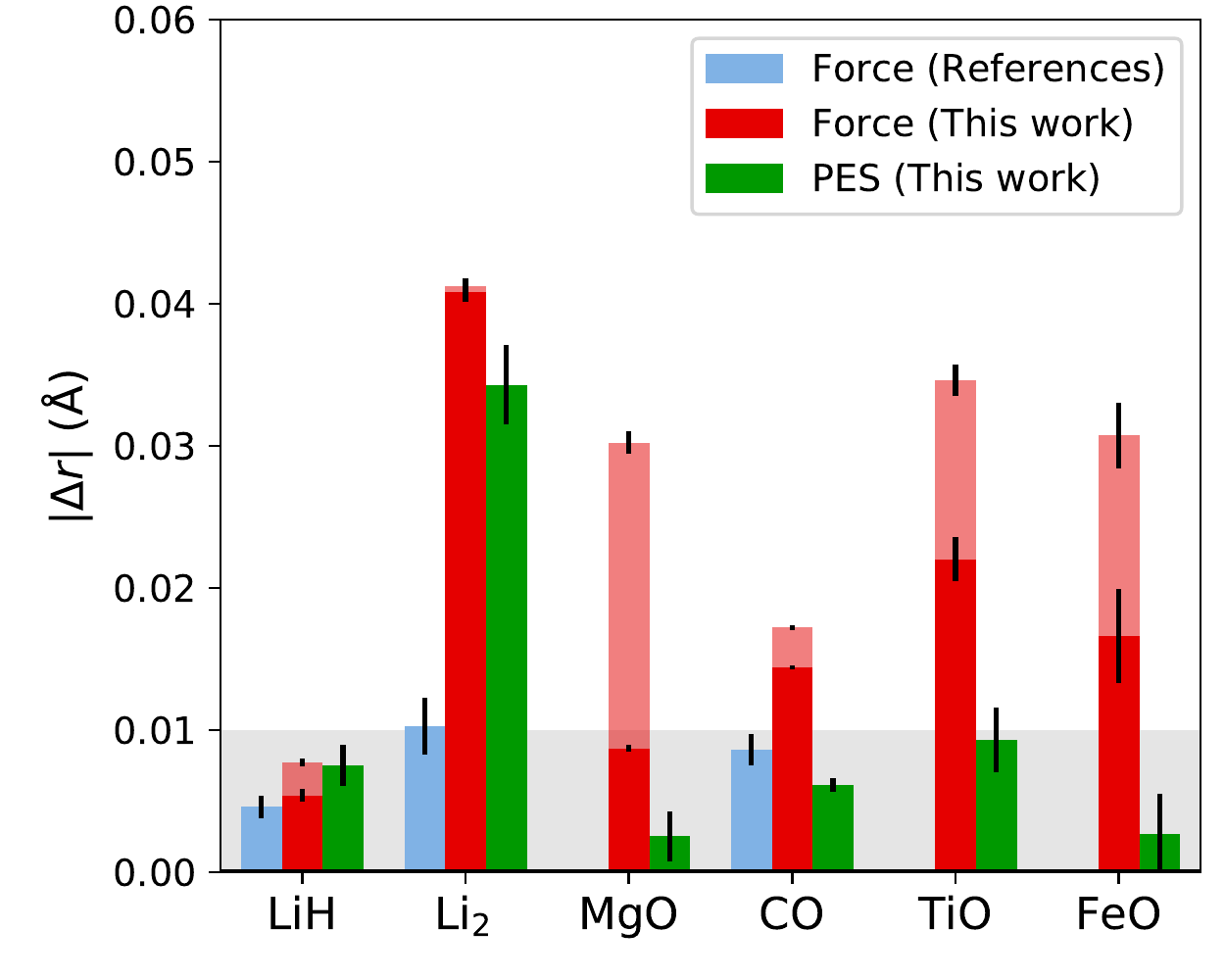}};
    \end{tikzpicture}
    \caption{Absolute deviations from experiment of estimated bond lengths $r$ from VMC (left) and DMC (right). Bars for each system from right to left represent QMC PES, QMC ZVZB force, and the median value out of available QMC force values from Refs.~ \onlinecite{Filippi2000,Assaraf2003, Casalegno2003,Chiesa2005,Lee2005,Badinski2007}. Gray marks acceptable accuracy. Solid and opaque color bars of the DMC forces correspond respectively to the better and worse results out of the left and right ionic forces, highlighting the net force induced.}
    \label{fig:R_comparison}
\end{figure*}

\section{Accuracy of QMC energies and forces}
\label{sec:zvzbforces}

In this section, we compare our results for VMC and DMC energies and ZVZB forces against earlier QMC works and other references to establish that our results are of representative quality. For each dimer molecule, let us consider the equilibrium bond length $r$ and vibrational frequency $\omega$ as quantities that represent the quality of the forces. Extracting these quantities is discussed in the supplemental material.  We consider sufficient accuracy\cite{Badinski2010} for practical applications, e.g. to structural optimization, to be $\sim 0.01 $ \AA{} for bond lengths.

Representative force curves for the CO dimer are shown in Fig.~\ref{fig:F_CO_curves}. The left and right VMC forces (on C and O ions, respectively) are equally large and have opposite signs, as expected on physical grounds, while the zero of force marks the predicted equilibrium geometry. The numerical gradient of the VMC PES is slightly more accurate than the ZVZB force, because the trial wavefunction does not represent a total variational minimum of the energy. Consistent with the observations in prior works \cite{Casalegno2003,Badinski2010,Moroni2014}, this is a direct manifestation of the self-consistency error present in mean-field based trial wavefunctions.  The mixed DMC forces are only slightly better than VMC, even though the DMC PES is substantially closer to experiment than the other estimates. Notably, the left and right DMC forces are slightly out of balance, meaning that the curves cross zero at different bond lengths and generally a net force is induced. This is due to differences in the approximated Pulay term. The behavior shown in Fig.~\ref{fig:F_CO_curves} is common to all dimer molecules studied in this work.

We compare the accuracy of our results for VMC and DMC forces to experimental reference data \cite{CRChandbook,Ram1999,Allen1996} and benchmark \textit{ab initio} calculations with CCSD(T) \cite{NIST2019,Miliordos2010,Sakellaris2011}. Deviations from the experimental bond lengths are shown in Fig.~\ref{fig:R_comparison} (numerical values available in the supplemental material). The mean absolute deviation (MAD) of bond lengths based on the VMC PES is $0.0223(3)$ \AA{}, which is slightly better than $0.0282(2)$ \AA{}, based on the VMC forces.  The DMC PES is by far the most accurate with MAD of bond-length being $0.0104(9)$ \AA{}. The MAD of mixed DMC forces is $0.0225(4)$ \AA{}, falling short of the accuracy of the DMC PES by more than a factor of two.   The mixed DMC forces only represent a slight improvement over the VMC forces and in fact give a performance almost the identical to the VMC PES. In addition, the performance of DMC forces is imbalanced between disparate atomic species in the dimer (an induced net force) due to the Reynolds approximation. The hybrid estimation (MAD: $0.0181(10)$ \AA) clearly improves upon mixed DMC forces, but not as significantly and systematically as in the original study \cite{Assaraf2003}.  While attempting to mitigate mixed-estimator error, the hybrid estimator amplifies the errors relating to the net force, which turns out to be significant.  While the effect of hybrid estimator remains positive overall, it does not present significant enough gains to overcome the deficiencies of the original estimates. 
Comparison to earlier works on QMC forces \cite{Filippi2000,Assaraf2003,Casalegno2003,Chiesa2005,Lee2005,Badinski2007}, where available, is also presented in Fig.~\ref{fig:R_comparison}. For added clarity, only the reference values with median deviation (LiH and CO: Ref~\onlinecite{Lee2005}, Li$_2$:  Ref.~\onlinecite{Filippi2000}) is used in the figures, but all numerical values are tabulated and discussed in better detail in the supplemental material. Our observations are consistent with prior works that use equivalent single reference trial wavefunctions. The main differences come from using other pseudopotentials, advanced DMC estimators \cite{Chiesa2005,Badinski2007} and, in some cases, all-electron calculation \cite{Assaraf2003,Filippi2000,Chiesa2005}. No prior force-based studies exist including metal oxide dimers.  Based on the comparison, our VMC ZVZB forces are generally representative among prior works, whereas our DMC ZVZB forces are not as accurate as some of those obtained from advanced estimators, which is not surprising.

In harmony with earlier works, our results emphasize that highly accurate forces benefit from improved trial wavefunctions, especially with heavier elements with larger $Z_{\mathrm{eff}}$. Potential improvements include backflow, anti-symmetrized geminal products, and multideterminant expansions, although this option, taken in isolation, is somewhat unsatisfying since the DMC PES with easy to access mean field trial wavefunctions is already sufficiently accurate for most applications. Therefore we expect that an important route of progress remains through the search for improved DMC force estimators, such as those proposed in Refs.~\onlinecite{Badinski2010,Moroni2014} or those that have yet to be developed.  However, in order to target applications of QMC forces in systems containing elements with high $\zeff$, there remains another highly important question of statistical efficiency, which we will focus on next.

\section{Scaling of statistical cost}
\label{sec:efficiency}

Understanding the scaling of statistical costs are key in estimating the affordability of QMC observables. To target applications of QMC forces with heavier and more numerous atoms, it is desirable to extract the dependence of statistical cost on key parameters describing the physical system under study. In this section, we examine how the statistical cost of both forces and total energies scale with the effective valence charge, $\zeff$, and estimate the implications of such scaling on the affordability of larger scale applications. We stress that our concern here is with ZVZB force estimators in a standard VMC and DMC context.  There have been recent developments improving the overall scaling of all-electron VMC\cite{Feldt2021} and lattice regularized DMC \cite{Nakano2020a}, and while these are interesting and promising areas of research, it is beyond the scope of this work.

The statistical error bar, $\sigma_e$, of an estimated in quantity in QMC given $M$ samples may be related to the intrinsic variance, $\sigma^2$, of the underlying distribution by
\begin{align}
    \sigma_e^2 = \frac{\sigma^2\tau}{M},
\end{align}
where $\tau$ is the statistical autocorrelation time of the Markov chain.

The computational cost $\mathcal{C}$ required to obtain this target error bar is simply the product of the number of samples and the time required to produce a single sample, $T_s$, which for the purposes of scaling analysis may be rendered as
\begin{align}
    \mathcal{C} &= MT_s \nonumber \\
                &\propto \sigma^2 \tau T_s.
\end{align}

For elemental molecules and solids, the cost depends on both the total number of atoms, $N_a$, and the elemental effective valence charge, $\zeff$, in a simple way that we take advantage of in constructing a cost model based on the explicit study of atoms and homo-nuclear dimers. The asymptotic scaling of the Monte Carlo step time, $T_s$, is known to be proportional to cube of the electron count, or $T_s\propto N_e^3=N_a^3\zeff^3$.  This same scaling has also been established for recent QMC force evaluation algorithms \cite{Filippi2016}. The variance of the QMC total energy is known to be extensive and this property has also recently been shown for variance of individual force components \cite{Assaraf2014}, or $\sigma^2(Na,\zeff)=N_a\sigma^2(\zeff)$.  The full scaling is actually a weighted sum between $N_e^2$ and $N_e^3$ contributions, and for many systems of interest the effective exponent may be considered to lie between 2 and 3.  While we discuss primarily the conservative case, data for other scaling scenarios are provided in the supplemental material.  The statistical autocorrelation time depends on the phase space being explored.  Since the type of Monte Carlo steps being considered include the motion of all electrons, the autocorrelation time for elemental ideal gases, polymers, and solids of various sizes is largely independent of the system size, or $\tau(N_a,\zeff)=\tau(\zeff)$. Incorporating all of these factors, we arrive at the following cost model
\begin{align}\label{eq:cost_model}
    \mathcal{C}(N_a,\zeff) \propto N_a^{\alpha+1} \zeff^\alpha \sigma^2(\zeff)\tau(\zeff),
\end{align}
where $\alpha=2$ and $\alpha=3$ correspond to the optimistic and conservative cases, respectively.
With this model, estimating the cost scaling for a range of systems is therefore reduced to obtaining $\zeff$ scaling relationships for $\tau$ and $\sigma^2$, which may reasonably be obtained from atoms and homonuclear dimers.  

In the next section, we compare our observed cost scaling relationship for the total energy to the one put forward for pseudopotential systems some time ago based on general arguments \cite{Hammond1987}.  We then obtain similar relationships for tail regression based force estimators, both with and without space warp, and discuss implications for the computational affordability of heavy element force studies.  Studies of this type are of great interest to the community, but have yet to be reported.

\begin{figure}[t]
    \centering
    \begin{tikzpicture}
    \node at ( 4.5, 0) {\includegraphics[width=8cm]{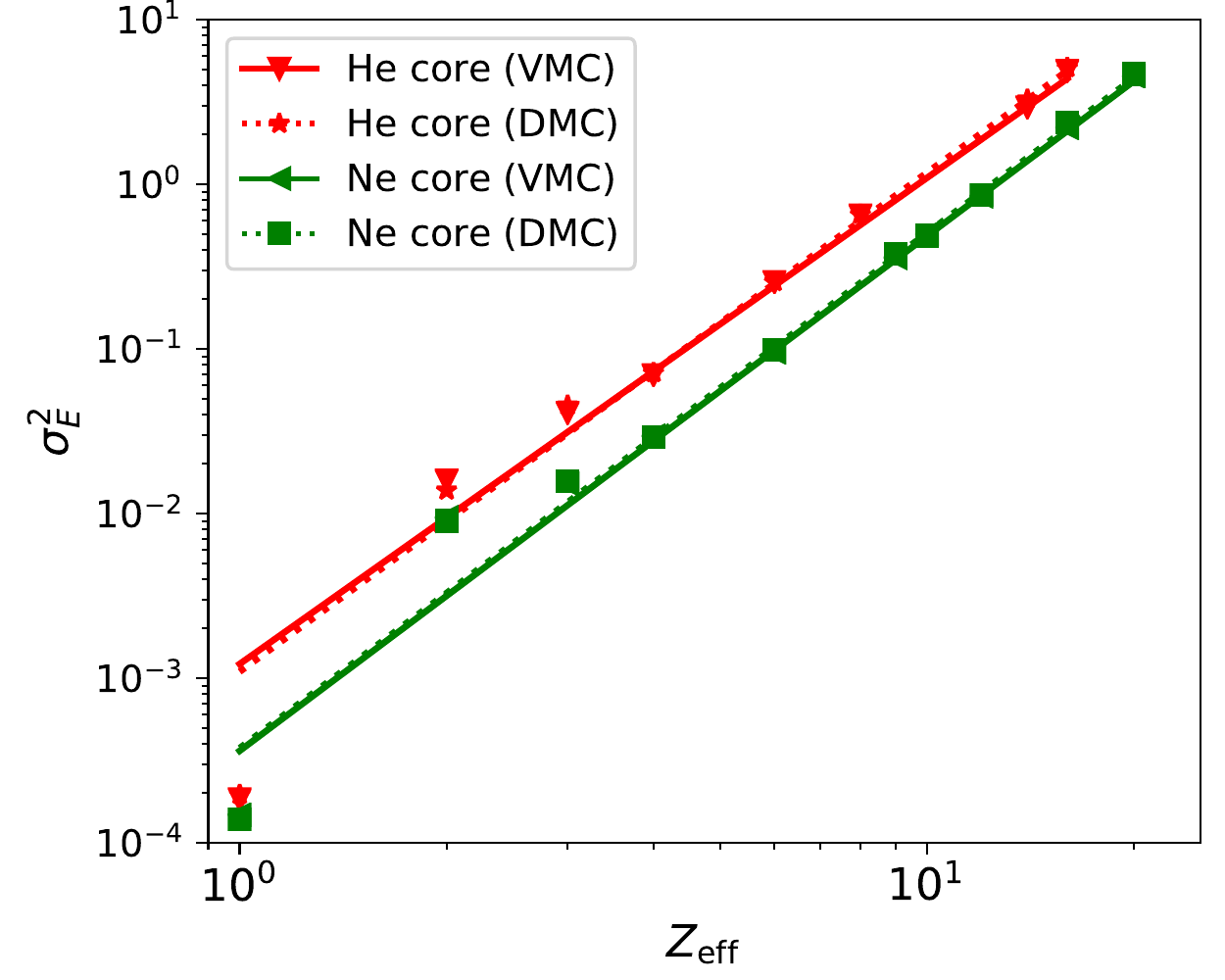}};
    \end{tikzpicture}
    \caption{Values and scaling fits of the intrinsic variance of VMC and DMC total energies for isolated atoms using pseudopotentials with either 2 (``He core'', red) or 10 (``Ne core'', green) electrons pseudized in the core. VMC and DMC results are fitted with power law scaling relations shown in solid and dashed lines respectively.}
    \label{fig:Evar_scaling}
\end{figure}

\begin{table}[h]
    \centering
    \begin{tabular}{lllll}
Quantity                & Core          & Method      & $a$            & $b$ \\ \hline \hline
$\sigma^2$              & He            & VMC         & 0.0012         & 3.0(5)     \\
                        & He            & DMC         & 0.0011         & 3.0(4)     \\
                        & Ne            & VMC         & 0.00036        & 3.1(2)     \\
                        & Ne            & DMC         & 0.00038        & 3.1(2)     \\
$\sigma^2 \tau$         & He            & VMC         & 0.00040        & 3.8(7)     \\
                        & He            & DMC         & 0.35           & 1.7(5)     \\
                        & Ne            & VMC         & 0.00024        & 3.3(3)     \\
                        & Ne            & DMC         & 0.20           & 1.9(2)     \\
    \end{tabular}
    \caption{Fitting parameters to $a\zeff^b$ of intrinsic $\sigma^2$ and autocorrelation renormalized variances $\sigma^2\tau$ of local energy from VMC and DMC simulations of isolated atoms with variable pseudized cores.  90\% confidence intervals of the exponents $b$ from Theil-Sen fits are given in parentheses. The prefactors $a$ are in the units of Ha$^2$. }
    \label{tab:fitdata_energy}
\end{table}

\subsection{Scaling of energy variance and cost}

The intrinsic variance of the total energy $\sigma^2_E$ is a well-known measure of trial wavefunction quality, and therefore a central observable in optimizing both the accuracy and statistical efficiency of a QMC simulation. Notably, $\sigma^2_E$ approaches zero as the trial wavefunction becomes exact. It is also expected to increase with $\zeff$, as the intrinsic energy scale of the semi-core region increases.

We have computed the intrinsic ($\sigma^2_E(\zeff)$) and autocorrelation adjusted ($\sigma^2_E(\zeff)\tau_E(\zeff)$) variances with VMC and DMC for a range of isolated atoms to establish empirical scaling relationships of the form $a\zeff^b$.  Helium- and neon-core potentials are treated separately.  For the helium-core series we consider elements spanning from Li to Ar ($1\le \zeff\le 16$).  A slightly larger range of $\zeff$ is covered by our selected neon core elements, which span from Na to Zn ($1\le \zeff\le 20$). 
Scaling parameters derived from robust Theil-Sen fits to the data are given in Table~\ref{tab:fitdata_energy}.  The per atom energy variance data, as well as the corresponding scaling fits, are plotted in Fig.~\ref{fig:Evar_scaling}. Numerical data for each atom can be found in Tables VI and VII in the supplemental material. The intrinsic variances for VMC and DMC agree very closely and yield similar scaling fits. The variance shifts substantially downward at fixed $\zeff$ for Ne core potentials versus the lighter He core counterparts, giving an average variance reduction of about a factor of three.  The trends with each core size agree, however, that the variance scales approximately as $\sigma_E^2 \sim \zeff^{3}$. 

By including the effects of autocorrelation, we may estimate the scaling of the computational cost, and compare our empicial findings with prior estimates made on the basis of geometric scaling arguments.  Under the optimistic assumption of $T_S\propto N_a^2\zeff^2$, Hammond \textit{et al.}\cite{Hammond1987} , estimated the cost scaling for isolated atoms ($N_a=1$) to be $\mathcal{C}\propto \zeff^{3.4}$ based on rational assumptions derived from atomic mean-field data alone.  This relationship also assumed a significant role for the statistical autocorrelation time, which is always appropriate to DMC since the diffusion occurs over small length scales.  Following the same assumptions (the optimistic cost model), we may estimate the DMC cost as $\mathcal{C}\propto \zeff^2\sigma^2_E(\zeff)\tau(\zeff)$ and we find scaling relations of $\mathcal{C}\propto \zeff^{3.7(5)}$ for He core potentials, and $\mathcal{C}\propto \zeff^{3.9(2)}$ for Ne core potentials.  Our findings generally confirm the rationalized scaling relationship of Hammond \textit{et al.}\cite{Hammond1987} for DMC.  In the case of VMC, we generate the Markov Chain using a number of Monte Carlo steps in between the evaluation of observables like the total energy and forces.  This is commonly done in practical calculations specifically to save computational costs in the observable evaluation and it also reduces or eliminates the autocorrelation, which is already small in VMC due to the large walker step distances enabled by the bare application of Metropolis Monte Carlo.  
We confirm that this procedure saturates the low auto-correlation limit, and we do not observe significant autocorrelation in the energy regardless of $\zeff$.

\begin{table}[b]
    \centering
    \begin{tabular}{lllll}
Quantity                & Core          & Method      & $a$            & $b$ \\ \hline \hline
$\sigma^2$              & He            & VMC         & 0.0017         & 5.5(6)     \\
                        & He            & DMC         & 0.0022         & 5.9(9)     \\
                        & Ne            & VMC         & 0.00010        & 5.5(4)     \\
                        & Ne            & DMC         & 0.000096       & 5.6(4)     \\
$\sigma^2 \tau$         & He            & VMC         & 0.0020         & 6.2(10)     \\
                        & He            & DMC         & 0.068          & 4.7(6)     \\
                        & Ne            & VMC         & 0.000073       & 5.7(5)     \\
                        & Ne            & DMC         & 0.026          & 4.3(3)     \\
    \end{tabular}
    \caption{Fitting parameters to $a\zeff^b$ of intrinsic $\sigma^2$ and autocorrelation renormalized variances $\sigma^2\tau$ of ZVZB forces from VMC and DMC simulations of isolated atoms with variable pseudized cores.  90\% confidence intervals of the exponents $b$ from Theil-Sen fits are given in parentheses. The prefactors $a$ are in the units of (Ha/Bohr)$^2$. }
    \label{tab:fitdata_zvzb}
\end{table}

\begin{figure}[t]
    \centering
    \begin{tikzpicture}
    \node at ( 4.5, 0) {\includegraphics[width=8cm]{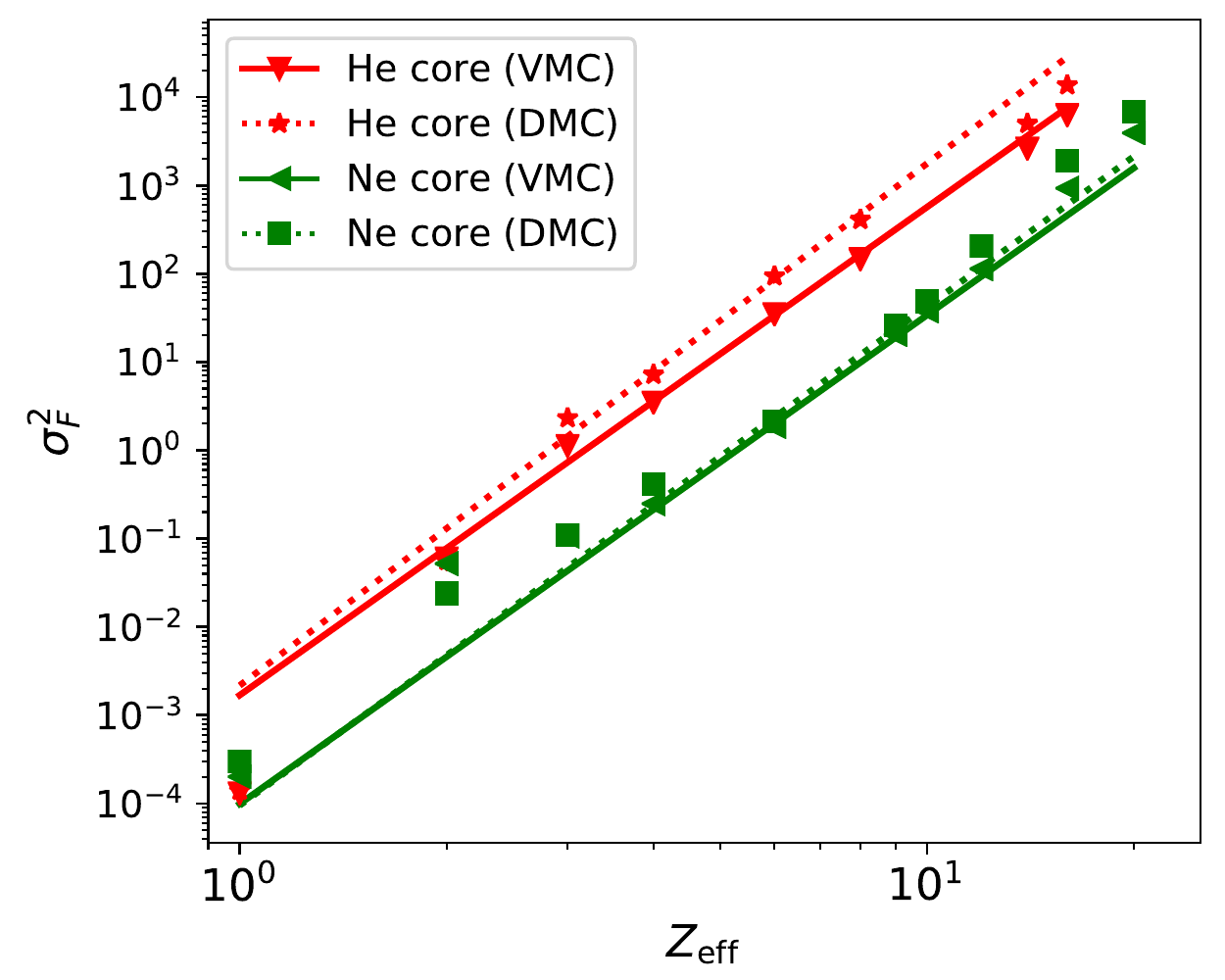}};
    \end{tikzpicture}
    \caption{Values and scaling fits of the intrinsic variance of VMC and DMC ZVZB forces for isolated atoms using pseudopotentials with either 2 (``He core'', red) or 10 (``Ne core'', green) electrons pseudized in the core. VMC and DMC results are fitted with power law scaling relations shown in solid and dashed lines respectively.}
    \label{fig:Fvar_scaling}
\end{figure}

\subsection{Cost scaling of ZVZB forces}

Having confirmed the expected scaling behavior of the QMC energy variance with respect to $\zeff$, we are now in a position to investigate the cost scaling behavior of the ZVZB force estimators.  We first consider only the intrinsic variance of the ZVZB estimator without the space warp transformation.  Variances are computed using the tail regression estimation (TRE) method, although we expect these results to closely relate to other methods to control the formally infinite variance of the bare ZVZB estimator, such as the use of guiding functions.

In Fig. \ref{fig:Fvar_scaling}, we show the scaling of the intrinsic variance of TRE based ZVZB force estimator as a function of $\zeff$ for the same set of atoms used in the total energy calculations.  It is immediately clear that the slopes of the variance versus $\zeff$ curves corresponding to different pseudopotential core sizes are extremely consistent for VMC and DMC, implying that much like the total energy, the rate of increase of the force variance with $\zeff$ generalizes for the ZVZB estimators. In both cases, the exponent of $\zeff$ is at least $5.5(5)$, which is noticeably higher than the $\zeff^3$ observed for the energy variance.  The scaling prefactor however is noticeably different.  As expected, and similar to the total energy, the force variance is highest for He core potentials, while the prefactor of the Ne core force variance drops.  However, the gain in variance reduction for Ne core over He core potentials is much larger than for the total energy.  While the variance reduced by about a factor of three for the energy, it is closer to a factor of 20 for the forces.  This gain is particularly important in the context of 3d transition metal elements, as this additional reduction in cost places them closer to current affordability. It is worth noting that the intrinsic variance scaling of the straightforward ZVZB forces will place a cap on the maximum system size that can be treated affordably.  Even in the absence of any significant difference in evaluation costs between energies and forces, which is a fair assumption given the work described in Refs.  \onlinecite{Sorella2010,Filippi2016}, the maximum system size will necessarily be smaller than the sizes reachable with energy-only QMC.

\begin{figure}[t]
    \centering
    \begin{tikzpicture}
    \node at ( 4.5, 0) {\includegraphics[width=8cm]{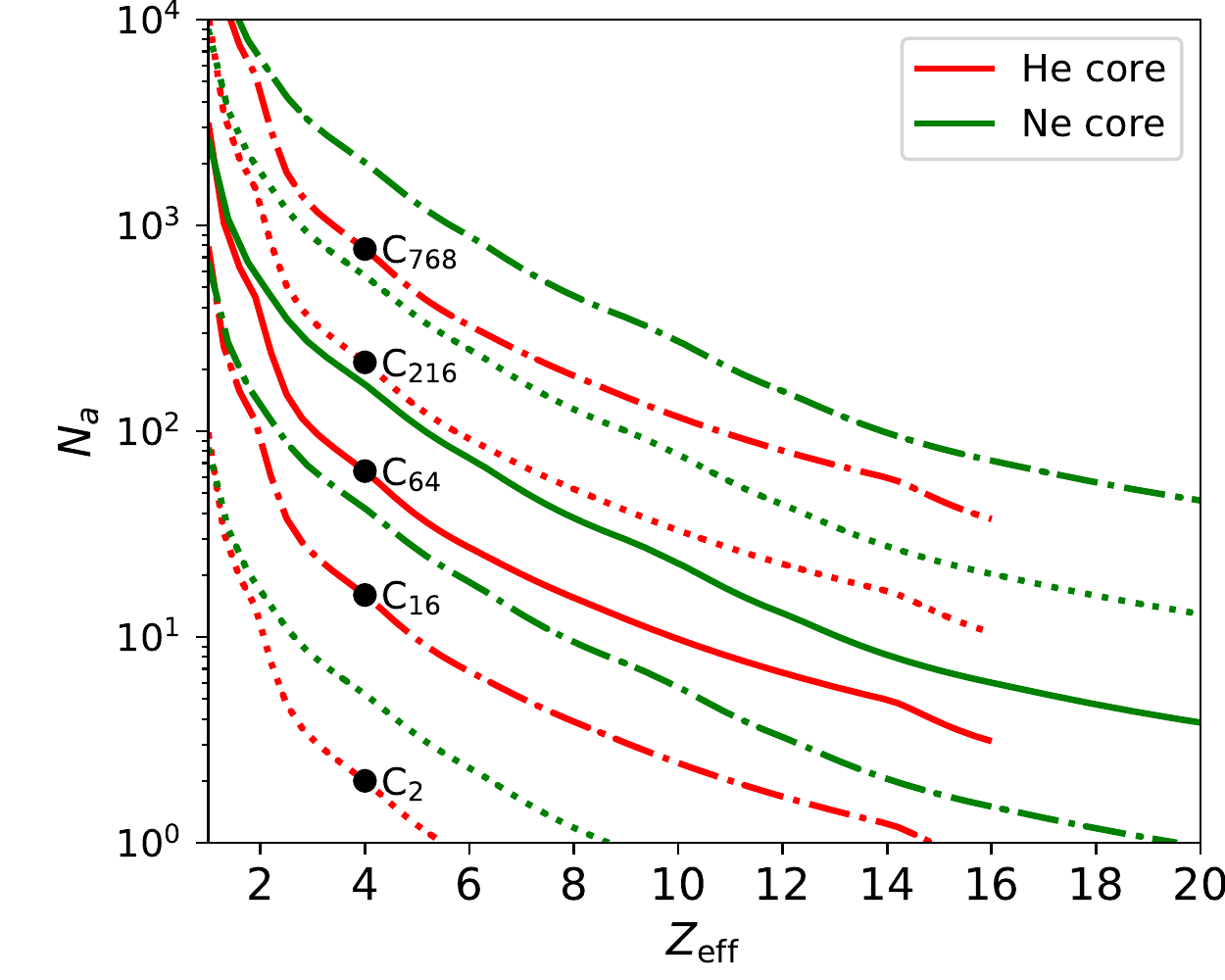}};
    \end{tikzpicture}
    \caption{Cost isocontours (Eq. \ref{eq:cost_model}) as the function of the number of atoms $N_a$ and $\zeff$, based on the $\zeff$ scaling behavior VMC ZVZB forces of isolated atoms. The cost is interpolated using Hermite splines using the observed values for He core (red) and Ne core (green) potentials separately. The cost contour values marked with black circles  correspond to 2, 16, 64, 216 and 768 carbon atoms (He-core), respectively. Adjacent He core and Ne core curves sharing a line style are contoured at identical cost values. Calculations performed at any ($N_a, \zeff$) along the paired contours are projected to be equally affordable.}
    \label{fig:cost_contour_zvzb}
\end{figure}

To see how affordable force calculations are as the system size, $N_a$,  and $\zeff$ are varied, we show the cost isocontours for VMC forces in Figure~\ref{fig:cost_contour_zvzb}. Red (green) lines are obtained from  the conservative cost model Eq.~\eqref{eq:cost_model} by using interpolated data of the intrinsic variances of He (Ne) core atoms. To facilitate the relative cost discussion, we have drawn isocontours through the cost associated with pure He-core carbon systems with atom counts of $N_a=2,16,64,216,768$ (dot, dash-dot, solid, dot, dash-dot lines respectively).  The partner green Ne-core isocontours are calculated using the same respective cost reference as the He-core carbon systems, and thus any $(N_a,\zeff)$ point on the paired contours represent identical computational affordability. The contours in Fig.~\ref{fig:cost_contour_zvzb}  demonstrate the strong tradeoff between system size and $\zeff$.  For example, if the forces for a 64 atom carbon calculation are computationally affordable with helium core potentials (solid red line), one could alternatively obtain forces with equal statistical resolution for a system containing several hundred Be atoms ($\zeff=2$), whereas a system containing more than ten Mg atoms ($\zeff=12$) might cost more.  Switching to a neon core potential (solid green curve) for Mg would bring systems about twice as large within reach. 

This overall scaling can be better understood by rearranging Eq.~\eqref{eq:cost_model} to find the dependence of $N_a$ on $\zeff$.  Assuming a force variance scaling relation $a\zeff^b$ and saturation of the autocorrelation time, we can show through simple algebra that the equally affordable system size, $N_a$, scales like $a^{1/4}\zeff^{-(3+b)/4}$ for the conservative cost model. Using $b=5.5(5)$ for VMC forces from Table~\ref{tab:fitdata_zvzb}, we see that the affordable system size scales roughly like $N_a\propto\zeff^{-2.1(1)}$.  Using an optimistic cost model with $T_s\propto N_a^2\zeff^2$ instead leads to $N_a\propto a^{1/3}\zeff^{-(2+b)/3}$, which for VMC forces is $N_a\propto\zeff^{-2.5(2)}$.  Interestingly this optimistic assumption for the Monte Carlo step time gives a more pessimistic predicted scaling relation for the system size affordability scaling.  For diffusion Monte Carlo, the effects of autocorrelation are much more significant with the cost  scaling like $\mathcal{C}\propto\zeff^{6.5(3)}$ on average. This slightly shifts the projected system size affordability scaling as $N_a\propto\zeff^{-1.9(1)}$ and $N_a\propto\zeff^{-2.2(1)}$ for the pessimistic and optimistic step time scenarios, respectively. In all cases, the affordability with respect to system size falls off rapidly with $\zeff$ with an exponent of two or slightly larger.  

This is a difficult scaling to cope with in practice, but one that still leaves many systems of modest size at higher $\zeff$ likely within reach with current computational resources.  If forces are affordable for a particular system size and $\zeff$, doubling $\zeff$ means we can only afford a quarter of the original system size.  There are gains to be had by reducing the variance prefactor $a$, but given the overall conservative $a^{1/4}$ scaling, this results in a modest gain in system size.  An order of magnitude improvement to the intrinsic force variance results in being able to simulate a system 78\% larger in size for example.  As we will show in the next section, this is possible through use of the space warp transformation.

\subsection{Impact of the space warp transformation on cost}

\begin{table}[b]
    \centering
    \begin{tabular}{lllll}
Quantity                & Core          & Method      & $a$          & $b$ \\ \hline \hline
$\sigma^2$              & He            & VMC         & 0.010        & 4.9(5)  \\
                        & He            & VMC-SW      & 0.0065       & 4.3(10)  \\
                        & He            & DMC         & 0.022        & 4.9(7)  \\
                        & He            & DMC-SW      & 0.013        & 4.2(8)  \\
$\sigma^2 \tau$         & He            & VMC         & 0.029        & 4.9(10)  \\
                        & He            & VMC-SW      & 0.0053       & 4.6(11)  \\
                        & He            & DMC         & 1.2          & 3.9(2)  \\
                        & He            & DMC-SW      & 2.1          & 3.0(4)  \\
                        
    \end{tabular}
    \caption{Fitting parameters to $a\zeff^b$ of intrinsic $\sigma^2$ and autocorrelation renormalized variances $\sigma^2\tau$ of VMC and DMC forces along the bond of homonuclear Helium-core dimers based on tail-regression ZVZB forces with or without space warp.  90\% confidence intervals of the exponents $b$ from Theil-Sen fits are given in parentheses.  The prefactors $a$ are in the units of (Ha/Bohr)$^2$.}
    \label{tab:fitdata_homonuclear}
\end{table}

\begin{figure}[t]
    \centering
    \begin{tikzpicture}
    \node at ( 4.5, 0) {\includegraphics[width=8cm]{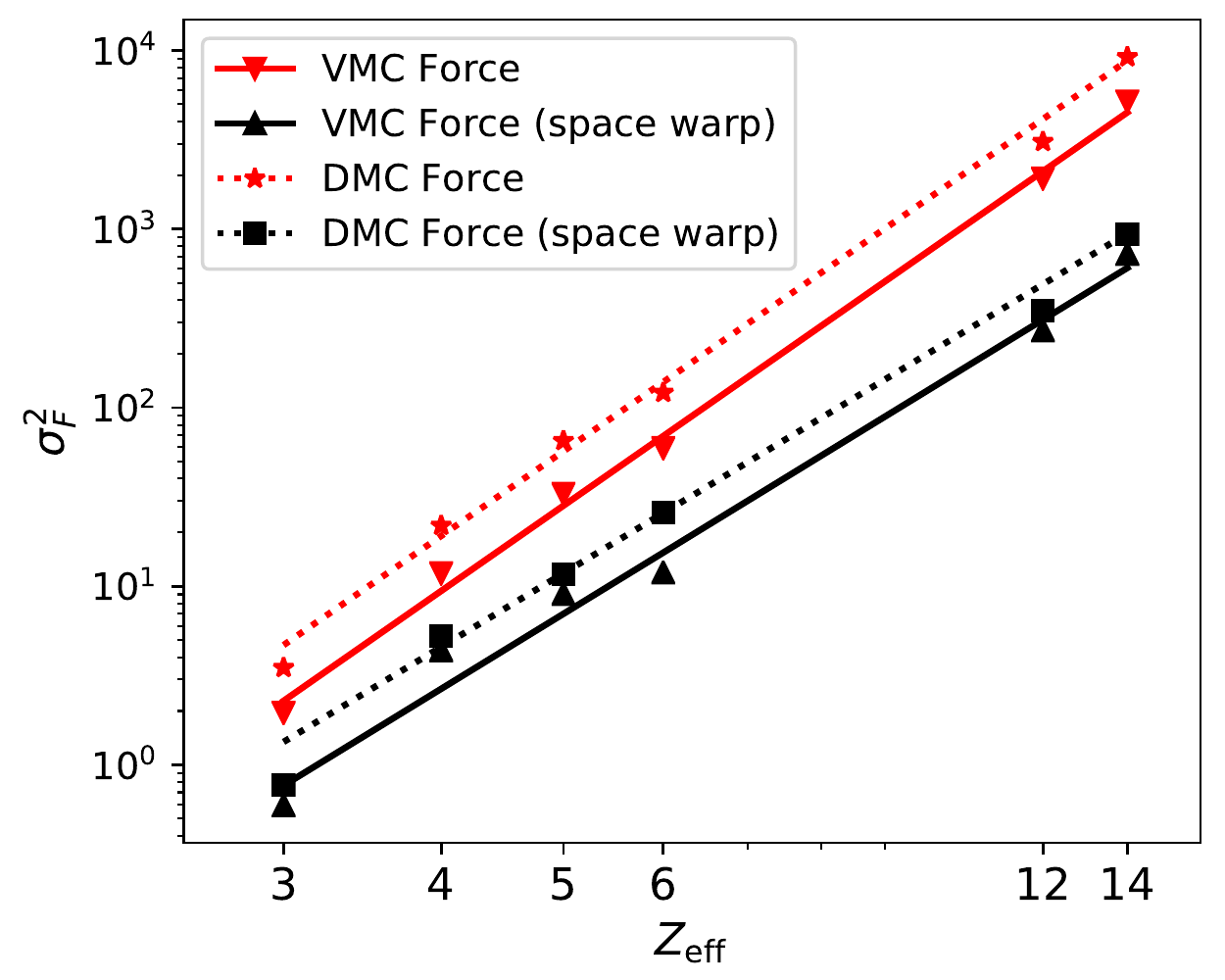}};
    \end{tikzpicture}
    \caption{Values and scaling fits of the intrinsic variance of VMC (solid) and DMC (dotted) forces along the bond direction of homonuclear dimers with (black) or without (red) space warp, using He core pseudopotentials with variable $\zeff$. VMC and DMC results are fitted with power law scaling relations shown in solid and dashed lines respectively.}
    \label{fig:Fvar_scaling_sw}
\end{figure}

Most current applications of QMC forces utilize the space warp transformation \cite{Filippi2000} to reduce the variance of the force estimates.  Here we investigate the variance properties of the regularized tail regression estimators when used on top of force samples obtained with the space warp transformation. Since the space warp transformation formally reduces the variance of atomic systems to zero\cite{Sorella2010}, here we obtain force variances and comparative statistical costs for a set of covalently bonded homonuclear dimers.  The choice of covalent bonding was made to mitigate difficulties in comparison across differing bonding types as the variations in bond lengths, e.g. between covalent and dispersively bonded atoms, result in variations in the space warp force variance. In particular, for species that are weakly bonded as dimers (e.g. metal atoms), the expanded bond lengths result in a much larger variance reduction for space warp forces since they are nearer to the formally zero variance atomic limit for space warp than would be seen in larger atomic clusters or solids.  With this in mind, we have chosen the B$_2$, C$_2$, N$_2$, O$_2$, Si$_2$, and S$_2$ molecules and He core pseudopotentials ($3\le\zeff\le 14$). In this selected setting, we can closely compare VMC forces with or without space warp in a fashion that is more representative of larger scale condensed phase systems that are the preferred target of production force applications.

\begin{figure}[t]
    \centering
    \begin{tikzpicture}
    \node at ( 4.5, 0) {\includegraphics[width=8cm]{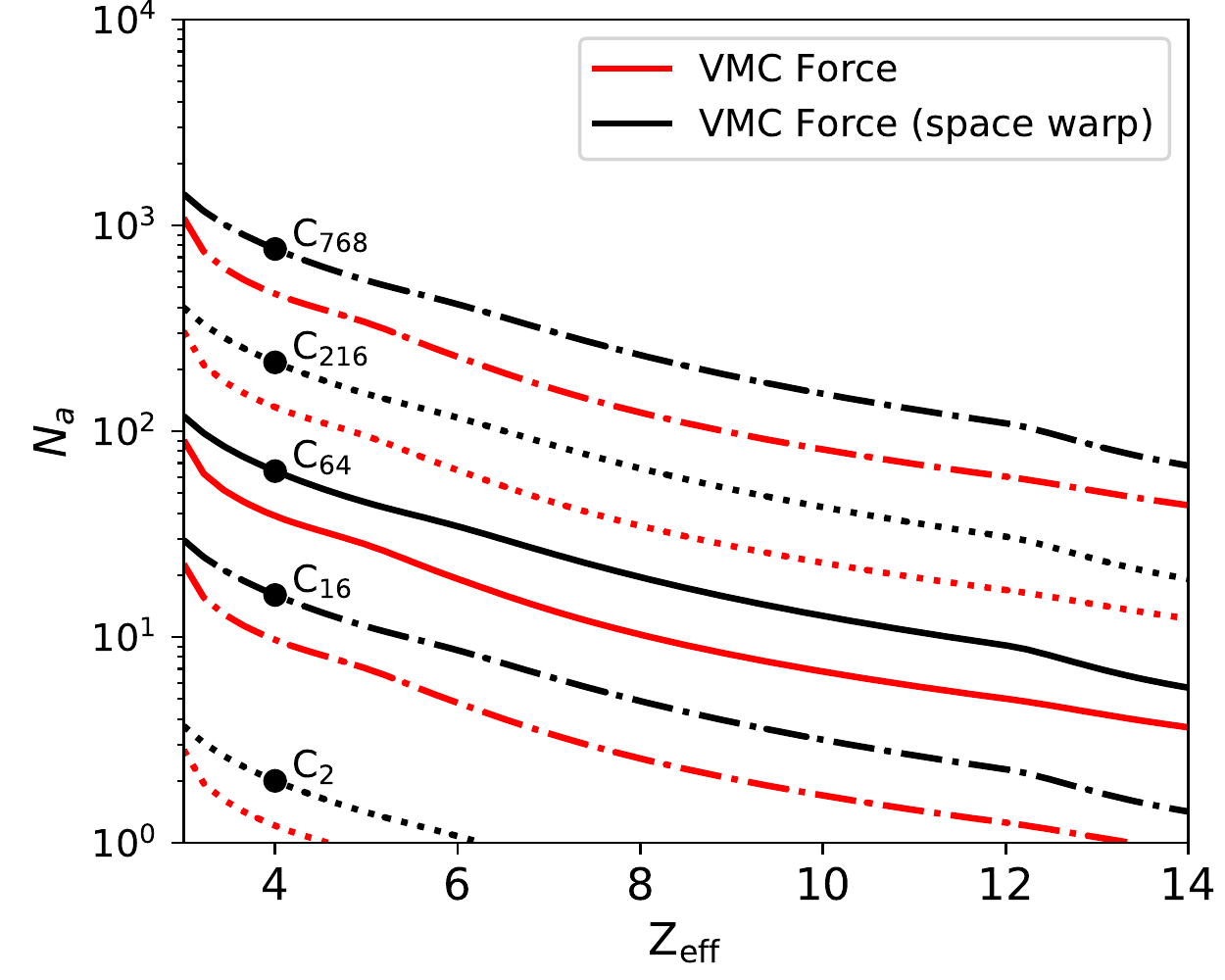}};
    \end{tikzpicture}
    \caption{Cost isocontours similar to Figure~\ref{fig:cost_contour_zvzb} are plotted for VMC forces with and without space warp. The data are based on homonuclear dimers from $\zeff=3$ up to $\zeff=14$ (He-core) as a function of the number of atoms ($N_a$) and $\zeff$. The cost contours for the forces with space warp are shown in black.  The cost contour values marked with black circles correspond to 2, 16, 64, 216 and 768 carbon atoms (with space warp). Adjacent black and red curves sharing a line style are contoured at identical cost values. Calculations performed at any ($N_a, \zeff$) along the paired contours are projected to be equally affordable.}
    \label{fig:cost_contour_zvzb_sw}
\end{figure}

Force variances for single atoms in each dimer from VMC and DMC simulations are plotted in Fig.~\ref{fig:Fvar_scaling_sw} as a function of $\zeff$ for tail regression estimators both without (red) and including the space warp transformation (black).  The forces in the dimers show approximately twice the variance as in the atomic setting, consistent with the expected extensivity of the force variance\cite{Assaraf2014}, while retaining a similar $\zeff$ scaling exponent (roughly  4.9(5) for dimers and  5.5(6)-5.9(9) for atoms with VMC and DMC respectively) as seen in Table~\ref{tab:fitdata_homonuclear}.  The space warp force estimates retain a high polynomial scaling with $\zeff$, though possibly with a slightly reduced exponent ( $\approx$ 4.3(1.0) for VMC and $\approx$ 3.0(4) for DMC), demonstrating a consistent improvement over tail regression alone, with a growing advantage at higher $\zeff$.  Using space warp reduces the statistical variance of the ZVZB forces by a factor of three for the lightest elements (B,C) and a factor of 7 for the heaviest elements (Si, S).

Similar to the atomic case, we assess isocontours of the projected computational cost for the VMC forces with (black) or without (red) space warp now for dimers as shown in Fig.~\ref{fig:cost_contour_zvzb_sw}.  As before, five selected isocontours are selected for carbon-based systems of various sizes, since a significant number of QMC force applications to date have featured systems of this type.  For example, in the recent work by Nakano \textit{et al.} \cite{Nakano2021}, the phonon spectrum of diamond was computed via the frozen phonon method and space-warp-based forces using 64 atom supercells for the majority of the calculations as well as a calculation of a 216 atom supercell as a convergence check.  As can be seen in Fig.~\ref{fig:cost_contour_zvzb_sw}, the addition of space warp is essential to reach these system sizes.  Without space warp, we estimate  that system sizes on the order of only 130 atoms could be afforded instead.  The cost advantages of space warp carry through to larger $\zeff$ as well.  On this same cost contour, which is representative of the scale accessed in recent works,  forces with the same statistical resolution for He core systems containing \textit{e.g.} 64 oxygen or 22 Mg atoms could be obtained without the space warp transformation (conservative cost model).  Including space warp, the estimated accessible system sizes for increase to 116 oxygen or 42 Mg atoms.  This underscores the importance of the space warp transformation, and also any future variance reduction techniques, in reaching real applications of QMC forces to higher $\zeff$ systems in an affordable manner.

\section{Conclusions}
\label{sec:conclusions}

We have calculated VMC and DMC forces on selected atoms and dimers using the standard ZVZB force estimators and Slater-Jastrow trial wavefunctions, with an aim to assess the performance and statistical properties of these force estimators versus the pseudopotential effective charge, $\zeff$. Our results for VMC, mixed DMC, and hybrid DMC ZVZB forces are of similar quality to prior works, as judged on the basis of derived molecular properties such as bond lengths and vibration frequencies of dimers.  The performance of VMC ZVZB forces for mean field wavefunctions rapidly degrade in quality as $\zeff$ increases, implying that techniques such as orbital optimization will need to become routine to obtain accurate VMC forces. Clearly improvements to DMC force estimators are desirable to recover forces of similar quality to the DMC PES, which is already of sufficient quality for most applications at the single determinant trial wavefunction level.  It will be of significant interest to see if DMC estimators that are more advanced than the simple mixed and hybrid DMC estimators considered here, perhaps such as those proposed in Refs. \onlinecite{Badinski2010,Moroni2014}, can reproduce the quality of the DMC PES for the range of $\zeff$ we have studied.

Through a careful accounting of the statistical variance, we have established that the computational cost ZVZB QMC forces using recently introduced tail regression techniques, and both with and without the space warp transformation, approximately obey power laws in $\zeff$.  We find that the cost of ZVZB forces grows more rapidly than that of the total energy, with an overall scaling of approximately $\zeff^{6.5(3)}$ for DMC forces versus the $\zeff^{3.8(3)}$ for the total energy under optimistic assumptions. We have also established an affordability relationship between system size and $\zeff$ that is insensitive to pseudopotential core size and optimistic or pessimistic assumptions about algorithmic time complexity.  In general we find that, at constant computational cost and constant statistical resolution, the accessible system size scales approximately as $\zeff^{-2}$ which serves to inform the scope of QMC force applications that may reasonably be approached with fixed computational resources available either today or in the future. These results underscore the challenge of obtaining QMC forces affordably for systems containing many high $\zeff$ atoms, such as transition metal oxide solids, and the general importance of continued development of variance reduction techniques.

\section{Acknowledgements}

RC would like to thank Ye Luo for help with the 3-body Jastrow gradient implementation. This research has been provided by the US Department of Energy, Office of Science, Basic Energy Sciences, Materials Sciences and Engineering Division, as part of the Computational Materials Sciences Program and Center for Predictive Simulation of Functional Materials. This research used resources of the Compute and Data Environment for Science (CADES) at the Oak Ridge National Laboratory, which is supported by the Office of Science of the U.S. Department of Energy under Contract No. DE-AC05-00OR22725.

This manuscript has been authored in part by UT-Battelle, LLC, under contract DE-AC05-00OR22725 with the US Department of Energy (DOE). The US government retains and the publisher, by accepting the article for publication, acknowledges that the US government retains a nonexclusive, paid-up, irrevocable, worldwide license to publish or reproduce the published form of this manuscript, or allow others to do so, for US government purposes. DOE will provide public access to these results of federally sponsored research in accordance with the DOE Public Access Plan (http://energy.gov/downloads/doe-public-access-plan).

Sandia National Laboratories is a multimission laboratory managed and operated by National Technology \& Engineering Solutions of Sandia, LLC, a wholly owned subsidiary of Honeywell International Inc., for the U.S. Department of Energy’s National Nuclear Security Administration under contract DE-NA0003525. This paper describes objective technical results and analysis. Any subjective views or opinions that might be expressed in the paper do not necessarily represent the views of the U.S. Department of Energy or the United States Government.

The data that support the findings of this study are openly available in The Materials Data Facility at [link to be provided upon acceptance].

\section{Supplemental material}

Reference to supplemental material. 

\section{References}
\bibliography{references}

\end{document}


\title{Supplemental material: Towards Quantum Monte Carlo Forces on Heavier Ions: Scaling Properties}
\author{Juha Tiihonen}
\affiliation{ 
Materials Science and Technology Division, Oak Ridge National Laboratory, Oak Ridge, TN 37831, United States of America
}%
\author{Raymond C Clay III}
\affiliation{ 
Sandia National Laboratories, Albuquerque, NM 87185, United States of America
}%
\author{Jaron Krogel}
\affiliation{ 
Materials Science and Technology Division, Oak Ridge National Laboratory, Oak Ridge, TN 37831, United States of America
}

\date{\today}

\maketitle

\section{Tail-regression estimation}

To analyze heavy-tailed QMC force data, we use a tail-regression estimator toolbox (TREAT) from Ref.~\cite{Rios2019}. Detailed theoretical foundations and advanced techniques for tail-regression are given in the original publication. For the purposes of this work, the task is to provide or estimate four essential parameters for left and right tail of the form \cite{Rios2019}
\begin{align}
    P(A) = \sum_{n=0}^N c_n |A-A_0|^{-\mu -n \Delta}.
\end{align}
The parameters corresponding to $\mu$, $\Delta$ and $N+1$ are called \textit{mu}, \textit{dmu}, and \textit{nparams}, respectively. A fourth parameter, \textit{mlogq}, is the negative logarithm of the quantile of input data defining the axial cutoff between samples and the fit. With these parameters set, TREAT can perform the regression and provide statistical analysis of the data. 

\begin{figure}[h]
    \centering
    \begin{tikzpicture}
    \node at ( 0, 0) {\includegraphics[width=8.0cm]{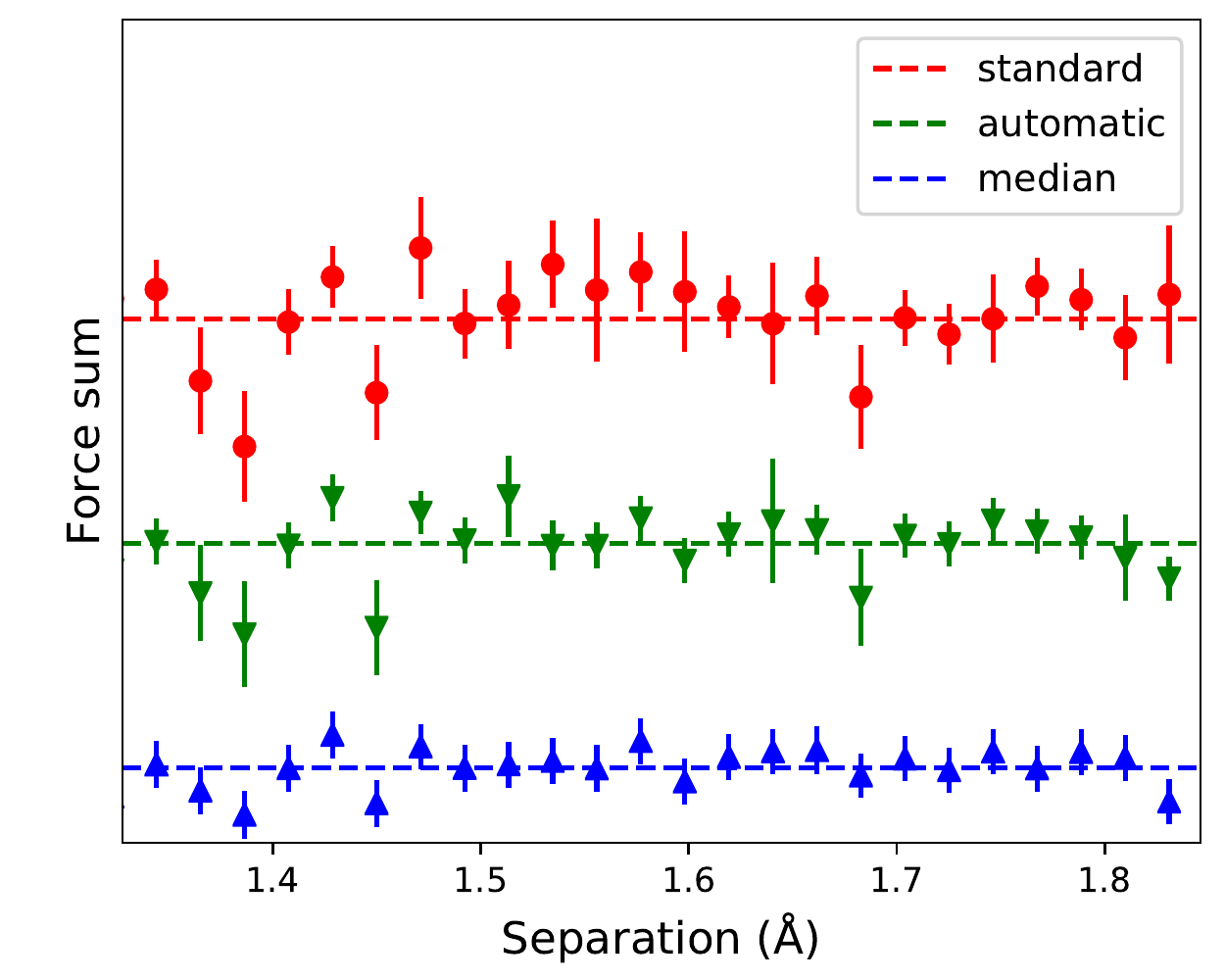}};
    \end{tikzpicture}
    \caption{Sum of forces in TiO dimer using VMC with PBE0 orbitals and different error estimation methods: standard, TREAT (automatic) and TREAT (median). Dashed lines indicate the targeted zeros of forces.}
    \label{fig:TiO_treat}
\end{figure}

Let us briefly outline three approaches or \textit{modes} for assessing these parameters: \textit{manual}, \textit{automatic} and \textit{median}. In the \textit{manual} mode, one supplies them by hand, thus enabling enforcement of symmetries, asymptotic behavior and other theoretical constraints to the numeric data. However, such information is in practice unavailable or difficult to apply to QMC data that has been \textit{reblocked} to eliminate sample autocorrelation, or for other practical reasons. Therefore, we mostly use \textit{automatic} assessment of parameters, which brackets common ranges of \textit{mu} and \textit{nparams} to provide as accurate regression of the tail as possible, requiring minimal expert supervision.  The automatic assessment is used exclusively to assess isolated atoms and dimers at the equilibrium. For dimers along the binding curve we use so-called \textit{median} mode to increase robustness of the auto-assessment. In the median mode TREAT is run three times: first to assess median \textit{mu} and \textit{dmu} over the binding curve, second to assess median \textit{mlogq} and \textit{nparams}, and finally fitting and evaluating the tail. The stabilizing effects of automatic and median modes are illustrated in Fig.~\ref{fig:TiO_treat} with the VMC net force of TiO. The standard means and error bars have large variability due to occasional spikes in the data. Tail-regression estimates in the automatic mode are usually more uniform, but sometimes fail the assessment due to insufficient data. The median mode has a steadier performance.

\begin{figure}[h]
    \centering
    \begin{tikzpicture}
    \node at (0, 0) {\includegraphics[width=8cm]{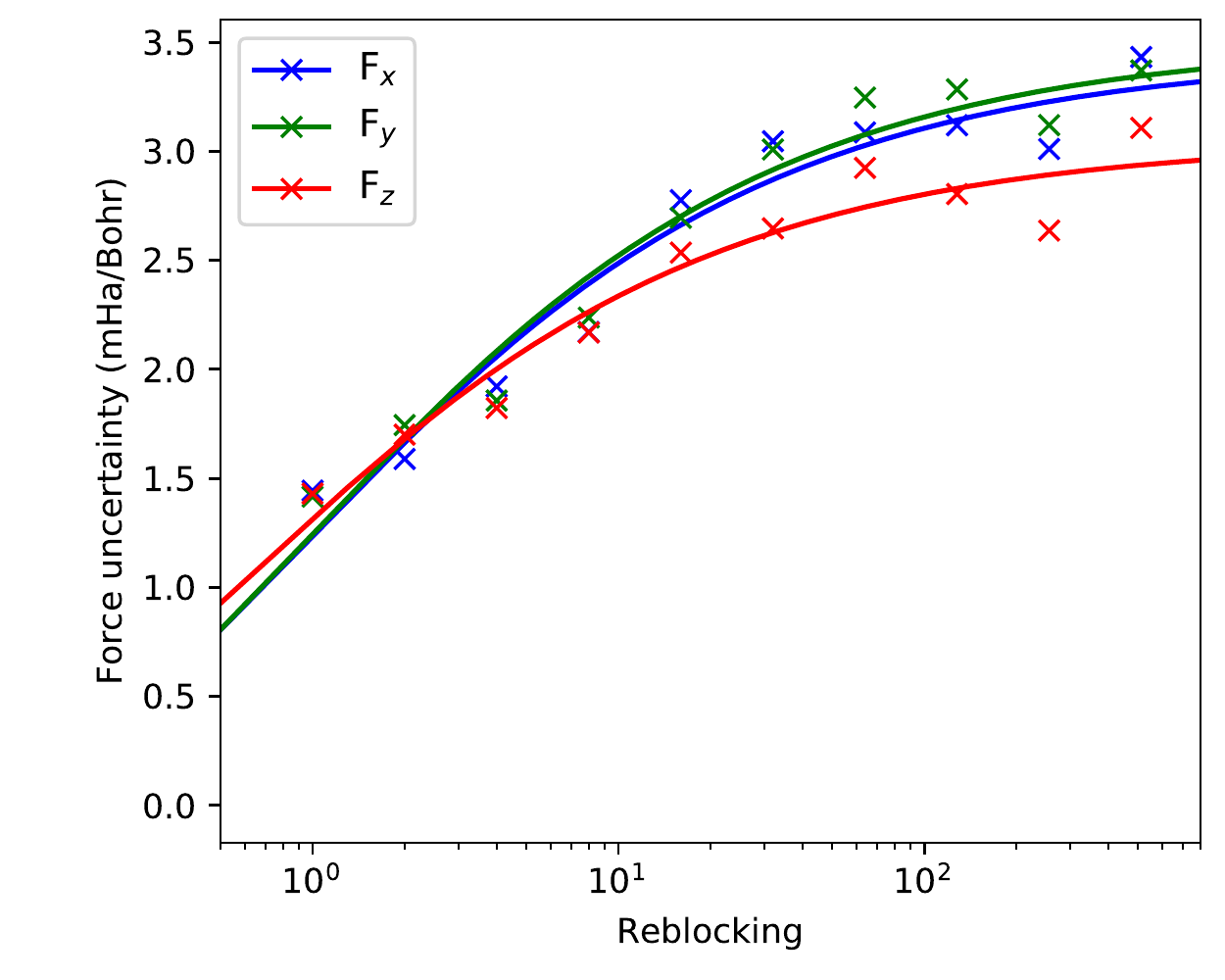}};
    \end{tikzpicture}
    \caption{Example of converging the VMC ZVZB-TRE uncertainties of Ti atom with reblocking. The exponential form usually fits well the transient region $R < \tau$, but is also resilient to large uncertainties of $\sigma_e$ at high $R$.}
    \label{fig:reblocking}
\end{figure}

The sample autocorrelation time $\tau$ is a measure of how many steps apart the samples become effectively uncorrelated. Therefore, the true statistical uncertainty, assuming standard distribution, is given by
\begin{equation}
    \sigma_e = \sqrt{\frac{\tau}{N-1}} \sigma.
\end{equation}
Estimation of $\tau$ in complex simulations can be challenging, and while TREAT can diagnose $\tau$ in terms of the standard estimator of uncertainty, it does not rectify the tail-regression uncertainty but assumes uncorrelated samples. In this work we use reblocking technique to estimate the autocorrelation. That is, starting from a relatively large set of input data (\textit{e.g.} 1-8 $\times 10^6$ samples with $36$ cores for VMC and 1-5 $\times 10^5$ samples over $1000$ walkers for DMC), we produce a series of data sets taking averages of $R$ subsequent samples, where $R$ have been chosen to be powers of 2. We then fit the uncertainties $\sigma_e$ obtained with TREAT from the reblocked data as a function of $R$ to a simple form
\begin{equation}
    \sigma_e(R) = a \exp(-bR^{-1/2}),
\end{equation}
where $a$ and $b$ are positive coefficients. The fit converges to $a$ with $R\rightarrow \infty$, as shown in Fig.~\ref{fig:reblocking}, giving the statistically true uncertainty. Estimate of the original sample auto-correlation is given by
\begin{equation}
    \tau = \frac{a}{\sigma_e(R=1)},
\end{equation}
which is used to estimate intrinsic variances from TREAT uncertainties.

\section{Curve fitting and integrated quantities}

Curve fitting is used to study three integrated quantities based on binding curves of dimers: equilibrium bond length $r$, vibrational frequency $\omega$ and self-consistency of the force $\Delta$, that is, the deviation from the PES gradient. The first two are done separately by fitting either energies, $E_{\mathrm{fit}}$ or forces, $F_{\mathrm{fit}}$. The equilibrium bond length is where $F_{\mathrm{fit}}=0$ ($\mathrm{d}E_{\mathrm{fit}}/\mathrm{d}r = 0$). The vibrational frequency depends on the force constant $k$, which is the derivative of $F_{\mathrm{fit}}$ or the second derivative of $E_{\mathrm{fit}}$ at the equilibrium. The fitted range is approximately $r_{\mathrm{expt}} \pm 20\%$ for each dimer. The range of separations used in the calculation of $|\bar{\Delta}|$ is $r_{\mathrm{expt}} \pm 15\%$, because the residual fitting error of $\tfrac{\mathrm{d} E_{\mathrm{fit}}}{\mathrm{d}r}$ tends to grow near the boundaries. 

The functional form of the fit can be important and has been studied in the past, \textit{e.g.}, in Refs.~\onlinecite{Badinski2008a,Zen2012,Luo2014}. Choosing the optimal fitting procedure depends on many parameters, such as the grid point density, spatial range of the grid, and the purpose of the fit.  We will use order $n=6$ polynomials to fit to the PES and $n=5$ to forces, although using a different forms, including the Morse potential, might lead to modest but insubstantial changes in the results. 

Error estimation of the integrated quantities is done by using the resampling technique. A number of independent random samples (1000) are generated based on the means and errorbars at each grid point, and the fluctuation propagated to the integrated quantities defines their statistical uncertainties.

\section{Validity of self-consistent VMC forces}

\begin{figure}[h]
    \centering
    \begin{tikzpicture}
    \node at (0, 0) {\includegraphics[width=8cm]{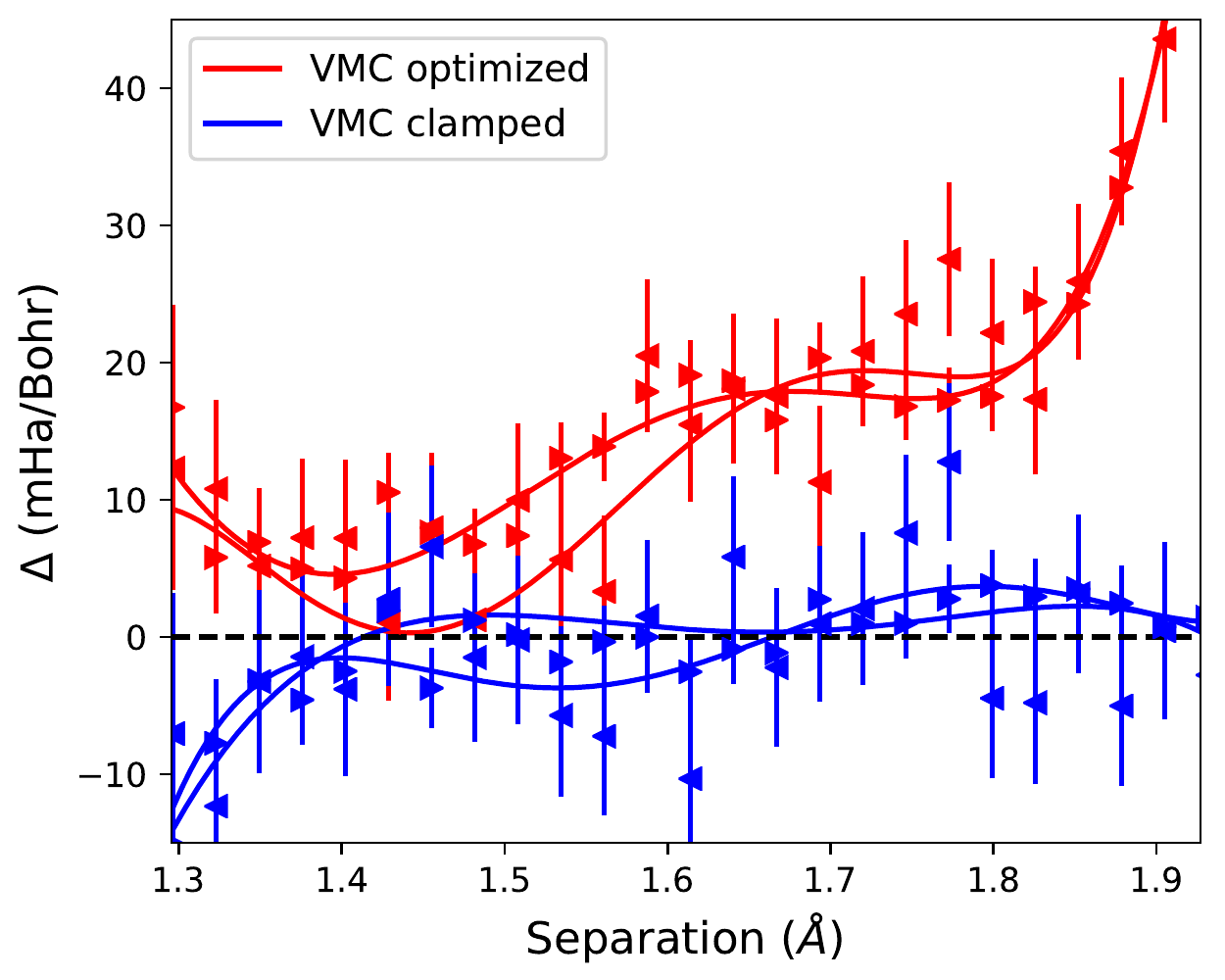}};
    \end{tikzpicture}
    \caption{Self-consistency error of the VMC ZVZB forces of FeO based on optimized 
    (red) or clamped (blue) trial wavefunctions. The distinct values of the forces on Fe and O are indicated by $\blacktriangleleft$ and $\blacktriangleright$, respectively.}
    \label{fig:FeO_SC_clamped}
\end{figure}

Two curves of VMC ZVZB self-consistency error $\Delta$ of FeO are presented in Fig.~\ref{fig:FeO_SC_clamped}. The first curve (red) is such as elsewhere in this work, where the Jastrow factor of the trial wavefunction is separately optimized in each geometry. This leads to systematic inconsistencies between forces and energies, due to missing implicit derivatives. The second curve (blue) is based on a clamped trial wavefunction, \textit{i.e.}, with orbitals and Jastrow parameters optimized at equilibrium separation and held fixed in other geometries. The lack of systematic errors beyond statistical fluctuation validates the implementation of the forces.

\FloatBarrier

\section{Numerical data for estimated bond lengths and vibrational frequencies}

\begin{figure*}[t]
    \centering
    \includegraphics[width=8cm]{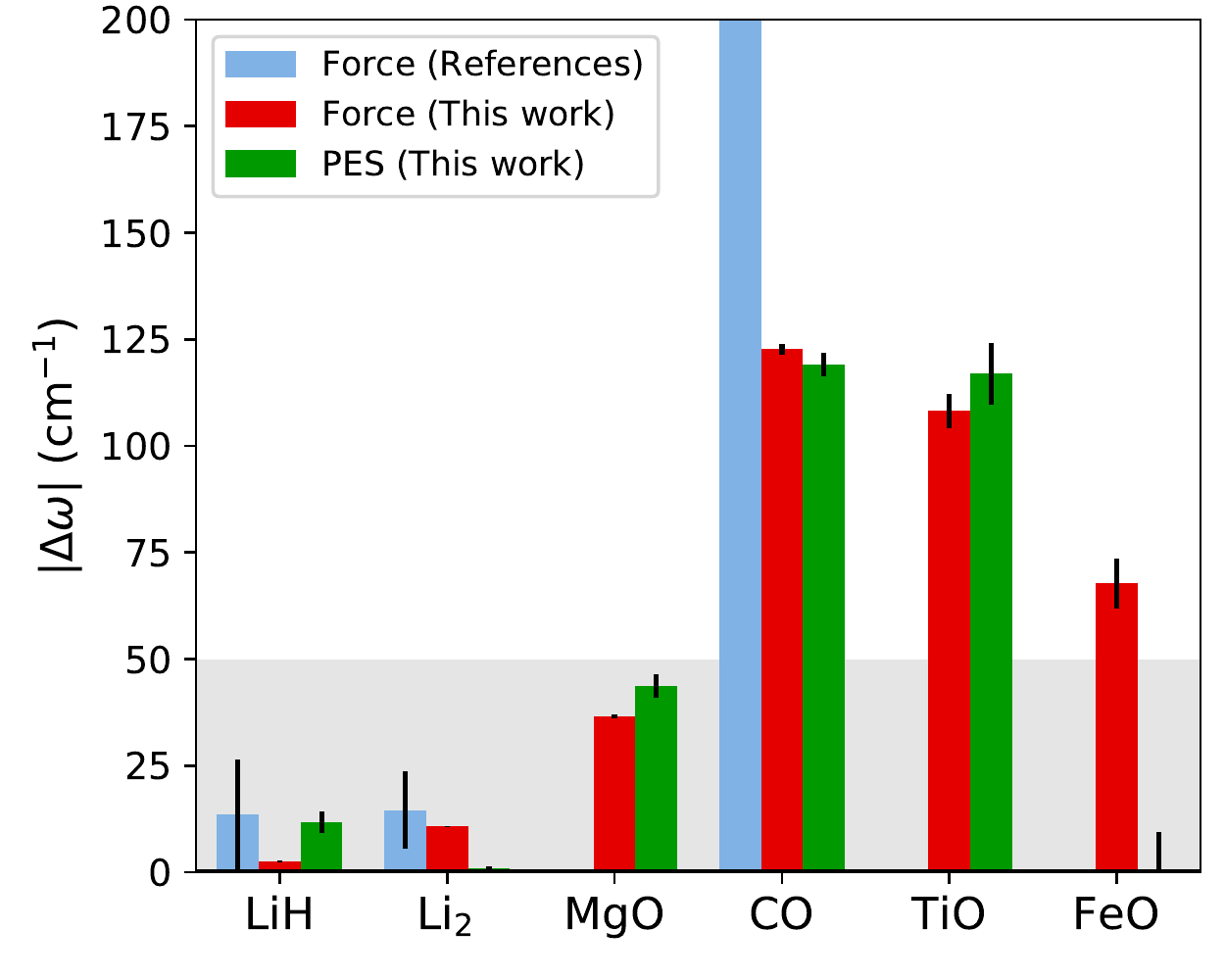}
    \includegraphics[width=8cm]{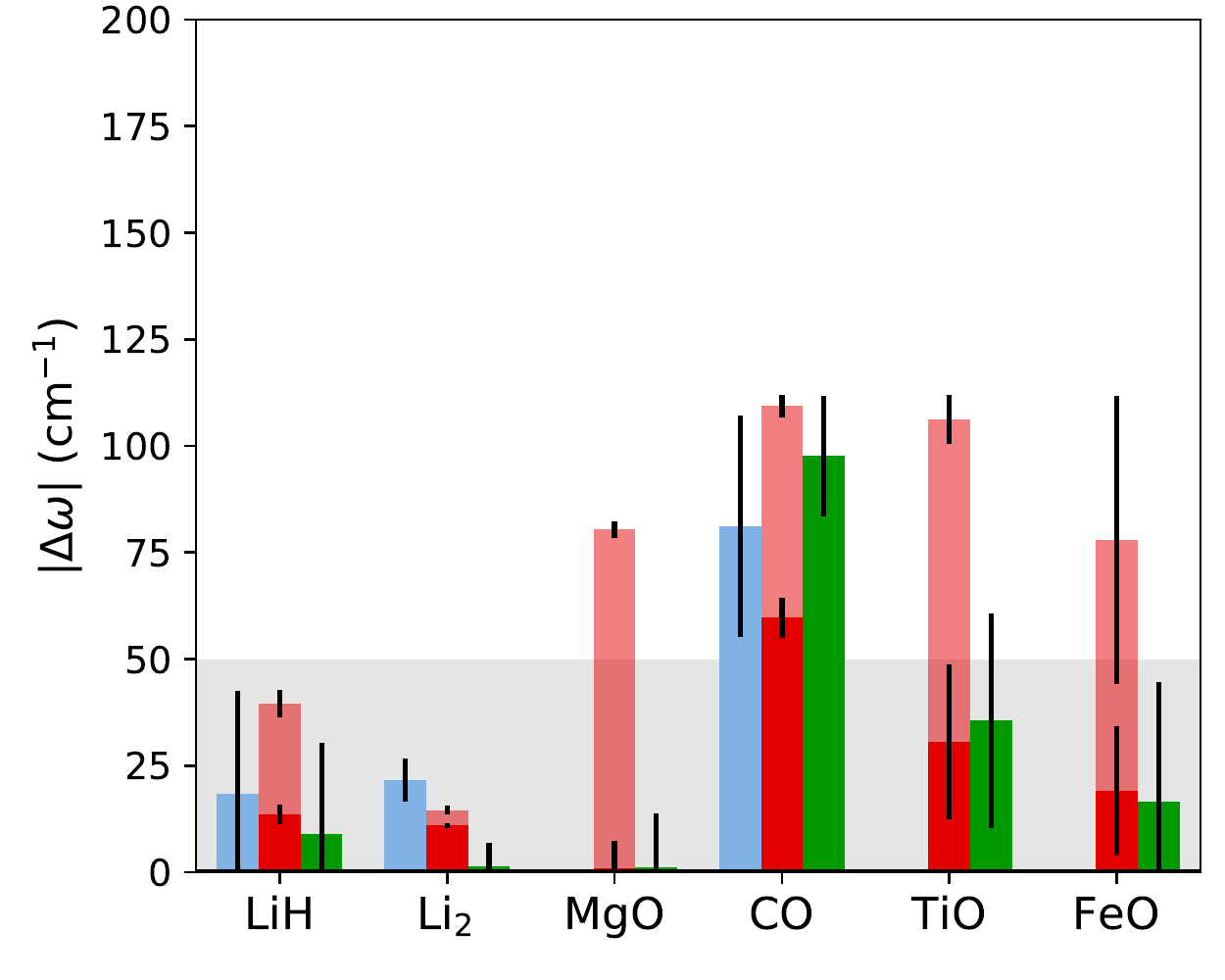}
    \caption{Absolute deviations from experiment of estimated vibrational frequencies $\omega$ from VMC (top) and DMC (bottom). Bars for each system from right to left represent QMC PES, QMC ZVZB force, and the median value out of available QMC force values from Refs.~ \onlinecite{Filippi2000,Assaraf2003, Casalegno2003,Chiesa2005,Lee2005,Badinski2007}. Solid and opaque color bars of the DMC forces correspond respectively to the better and worse results out of the left and right ionic forces, highlighting the net force induced. Gray marks acceptable accuracy. }
    \label{fig:W_comparison}
\end{figure*}

Table~\ref{tab:rwdata} contains numerical data for the estimated bond lengths $r$ and vibrational frequencies $\omega$ based on VMC and DMC (PBE0 orbitals) energies and ZVZB forces including hybrid estimates from this work, but also benchmark CCSD(T) calculations from Refs.~\cite{NIST2019,Miliordos2010,Sakellaris2011} and experimental results from Refs.~\cite{CRChandbook,Ram1999,Allen1996}. Deviations of the bond lengths from experiment are discussed and presented in Sec. IV of the main article. Deviations of the vibrational frequencies are presented in Fig.~\ref{fig:W_comparison}, including median estimates from available prior works of QMC forces. Detailed comparison of the prior QMC works is given later in the supplemental material.

\begin{table*}[t]
    \centering
    \begin{tabular}{llllllll}
    & Method          & $r_E$           & $r_F$ (left)    & $r_F$ (right)   & $\omega_E$      & $\omega_F$ (left) &  $\omega_F$ (right)  \\ \hline\hline
LiH & CCSD(T)\cite{NIST2019}         & 1.5963          &                 &                 & 1412.0          &                 &                   \\
    & PBE0            & 1.61106         &                 &                 & 1377.1          &                 &                   \\
    & VMC             & 1.6002(2)       & 1.59750(5)      & 1.59746(8)      & 1417(3)         & 1408.0(0.3)     & 1408.3(0.4)       \\
    & DMC             & 1.602(2)        & 1.6026(3)       & 1.6003(5)       & 1400(30)        & 1389(2)         & 1392(3)           \\
    & hybrid          &                 & 1.6073(6)       & 1.6036(10)      &                 & 1366(4)         & 1375(5)           \\
    & Expt\cite{CRChandbook}            & 1.5948994       &                 &                 & 1405.5          &                 &                   \\
    \\
Li$_2$ & CCSD(T)\cite{NIST2019}         & 2.667           &                 &                 & 345.0           &                 &                   \\
    & PBE0            & 2.75263         &                 &                 & 333.9           &                 &                   \\
    & VMC             & 2.7113(3)       & 2.71630(8)      & 2.71522(8)      & 350.5(0.4)      & 340.56(6)       & 341.32(7)         \\
    & DMC             & 2.708(3)        & 2.7145(7)       & 2.7141(7)       & 350(6)          & 340.4(0.5)      & 338.8(0.5)        \\
    & hybrid          &                 & 2.7123(14)      & 2.7123(14)      &                 & 338.3(1.0)      & 336.8(1.1)        \\
    & Expt\cite{CRChandbook}            & 2.6732974       &                 &                 & 351.41          &                 &                   \\
    \\
MgO & CCSD(T)\cite{NIST2019}    & 1.7536          &                 &                 & 798.0           &                 &                   \\
    & PBE0            & 1.7308          &                 &                 & 812.7           &                 &                   \\
    & VMC             & 1.7219(5)       & 1.71864(11)     & 1.7185(6)       & 829(3)          & 821.7(0.4)      & 825(2)            \\
    & DMC             & 1.751(2)        & 1.7395(3)       & 1.7179(8)       & 784(13)         & 765.1(0.9)      & 805(4)            \\
    & hybrid          &                 & 1.7635(7)       & 1.720(2)        &                 & 705(2)          & 786(7)            \\
    & Expt\cite{CRChandbook}            & 1.7481687       &                 &                 & 785.21          &                 &                   \\
    \\
CO  & CCSD(T)\cite{NIST2019}            & 1.1325          &                 &                 & 2171.0          &                 &                   \\
    & PBE0            & 1.11852         &                 &                 & 2249.4          &                 &                   \\
    & VMC             & 1.11340(11)     & 1.11147(9)      & 1.1113(2)       & 2289(3)         & 2292.5(1.3)     & 2296(3)           \\
    & VMC-J3          & 1.11716(12)     & 1.11350(10)     & 1.1137(2)       & 2268(4)         & 2272(2)         & 2270(3)           \\
    & VMC-J3+oo.      & 1.11718(11)     & 1.11581(9)      & 1.1156(2)       & 2259(3)         & 2240.0(1.2)     & 2241(3)           \\
    & DMC             & 1.1222(5)       & 1.1139(2)       & 1.1111(2)       & 2270(20)        & 2258(3)         & 2279(3)           \\
    & DMC-J3          & 1.1218(6)       & 1.1143(2)       & 1.1143(3)       & 2190(20)        & 2241(3)         & 2249(4           \\
    & DMC-J3+oo.      & 1.1220(6)       & 1.1162(2)       & 1.1157(3)       & 2200(20)        & 2222(3)         & 2227(4)            \\
    & hybrid-J2       &                 & 1.1159(4)       & 1.1111(4)       &                 & 2230(5)         & 2268(6)           \\
    & hybrid          &                 & 1.1159(4)       & 1.1111(4)       &                 & 2230(5)         & 2268(6)           \\
    & Expt\cite{CRChandbook}            & 1.1283199       &                 &                 & 2169.81         &                 &                   \\
    \\
TiO & CCSD(T)\cite{Miliordos2010}       & 1.619002        &                 &                 & 1018.0          &                 &                   \\
    & PBE0            & 1.59687         &                 &                 & 1067.8          &                 &                   \\
    & VMC             & 1.5874(6)       & 1.5809(8)       & 1.5821(7)       & 1126(8)         & 1117(5)         & 1113(4)           \\
    & DMC             & 1.611(3)        & 1.598(2)        & 1.5857(12)      & 1040(30)        & 1063(8)         & 1115(6)           \\
    & hybrid          &                 & 1.619(4)        & 1.589(3)        &                 & 980(20)         & 1089(13)          \\
    & Expt\cite{Ram1999}             & 1.6203          &                 &                 & 1009.18         &                 &                   \\
    \\
FeO & CCSD(T)\cite{Sakellaris2011}         & 1.611001        &                 &                 & 929.0           &                 &                   \\
    & PBE0            & 1.60126         &                 &                 & 903.3           &                 &                   \\
    & VMC             & 1.6007(11)      & 1.583(3)        & 1.5792(12)      & 882(10)         & 962(13)         & 950(6)            \\
    & DMC             & 1.622(3)        & 1.603(4)        & 1.589(3)        & 900(30)         & 860(20)         & 940(12)           \\
    & hybrid          &                 & 1.622(10)       & 1.610(6)        &                 & 800(40)         & 900(30)           \\
    & Expt\cite{Allen1996}            & 1.6193997       &                 &                 & 882.0           &                 &                   \\
                        \hline\hline
    \end{tabular}
    \caption{Equilibrium bond lengths $r$ (\AA) and vibrational frequencies $\omega$ (cm$^{-1}$) calculated in this work, benchmark studies \cite{NIST2019,Miliordos2010,Sakellaris2011} and experiments \cite{CRChandbook,Ram1999,Allen1996}. All of our results are based on PBE0 trial wavefunctions with two-body Jastrow factor, except CO contains also results with three-body Jastrow factor with (J3+oo.) and without (J3) orbital optimization, which are discussed in later sections. Subscript $E$ indicates that the property was obtained from the PES; otherwise the values are fitted to ZVZB forces on left or right ions.}
    \label{tab:rwdata}
\end{table*}

\FloatBarrier

\section{Scaling of self-consistency error with $\zeff$}
\label{sec:self-consistency}

\begin{figure}[b]
    \centering
    \begin{tikzpicture}
    \node at (0, 0) {\includegraphics[width=8cm]{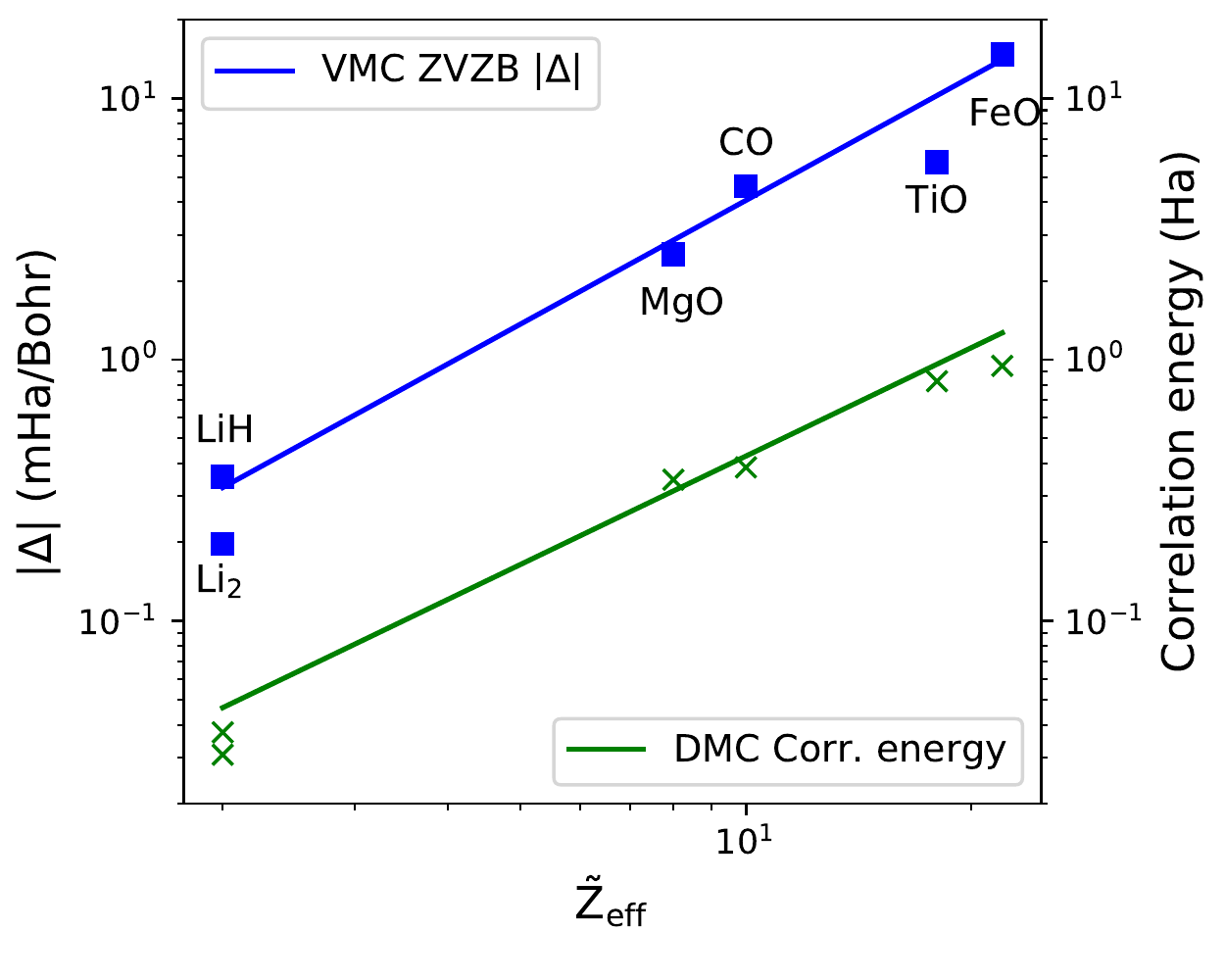}};
    \end{tikzpicture}
    \caption{Mean unsigned self-consistency error $|\bar{\Delta}|$ is plotted as a function of $\tilde{Z}_{\mathrm{eff}}=(Z^{\mathrm{left}}_{\mathrm{eff}}+Z^{\mathrm{right}}_{\mathrm{eff}})/2$ for VMC ZVZB dimer forces with PBE0 trial wavefunctions.  The correlation energy, as estimated from DMC, is plotted on a different scale to allow qualitative comparison with $|\bar{\Delta}|$. }
    \label{fig:SC_error}
\end{figure}

As seen and discussed earlier, the accuracy of both VMC and DMC forces suffer from inconsistency with the PES due to Jastrow optimization without orbital rotations. Let us boil this effect down to mean unsigned self-consistency error
\begin{align}
\label{eq:meanselfconsistency}
    |\bar{\Delta}_\lambda| = \frac{1}{\lambda_2-\lambda_1} \int_{\lambda_1}^{\lambda_2} |\Delta(\lambda)| \,\mathrm{d}\lambda,
\end{align}
where $\Delta_\lambda$ is averaged over an arbitrary range, here $r_\mathrm{expt} \pm 15\%$. Scaling of $|\bar{\Delta}_\lambda|$ with the mean $\zeff$ of the selected dimers, $\tilde{Z}_{\mathrm{eff}}=(Z^{\mathrm{left}}_{\mathrm{eff}}+Z^{\mathrm{right}}_{\mathrm{eff}})/2$, is presented in Fig.~\ref{fig:SC_error}. Fitting the data to $a \zeff^b$ shows that the self-consistency error of VMC forces is characterized reasonably well by the polynomial scaling relationship, behaving like $\tilde{Z}_{\mathrm{eff}}^{1.6(3)}$.  Similar exponents for UHF-based trial wavefunction yields $\tilde{Z}_{\mathrm{eff}}^{2.1(10)}$, because the effect of Jastrow factor is even higher. With DMC forces the PBE0-based self-consistency error scales with 1.7(13). 

The similarity in self-consistency scaling among VMC and DMC forces suggests a commonality in the missing information from implicit wavefunction parameter derivatives which cannot be captured by any of these force estimators.  Interestingly, the degree of missing information trends in a similar fashion to the correlation energy, suggesting that the energetics of each physical system sets the scale of the expected self-consistency error. The correlation energy, as estimated from the energy difference between UHF and DMC at the equilibrium geometry, is shown according to the axes on the right of Fig.~\ref{fig:SC_error}. The Jastrow factor is designed to take into account the many-body correlation, and the similarity between both the self-consistency error of forces and the correlation energy is in harmony with the understanding that the effect originates from the Jastrow optimization.

The rapid growth of self-consistency error with $\tilde{Z}_{\mathrm{eff}}$ also highlights the added challenge  systems with heavier elements are likely to pose for ZVZB force calculations with pseudopotentials.  An increased self-consistency error in the mean-field trial wave function suggests that the improvement needed from more optimized forms of the wave function also grows for these systems.  This gives some strong limits on how accurate forces can be with a community standard workflow.

\FloatBarrier

\section{Mitigating the force self-consistency error}
\label{sec:grappling}

It is therefore imperative to find ways of mitigating the self-consistency error of QMC forces in ways that are both effective and computationally tractable. In the following we shall confine ourselves to getting the most accurate forces out of a single-determinant Slater-Jastrow wavefunction. We focus strictly on the CO dimer, because it is the simplest of the oxides to demonstrate all characteristics seen in the heavier ones: measurable VMC self-consistency error, large DMC self-consistency error, and a noticeable difference between the DMC forces on the different atoms in the dimer.

\begin{figure*}[h]
    \centering
  {\includegraphics[width=6cm]{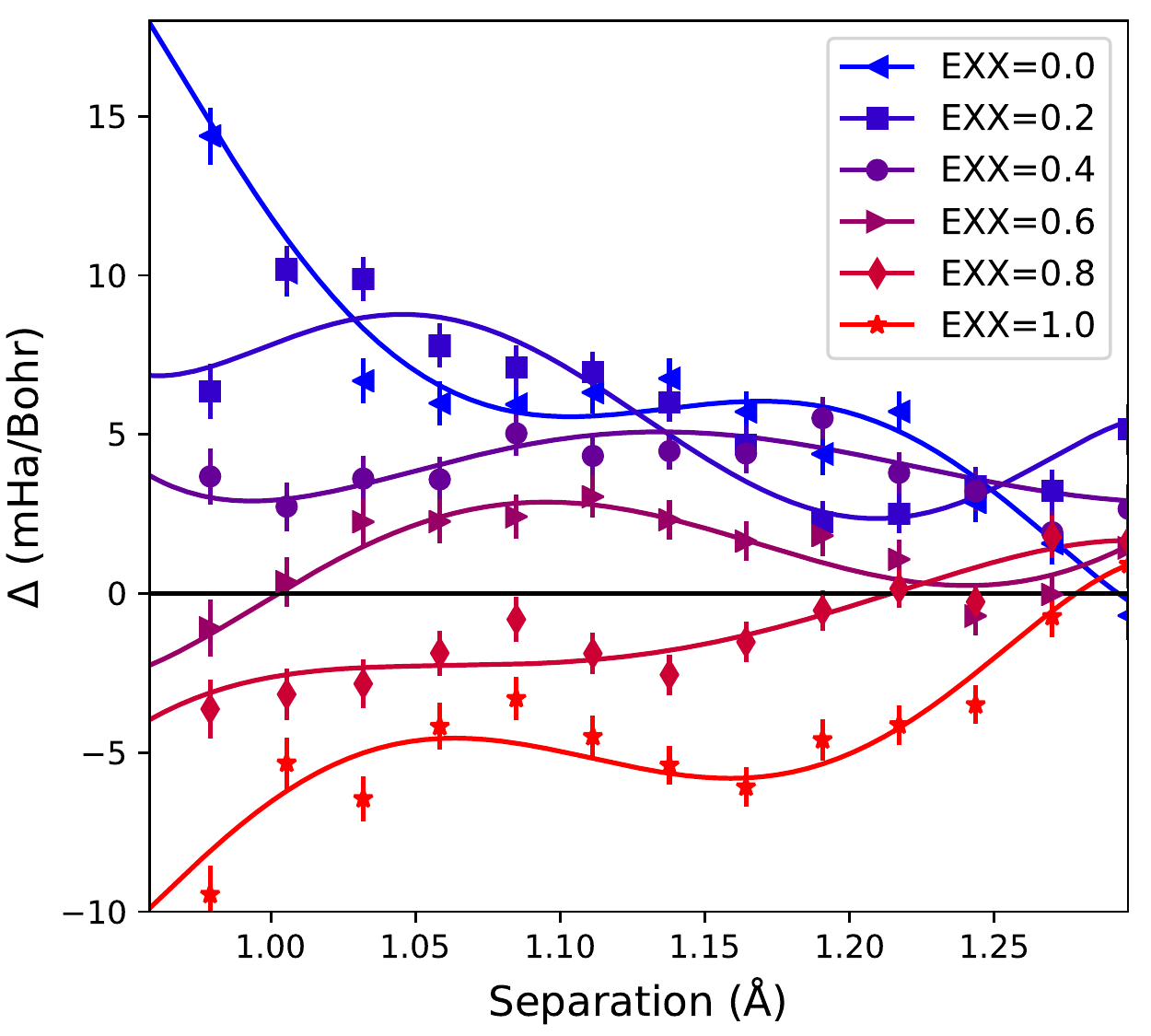}
   \includegraphics[width=6cm]{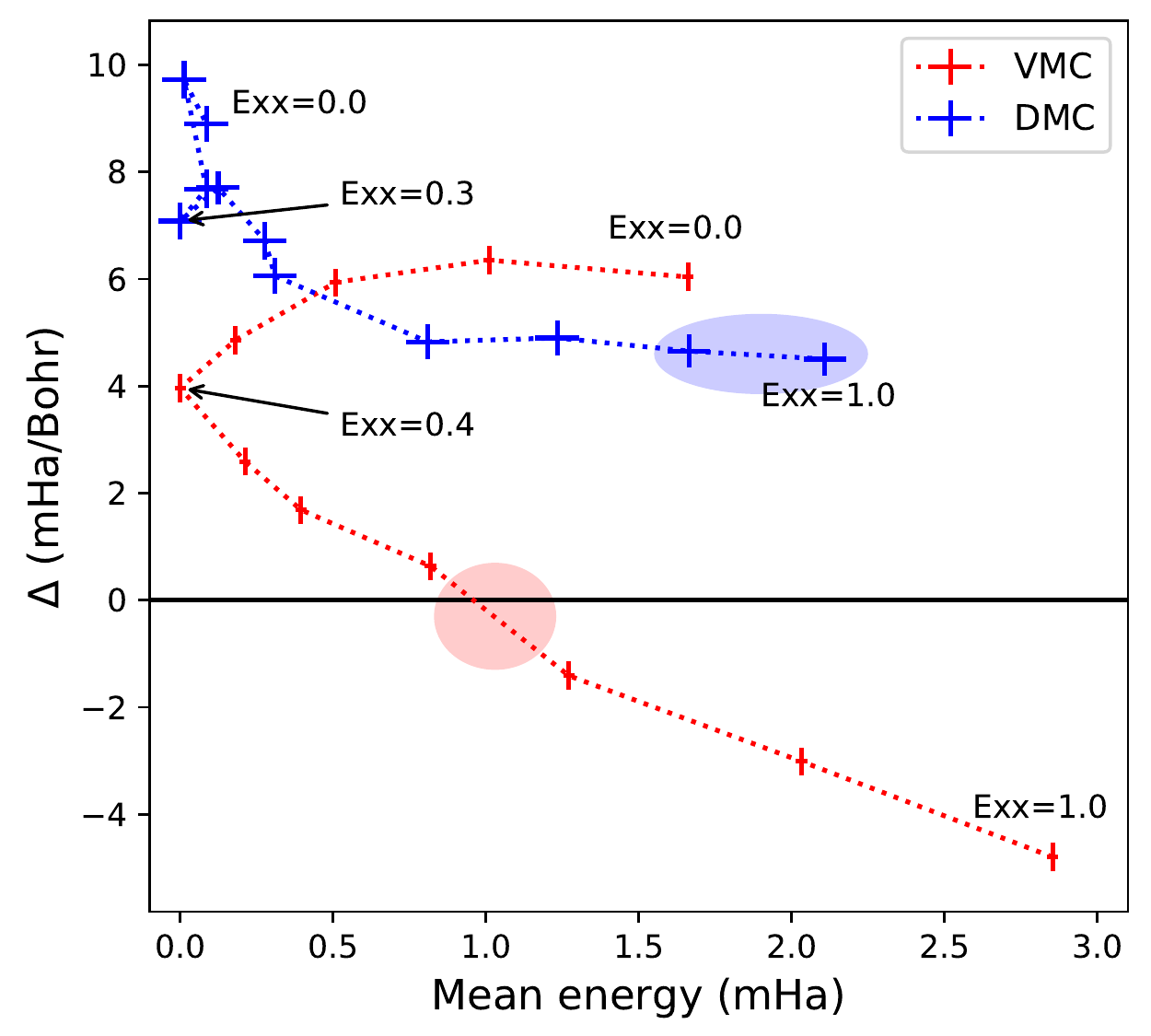}}
    \caption{(left) Self-consistency error $\Delta$ (mHa/Bohr) of the left force versus dimer separation $r$ (\AA) for the CO dimer. The different colors represent orbitals generated using different EXX fractions. (right) Mean signed self-consistency error of VMC (red), and mean net force of DMC (blue) plotted against respective mean energy differences from the minimum. }
    \label{fig:co_exx_comparison}
\end{figure*}

\begin{figure*}[t]
    \centering
 {\includegraphics[width=6.5cm]{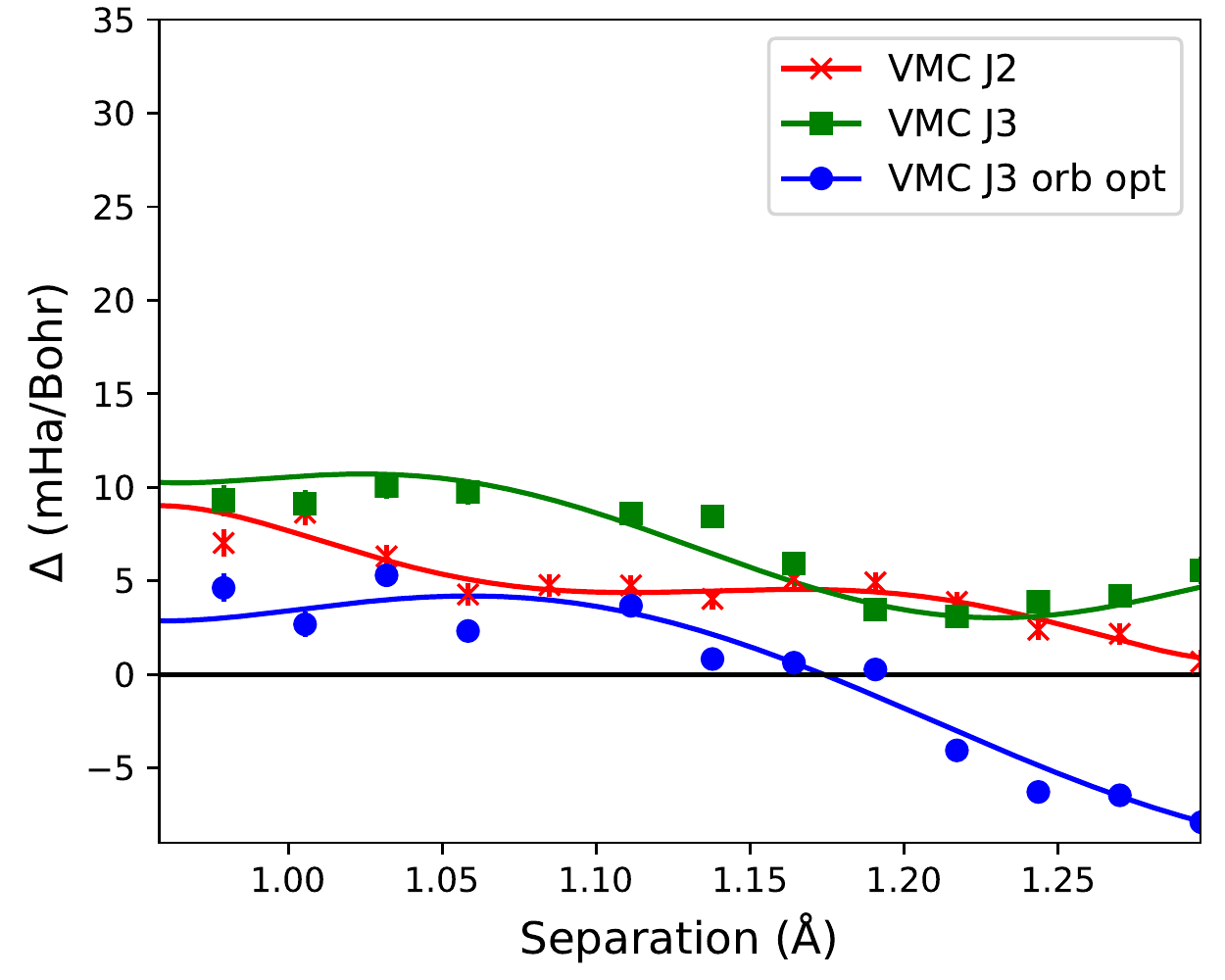}
 {\includegraphics[width=6.5cm]{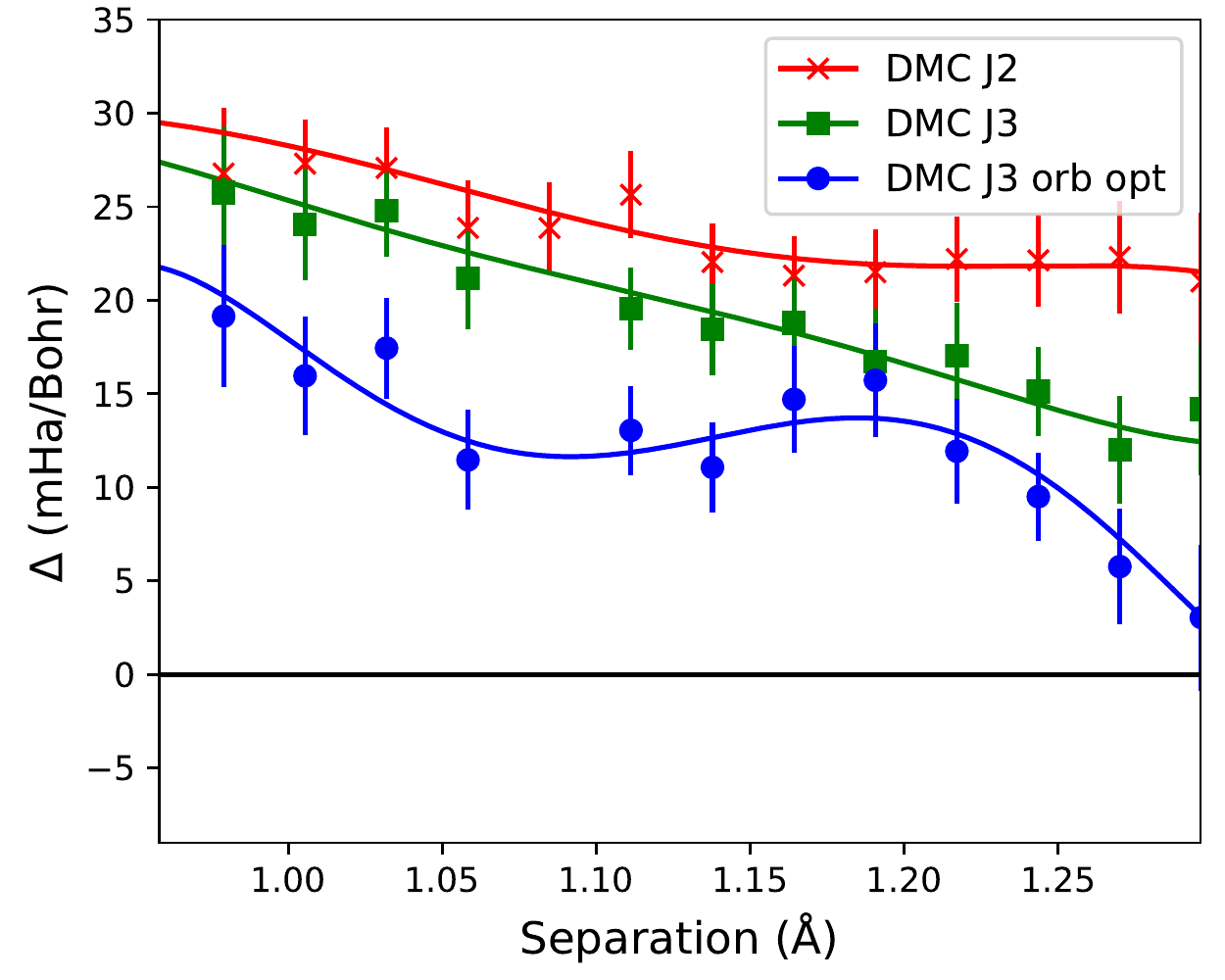}}
 }
    \caption{(Left) VMC self-consistency error $\Delta$ (mHa/Bohr) vs. dimer separation $r$ (\AA) for the CO dimer.  This was calculated using the following single determinant ans\"{a}tze: (red) 1 and 2-body Jastrows, (green) one, 2-body, and 3-body Jastrows, and (blue)  one, two, and 3-body Jastrows with optimized orbitals in a limited active space (56 orbitals).  (Right) Averaged DMC self-consistency error (mHa/Bohr) vs. dimer separation r (\AA). }
    \label{fig:co_orbopt_vqz_comparison}
\end{figure*}

A popular way to maximize the QMC performance is to perform a parameter scan of the mean-field orbitals. In particular, the optimal exact exchange fraction (EXX), scaling between pure PBE (EXX=0) and Hartree--Fock (EXX=1), can be sought to minimize the local energy for each individual system, and VMC and DMC separately. However, as seen in Figure~\ref{fig:co_exx_comparison}, while EXX scans can improve the SC errors, the EXX fraction corresponding to minimal self-consistency error does not correspond to a variational energy or discernable force variance minimum. The left panel of Fig.~\ref{fig:co_exx_comparison} shows that $\Delta$ can be minimized by varying the EXX fraction between pure UHF (EXX=1.0; pure red) and PBE (EXX=0; pure blue). However, based on the right panel, the VMC energy minimum is near EXX=0.4, while the minimal $\Delta$ (red circle) is around EXX=0.75. Similar comparison of the DMC energy to a measure of self-inconsistency (net force) leads to different EXX values. The conclusion is that EXX optimization by itself is not an efficient way to improve the performance of ZVZB forces.


Let us instead consider improvements in the forces resulting from improving the Jastrow factor and orbital re-optimization in the presence of the Jastrow. We compare the CO dimer curve at both the VMC and DMC level using three different types of trial wave functions: Slater-Jastrow with up to two body Jastrows using PBE0 orbitals (1+2J), Slater-Jastrow with up to three-body Jastrows using PBE0 orbitals (1+2+3J), and lastly a Slater-Jastrow with up to three-body Jastrows using  re-optimized orbitals using the PBE0 set as a starting point (1+2+3J+OO).  The aug-ccVQZ basis set for CO has 206 basis functions, and so a full spin-unrestricted optimization of these orbitals would introduce an additional 2030 parameters into the linear method optimization procedure\cite{Filippi2016,Assaraf2017,Flores2019}.  In order to reduce computational cost, we restrict orbital optimization to the 56 lowest energy orbitals, which introduces 510 optimizable parameters. Optimization within this reduced set does not guarantee a zero self-consistency error, but we will show that even within this restricted active space we are able to dramatically improve the accuracy of the VMC and DMC force estimators. 

We consider first the VMC self-consistency error.  The origin of this problem has been recognized for a long time\cite{Casalegno2003,Moroni2014}: the parameters defining the mean-field orbitals have an implicit ion dependence yet are not at a variational minimum of the QMC Hamiltonian.  Moroni \textit{et al.} \cite{Moroni2014} showed that the solution to this problem is orbital optimization.  In the left panel of Figure \ref{fig:co_orbopt_vqz_comparison}, we see that relative to the 1+2J wave function, the inclusion of a three-body Jastrow does little to mitigate the self-consistency error of VMC forces, and in fact arguably increases it at intermediate separations.  However, by performing orbital optimization on the 1+2+3J wave function--even in the restricted active space--we can significantly reduce the self-consistency error. While this behavior is expected, what is striking is the incredibly small energy differences between the 1+2+3J and orbital optimized 1+2+3J wave functions:  across the range of separations, orbital optimization recovers on the order of $0.2$ mHa, but can easily change the self-consistency error by a factor of 2.  Consistent with prior results, this suggests that orbital optimization remains an appealing route for reduction of self-consistency error.

We now consider the impact of the improved trial wavefunctions on DMC ZVZB forces. In the right side of Figure \ref{fig:co_orbopt_vqz_comparison}, we show the average DMC self-consistency error versus dimer separation. It is immediately noticeable that the self-consistency error is larger at the DMC level than it is for VMC.  The self-consistency error for the 1+2J wave function is between between 25-30mHa, 3-5 times larger on average than it is for VMC.  However, the DMC estimator does appear to be unambiguously improved by improving the wave function.  The 1+2+3J wave function reduces the SC error by 20-25\% across the dimer curve relative to 1+2J.  Upon performing orbital optimization on the 1+2+3J wavefunction, we further observe a consistent decrease of the SC error across the range of separations by between 5-8 mHa/Bohr, which is comparable to the improvement we saw at the VMC level.  Despite these gains, further improvements to the wavefunction are needed for the DMC forces to match the quality of the PES when it comes to derived quantities such as the bond length.  While both the VMC and DMC forces have improved in accuracy, the DMC forces still do not exceed the VMC forces in terms of absolute accuracy, since the deviation from experiment for the bond lengths are $0.01251(9)$ \AA{} for VMC forces, $0.0124(3)$ \AA{} for mixed DMC forces, versus $0.0063(6)$ \AA{} for the DMC PES.

In order to assess how the force imbalance depends on quality of trial wave function, we computed the net force at each dimer separation and performed an average over the whole dimer curve for each of the three considered wave functions.  We find that, overall, the force imbalance is a secondary source of error compared to the overall self-consistency error. The mean self-consistency error ranges between 5-18 mHa/Bohr with our best wavefunction, while the mean net forces per wavefunction over the dimer separation are only 7.1(3) mHa/Bohr for 1+2J, 1.6(4) mHa/Bohr for 1+2+3J, and 3.4(4) mHa/Bohr for 1+2+3J+OO.  The improvement of the Jastrow has also rectified the discrepancy in the derived bond length from left and right DMC forces, with the two now giving the same bond length to within $0.0001(4)$ \AA{}. Since the DMC fixed-node wave function is related to the trial wave function by a change of Jastrow factor, it is expected that improvements to the Jastrow factor within the trial wave function should substantially impact this error.  Orbital optimization, while bringing down the overall self-consistency error, actually worsens the force imbalance a small amount,  though still with a minimal impact on imbalanced bond length as they remain in agreement to $0.0005(4)$ \AA{}. 

As expected, the Reynolds approximation introduces an extremely strong dependence on the quality of the underlying trial wave function, which is not particularly satisfying given the greater insensitivity of the DMC PES to the wave function parameters than in VMC. While we were able to demonstrate a substantial mitigation of these errors by deploying better wave functions and performing orbital optimization, there's a good chance that in cases where the VMC and DMC PES's are not extremely similar to begin with, forces obtained from the Reynolds approximation might not mirror the true DMC PES accurately enough to significantly improve the force estimate over the one from VMC. On a practical side however, as long as the VMC ZVZB and DMC ZVZB with Reynolds forces are \textit{better} than those from DFT, they still have value.  In Table I (main text), we see that even for single Slater-determinant with two body-jastrow calculations, the VMC and DMC predicted bond lengths and vibrational frequencies are often very comparable to those derived from PBE0, and they are likely to be improved with orbital optimization and the inclusion of a 3-body Jastrow. For general applications where the accuracy of DMC is required, however,  significantly better approximations to the left Pulay should be employed.

\FloatBarrier

\section{Basis set convergence and equilibrium forces}

\begin{figure*}[h!]
    \centering
    \begin{tikzpicture}
    \node at (0, 0) {\includegraphics[width=13.5cm]{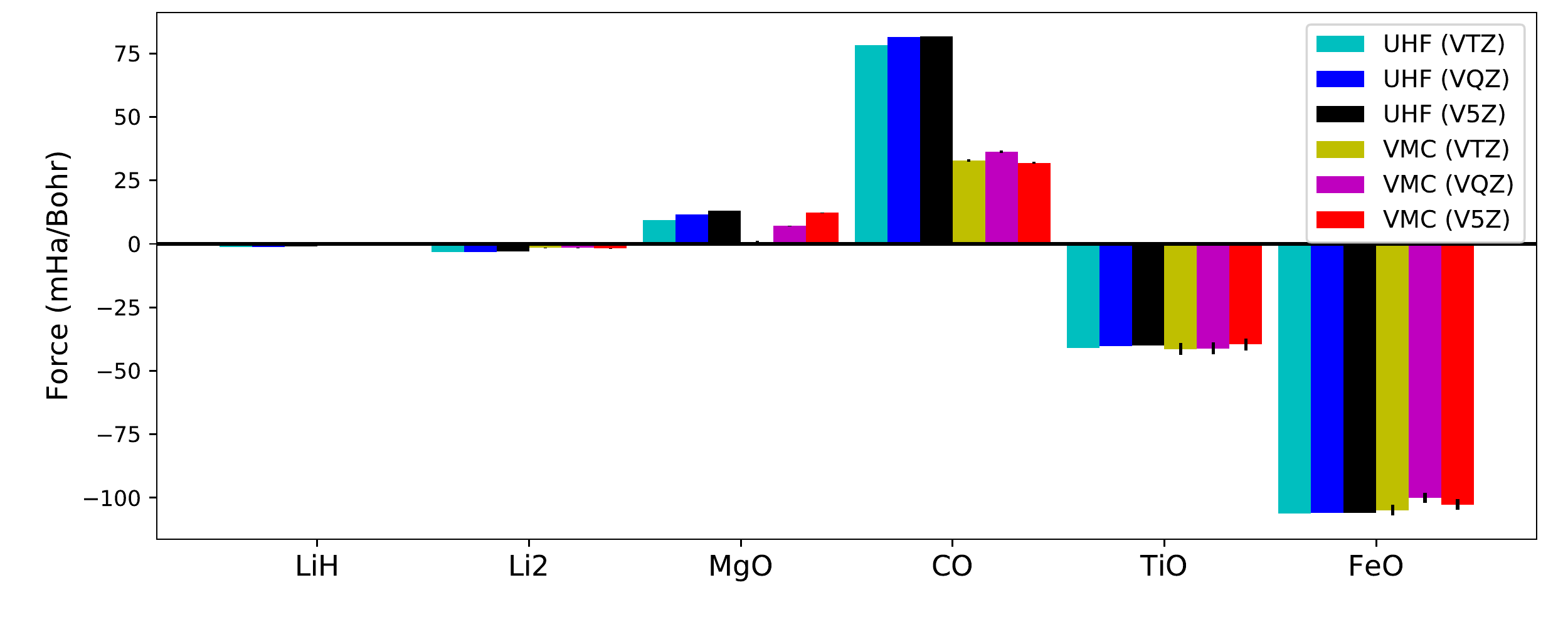}};
    \node at (0, -5.5) {\includegraphics[width=13.5cm]{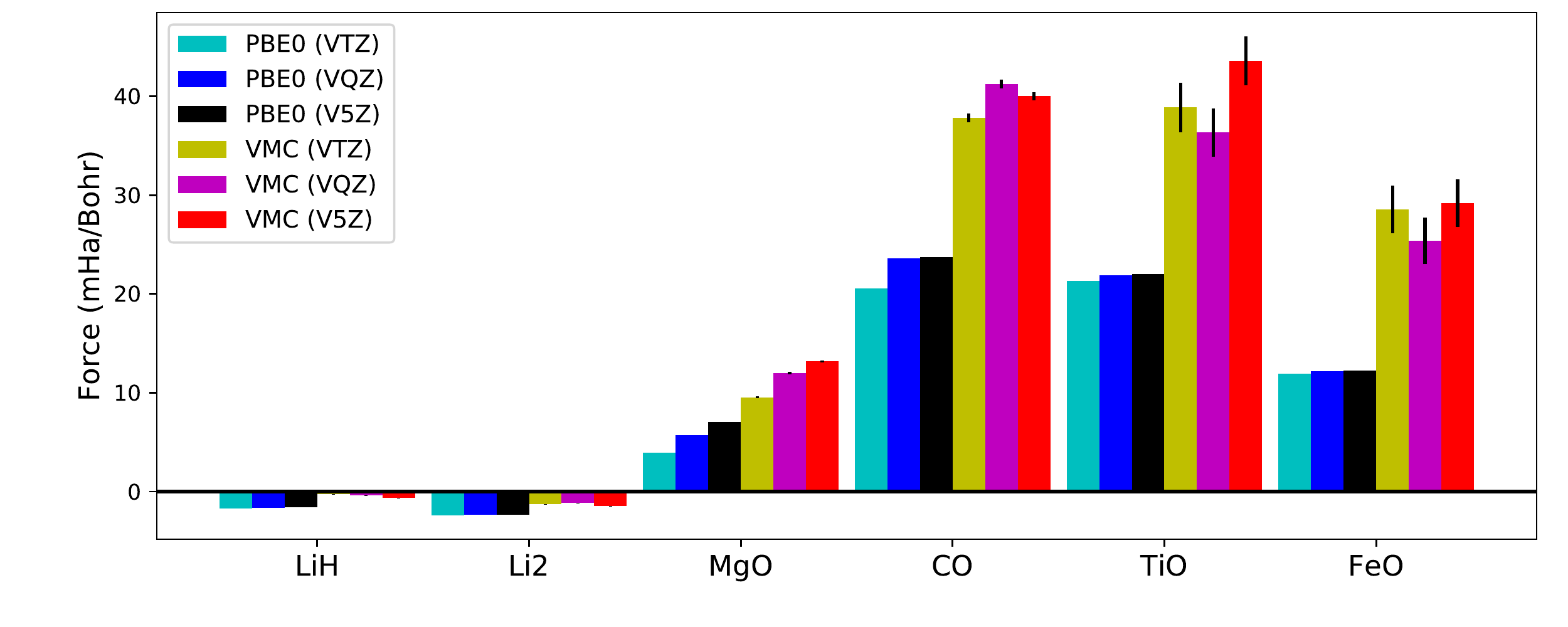}};
    \end{tikzpicture}
    \caption{Estimated ZVZB forces at experimental bond lengths computed with SCF and VMC using UHF (above) and PBE0 (below) orbitals with aug-VTZ, aug-VDZ and aug-V5Z basis sets. }
    \label{fig:Feqms}
\end{figure*}

The basis set convergence is studied in terms of force values at the reference geometry, $r_{\mathrm{expt}}$. Both PBE0 forces and the associated VMC forces are presented in Fig.~\ref{fig:Feqms} with three basis sets, VTZ, VQZ, and V5Z. Only the left forces are shown, because the VMC forces are manifestly symmetric apart from variance. The Jastrow factors of each dimer are optimized with V5Z and used with all the bases. Based on the data, we have chosen to use VTZ basis for all dimers, except MgO, where VQZ is chosen. 

\begin{figure*}[h!]
    \centering
    \begin{tikzpicture}
    \node at (0, 0) {\includegraphics[width=13.5cm]{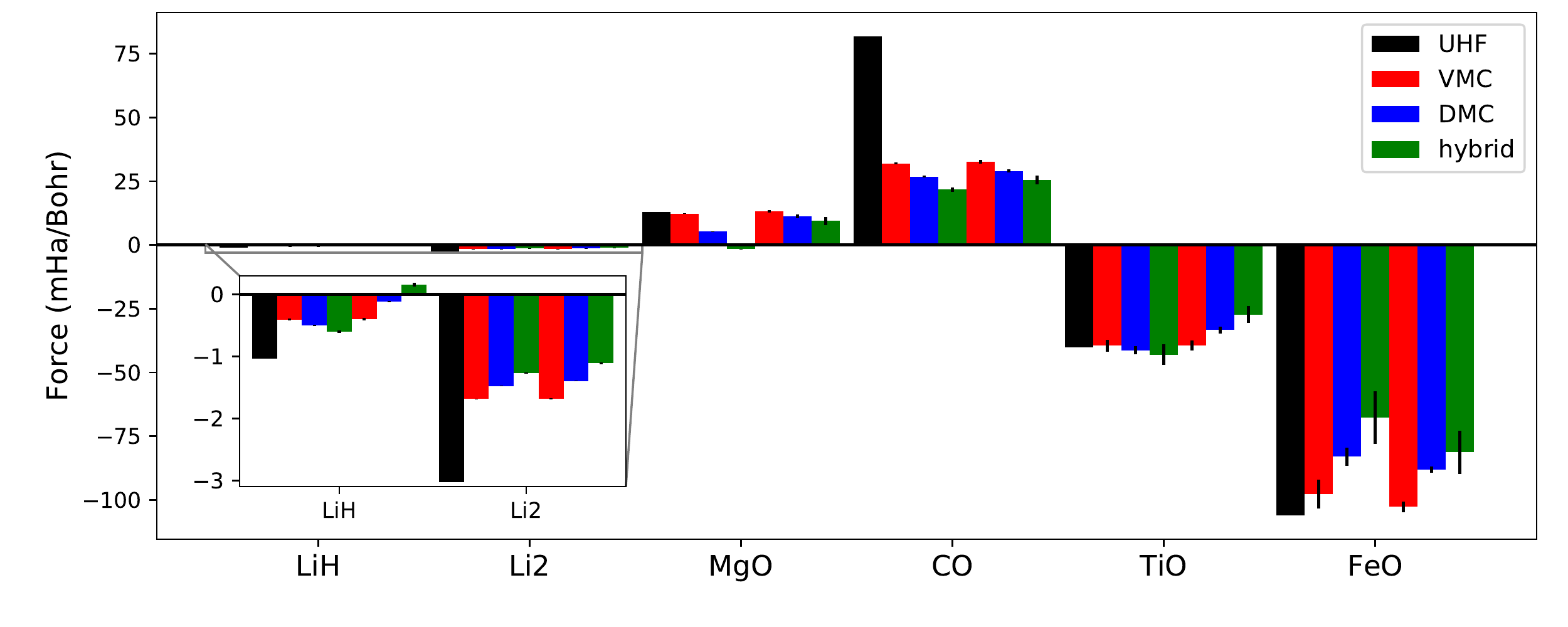}};
    \node at (0, -5.5) {\includegraphics[width=13.5cm]{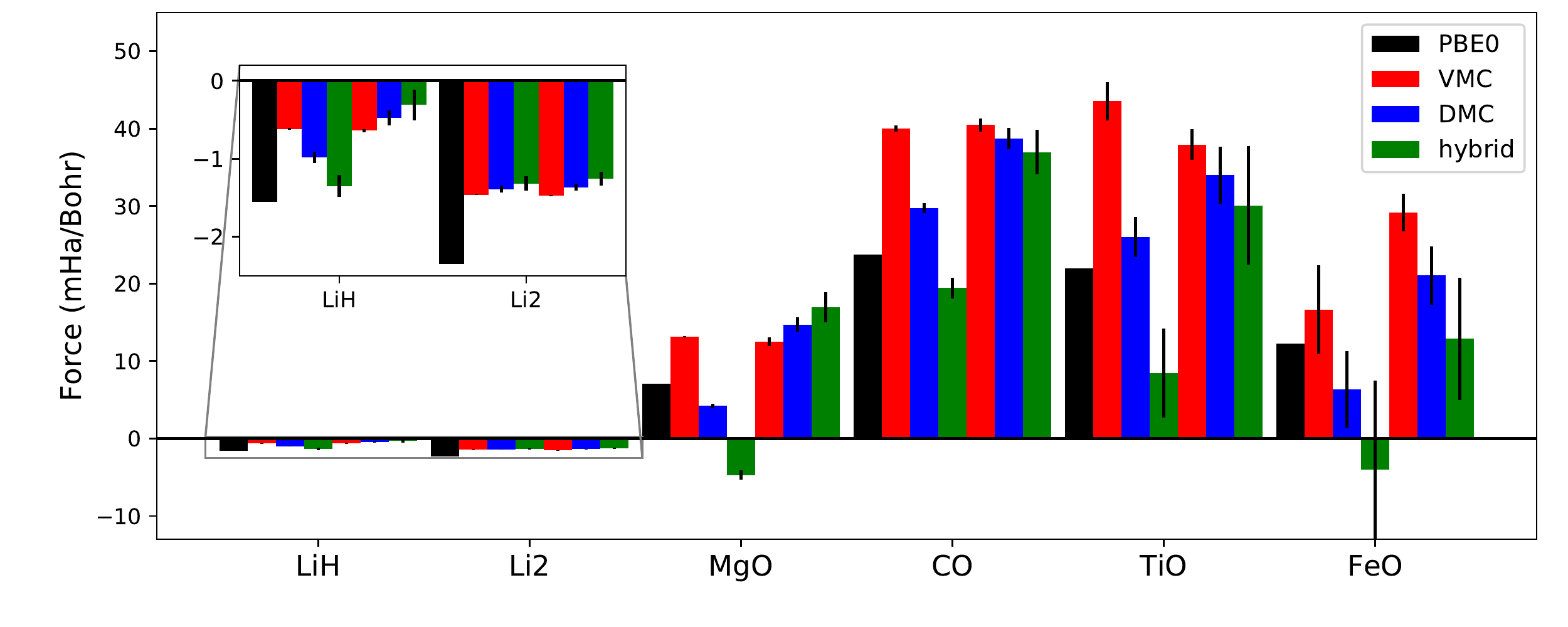}};
    \end{tikzpicture}
    \caption{Estimated ZVZB forces at experimental bond lengths computed with SCF, VMC, DMC and hybrid ZVZB estimators using UHF (above) and PBE0 (below) orbitals with V5Z basis set. }
    \label{fig:Feqms_dmc}
\end{figure*}

The data including SCF and DMC forces is tabulated in Table~\ref{tab:Feqms_uhf} for UHF determinants and Table~\ref{tab:Feqms} for PBE0 determinants. Comparison of mixed and hybrid DMC forces at equilibrium is given in Fig.~\ref{fig:Feqms_dmc} with the V5Q basis set. 

\begin{table}[]
    \centering
    \begin{tabular}{lllllllll}
    & Method          &                 & $F$ (left)      &                 &                 & $F$ (right)     &                  \\
    &                 & VTZ             & VQZ             & V5Z             & VTZ             & VQZ             & V5Z              \\ \hline\hline
LiH & UHF             & -1.1733635      & -1.1333974      & -1.0311665      &                 &                 &                   \\
    & VMC             & -0.390(11)      & -0.432(11)      & -0.405(13)      & 0.44(2)         & 0.41(2)         & 0.40(2)           \\
    & DMC             & -0.99(7)        & -0.99(7)        & -0.50(8)        & 0.62(10)        & 0.50(11)        & 0.12(9)           \\
Li2 & UHF             & -3.1202507      & -3.1409551      & -3.0274961      &                 &                 &                   \\
    & VMC             & -1.470(7)       & -1.587(7)       & -1.688(9)       & 1.474(7)        & 1.585(7)        & 1.685(8)          \\
    & DMC             & -1.37(5)        & -1.43(5)        & -1.48(5)        & 1.45(5)         & 1.40(5)         & 1.40(5)           \\
MgO & UHF             & 9.5078863       & 11.5212239      & 13.0436547      &                 &                 &                   \\
    & VMC             & 0.6(0.6)        & 7.16(9)         & 12.25(8)        & -0.7(0.5)       & -7.2(0.5)       & -13.2(0.5)        \\
    & DMC             & -6.3(0.3)       & -0.4(0.3)       & 5.3(0.3)        & -1.3(1.0)       & -9.3(0.8)       & -11.3(1.3)        \\
CO  & UHF             & 78.3546928      & 81.5848802      & 81.7178423      &                 &                 &                   \\
    & VMC             & 32.8(0.4)       & 36.3(0.4)       & 32.0(0.4)       & -35.1(0.8)      & -37.3(0.8)      & -32.6(0.8)        \\
    & DMC             & 28.1(0.7)       & 35.1(0.7)       & 26.8(0.7)       & -33.3(1.3)      & -38.5(1.0)      & -29.0(1.3)        \\
TiO & UHF             & -40.9783141     & -40.3456662     & -40.1295219     &                 &                 &                   \\
    & VMC             & -41(3)          & -41(3)          & -40(3)          & 38(2)           & 47(2)           & 39(2)             \\
    & DMC             & -44(5)          & -47(5)          & -41(5)          & 32(3)           & 39(4)           & 33(4)             \\
FeO & UHF             & -106.2975707    & -106.0764988    & -106.0129389    &                 &                 &                   \\
    & VMC             & -110(6)         & -103(6)         & -98(6)          & 105(3)          & 100(3)          & 103(3)            \\
    & DMC             & -105(6)         & -88(6)          & -83(6)          & 91(5)           & 83(4)           & 88(5)             \\
    \end{tabular}
    \caption{Equilibrium force values in mHa/Bohr using UHF trial wavefunctions and various basis sets.}
    \label{tab:Feqms_uhf}
\end{table}

\begin{table}[]
    \centering
    \begin{tabular}{lllllllll}
    & Method          &                 & $F$ (left)      &                 &                 & $F$ (right)     &                  \\
    &                 & VTZ             & VQZ             & V5Z             & VTZ             & VQZ             & V5Z              \\ \hline\hline
LiH & PBE0            & -1.7174509      & -1.6366382      & -1.5557202      &                 &                 &                   \\
    & VMC             & -0.248(10)      & -0.358(11)      & -0.617(14)      & 0.22(2)         & 0.34(2)         & 0.64(2)           \\
    & DMC             & -1.00(6)        & -0.93(5)        & -0.98(8)        & 0.59(10)        & 0.49(9)         & 0.47(10)          \\
Li2 & PBE0            & -2.3979264      & -2.3681756      & -2.3448929      &                 &                 &                   \\
    & VMC             & -1.254(6)       & -1.154(13)      & -1.460(7)       & 1.241(6)        & 1.142(14)       & 1.472(7)          \\
    & DMC             & -1.19(5)        & -1.21(13)       & -1.39(5)        & 1.24(5)         & 1.14(13)        & 1.36(5)           \\
MgO & PBE0            & 3.9364609       & 5.7371909       & 7.0522571       &                 &                 &                   \\
    & VMC             & 9.55(7)         & 12.0(0.1)       & 13.17(8)        & -8.2(0.6)       & -12.6(0.6)      & -12.5(0.6)        \\
    & DMC             & -0.5(0.4)       & 3.0(0.4)        & 4.2(0.3)        & -4.5(1.1)       & -11.9(1.0)      & -14.7(0.9)        \\
CO  & PBE0            & 20.5586222      & 23.5869345      & 23.7307699      &                 &                 &                   \\
    & VMC             & 37.8(0.4)       & 41.2(0.4)       & 40.0(0.4)       & -37.1(0.8)      & -41.4(0.8)      & -40.5(0.9)        \\
    & DMC             & 33.3(1.0)       & 36.6(0.9)       & 29.7(0.6)       & -41.6(1.0)      & -44.4(1.1)      & -38.7(1.4)        \\
TiO & PBE0            & 21.3299149      & 21.8679477      & 21.9946493      &                 &                 &                   \\
    & VMC             & 39(3)           & 36(3)           & 44(3)           & -36(3)          & -40(3)          & -38(2)            \\
    & DMC             & 18(5)           & 21(6)           & 26(3)           & -29(4)          & -31(5)          & -34(4)            \\
FeO & PBE0            & 11.9526571      & 12.2093923      & 12.2333751      &                 &                 &                   \\
    & VMC             & 36(6)           & 23(6)           & 17(6)           & -29(3)          & -25(3)          & -29(3)            \\
    & DMC             & 8(7)            & 3(6)            & 6(5)            & -22(3)          & -22(2)          & -21(4)            \\
    \end{tabular}
    \caption{Equilibrium force values in mHa/Bohr using PBE0 trial wavefunctions and various basis sets.}
    \label{tab:Feqms}
\end{table}

\FloatBarrier

\section{Energy and force curves for dimers}

\begin{figure}[h]
    \centering
    \begin{tikzpicture}
    \node at (-3.3, 4.8) {\includegraphics[width=6cm]{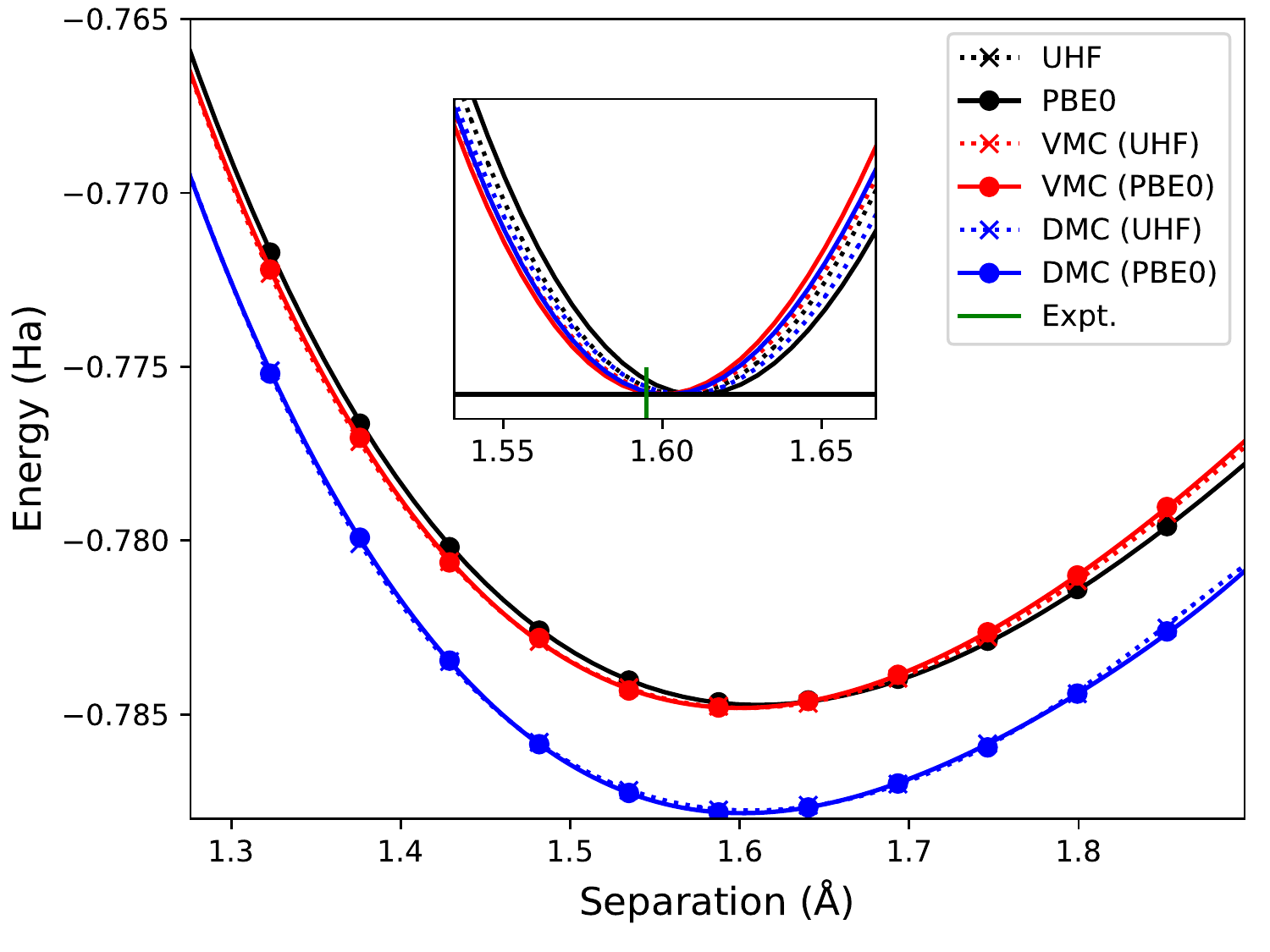}};
    \node at ( 3.3, 4.8) {\includegraphics[width=6cm]{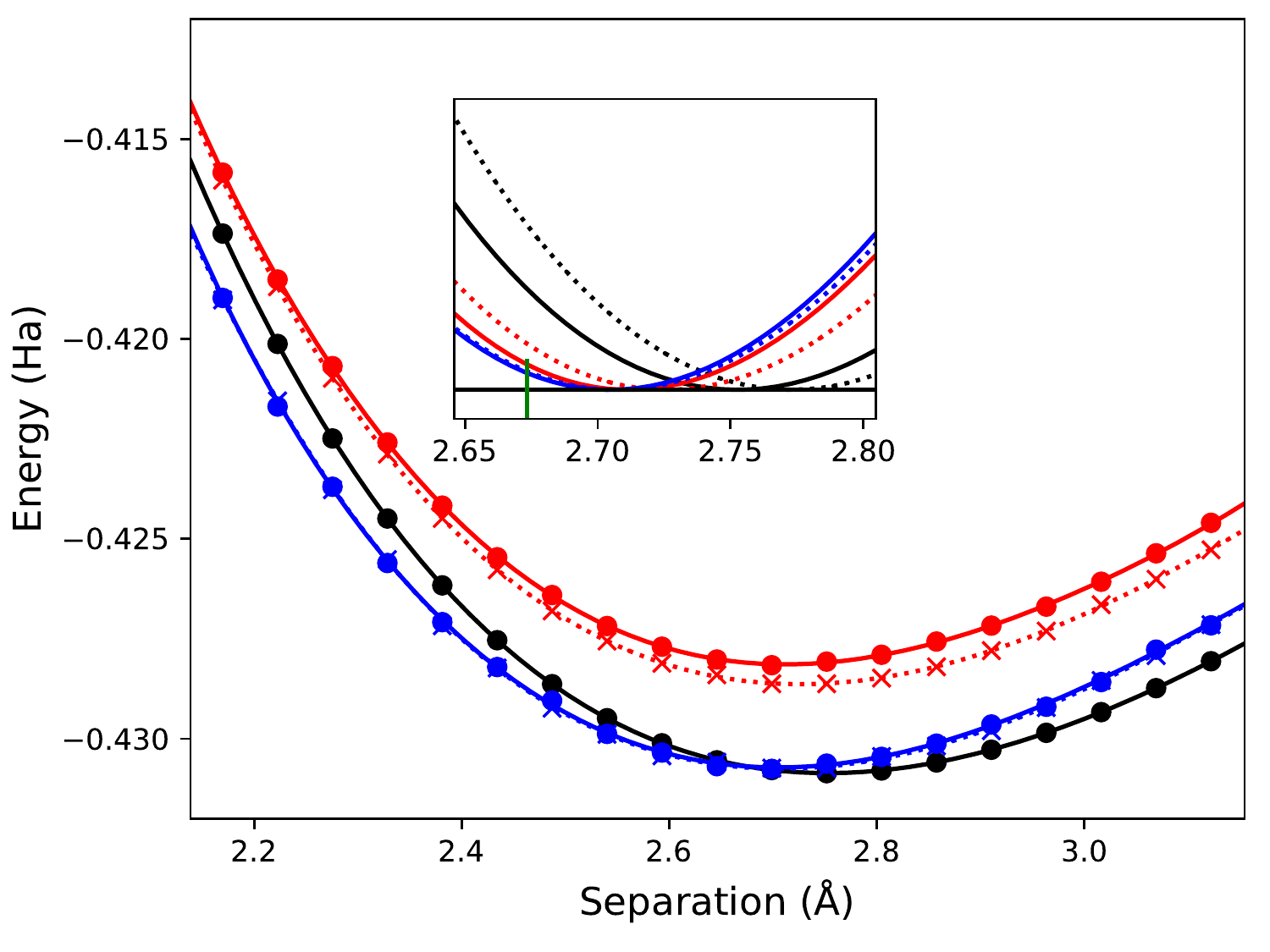}};
    \node at (-3.3, 0) {\includegraphics[width=6cm]{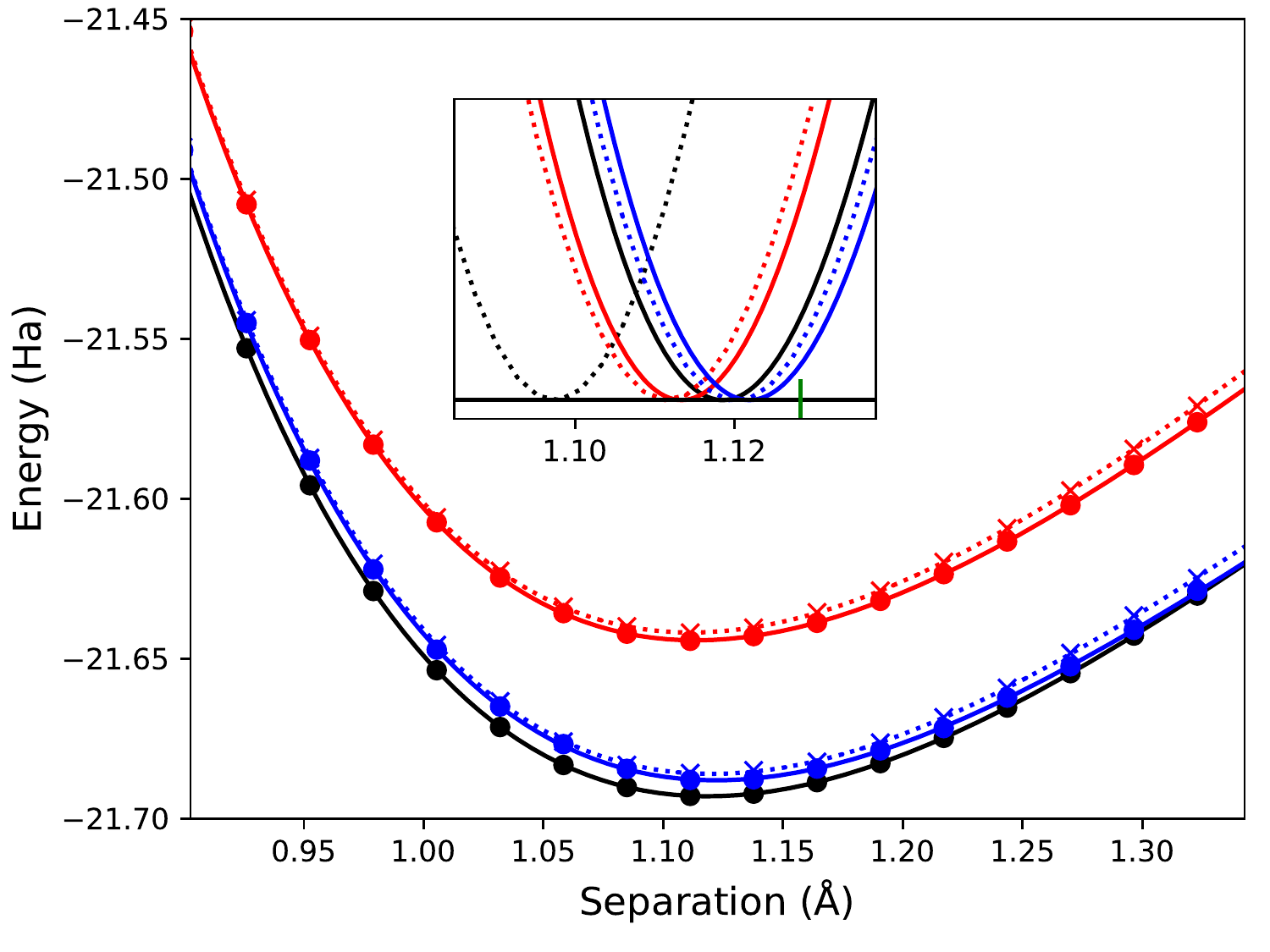}};
    \node at ( 3.3, 0) {\includegraphics[width=6cm]{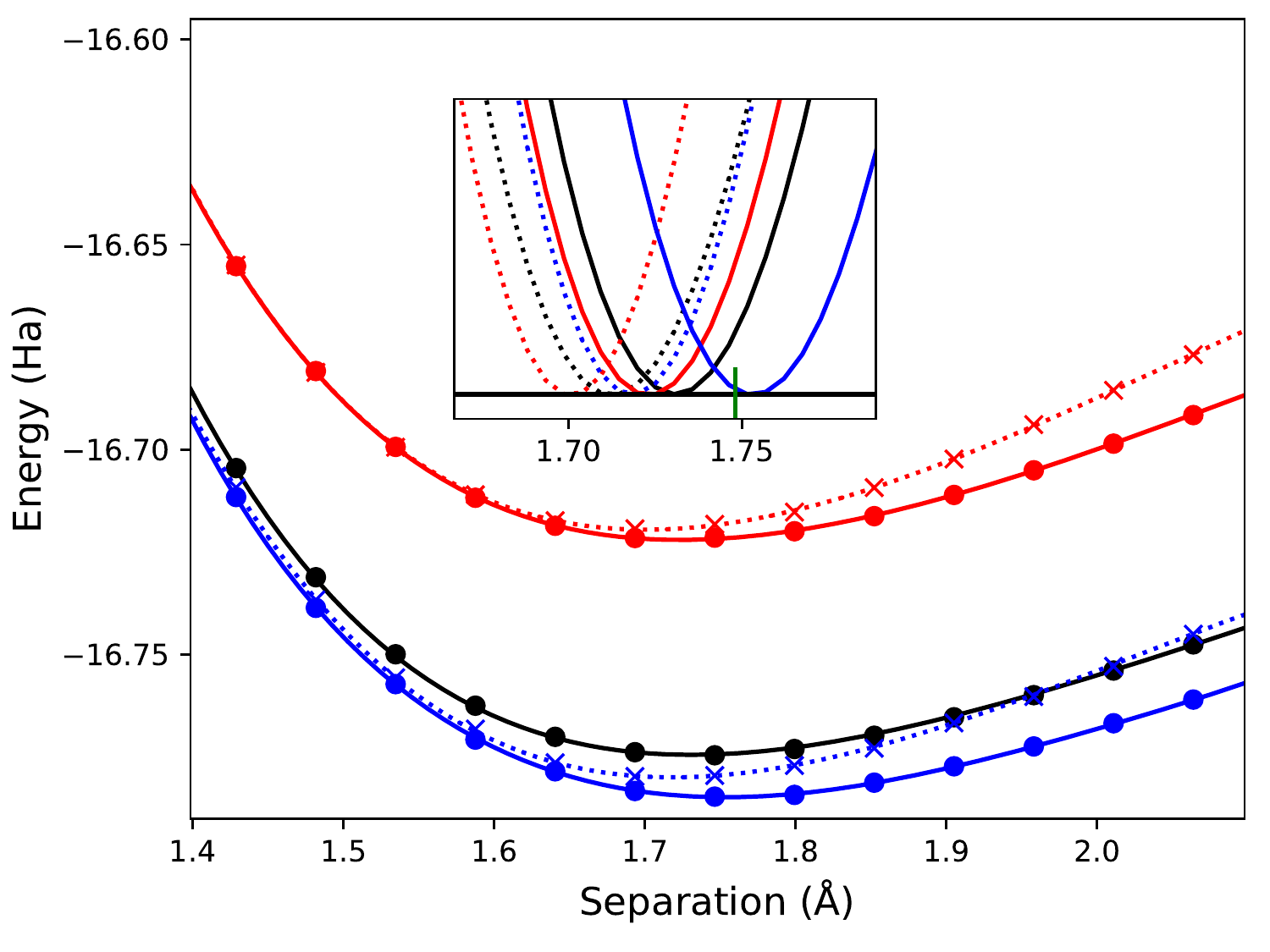}};
    \node at (-3.3,-4.8) {\includegraphics[width=6cm]{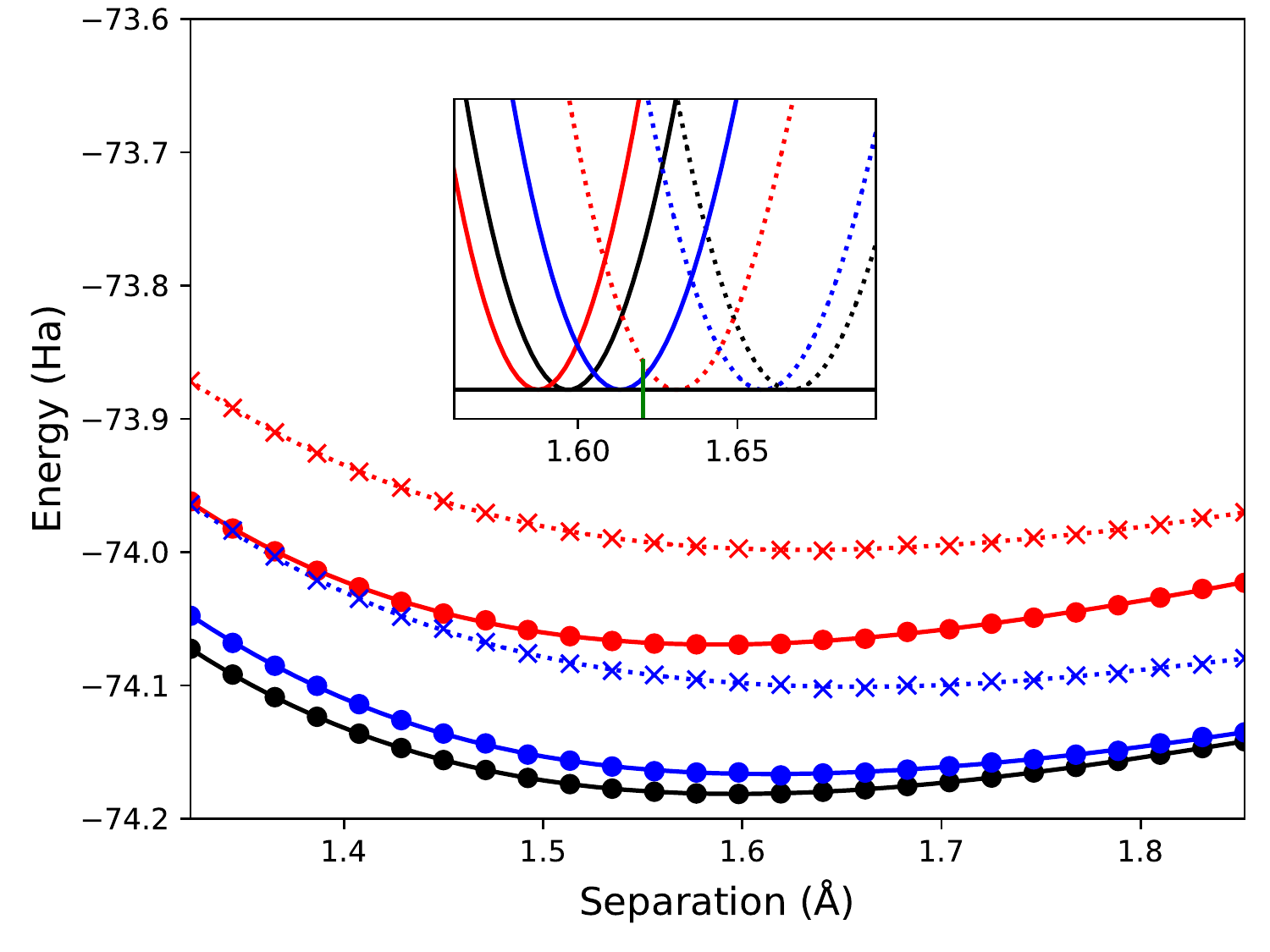}};
    \node at ( 3.3,-4.8) {\includegraphics[width=6cm]{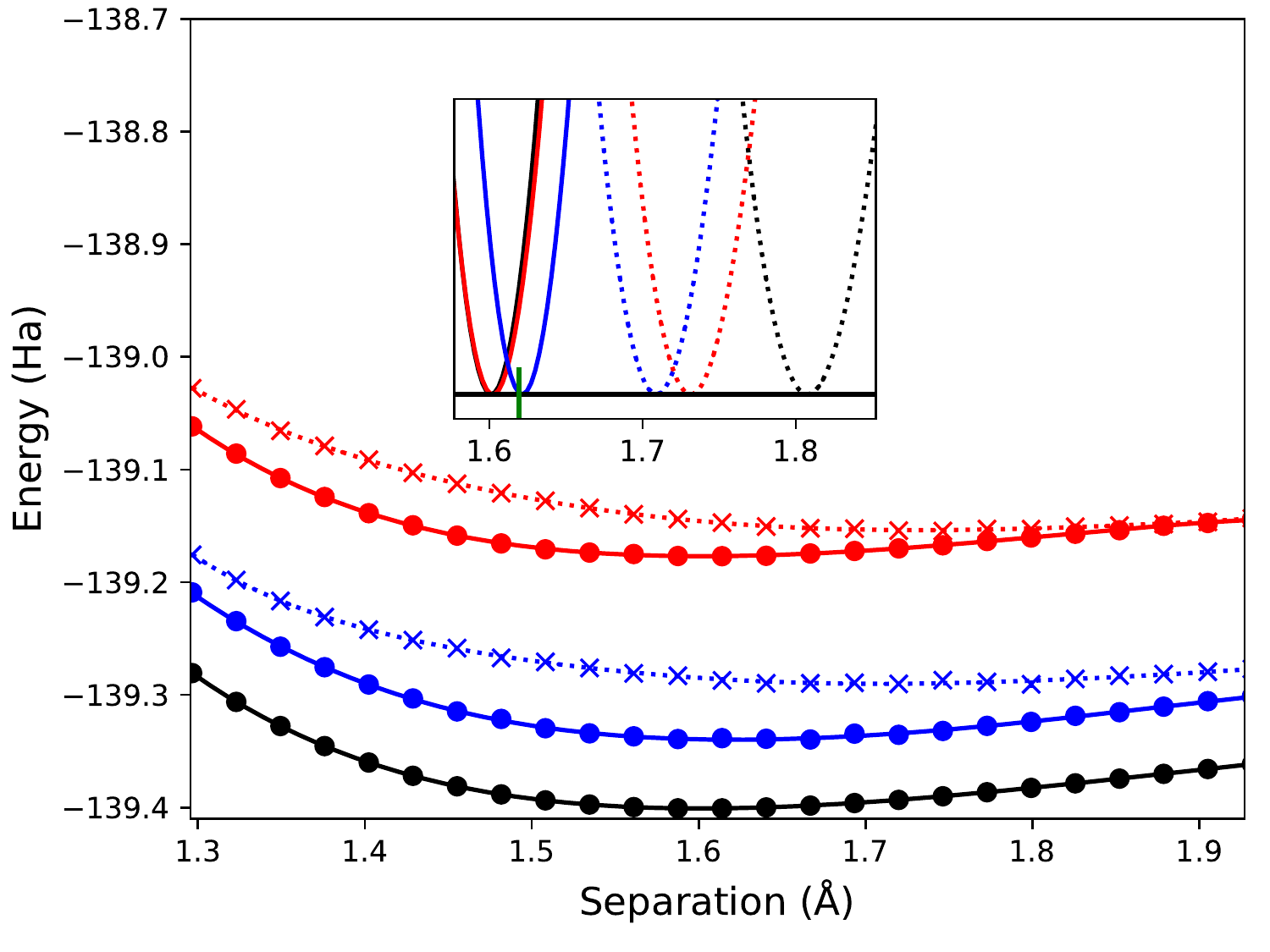}};
    \end{tikzpicture}
    \caption{Binding curves of dimers ranging $\pm 20 \%$ around experimental bond length (green marking). The solid (PBE0) and dotted (UHF) lines with black (SCF), red (VMC) and blue (DMC) color are 6-degree polynomial fits and markers are the QMC estimates. The absolute SCF (UHF) energy curve is often off the chart.}
    \label{fig:Ecurves}
\end{figure}

\begin{figure}[h]
    \centering
    \begin{tikzpicture}
    \node at (-3.3, 4.8) {\includegraphics[width=6cm]{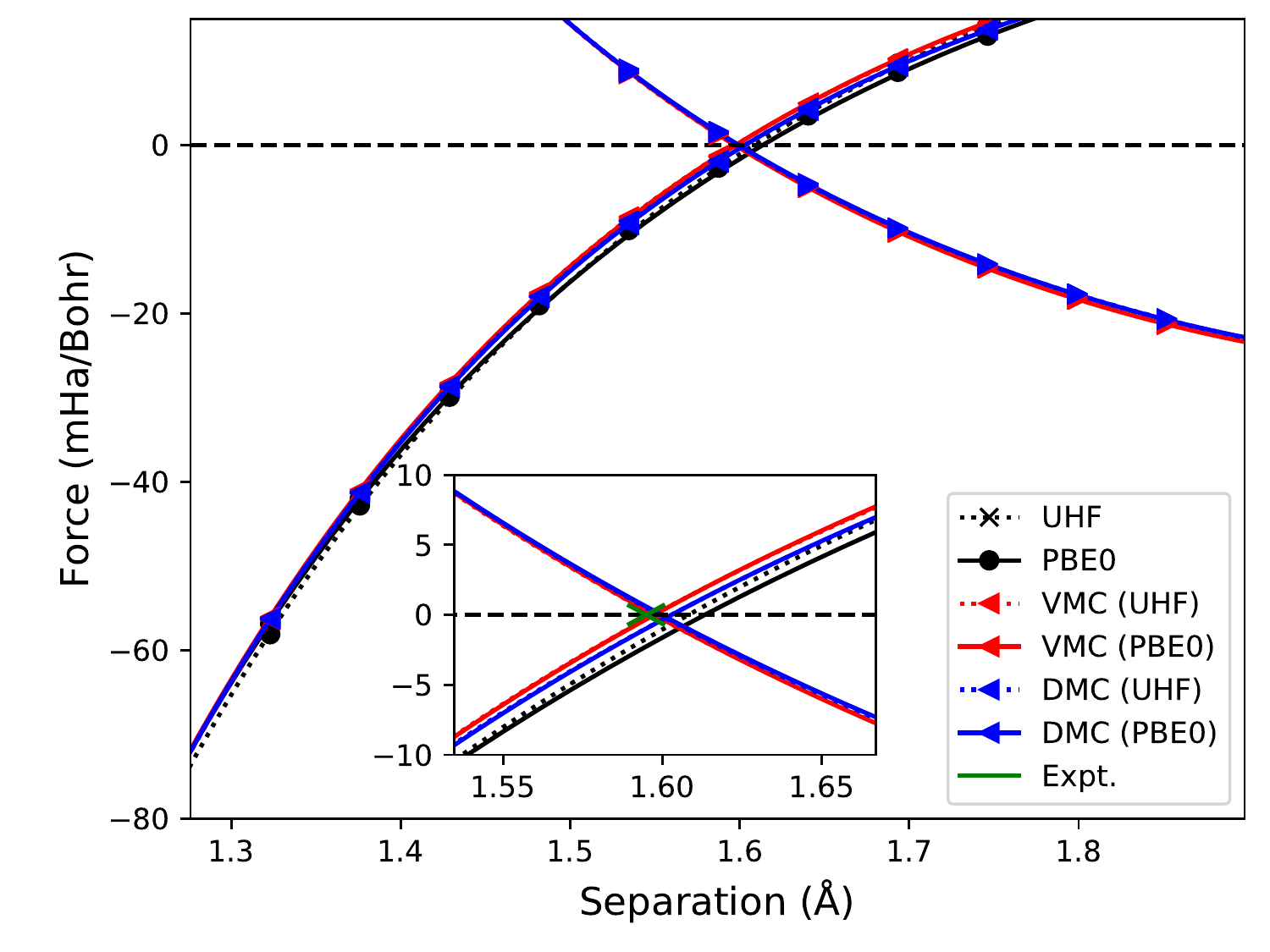}};
    \node at ( 3.3, 4.8) {\includegraphics[width=6cm]{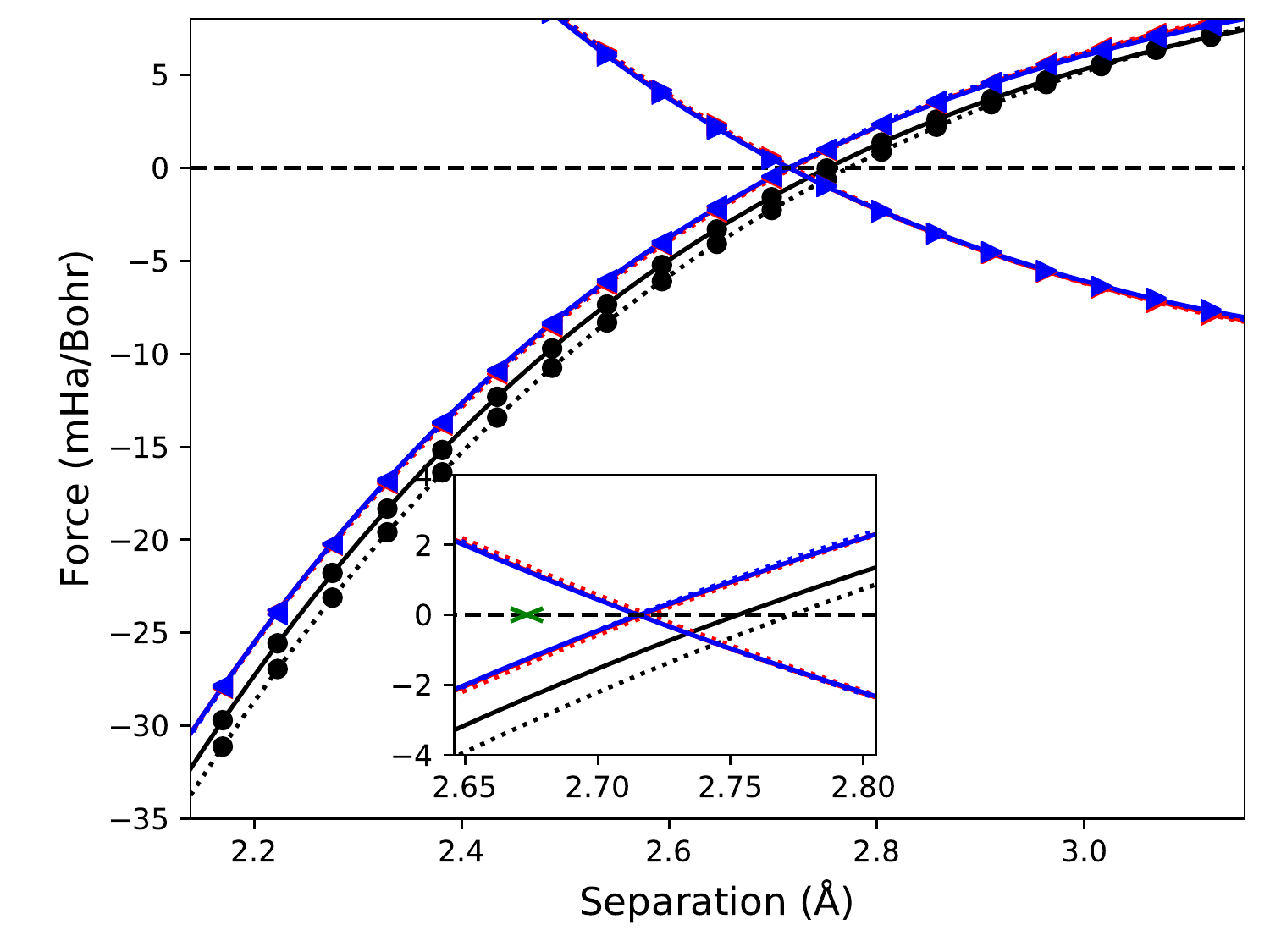}};
    \node at (-3.3, 0) {\includegraphics[width=6cm]{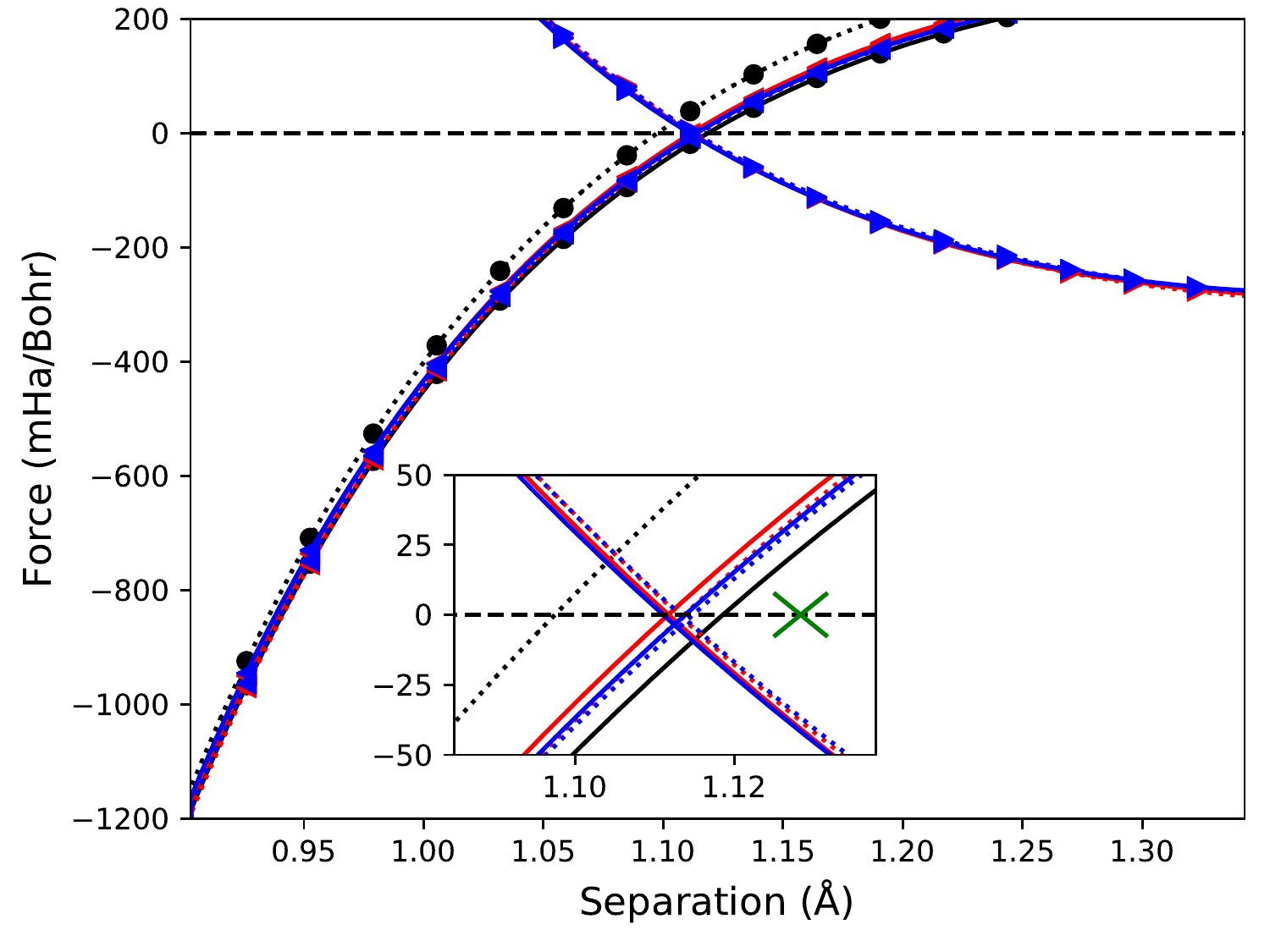}};
    \node at ( 3.3, 0) {\includegraphics[width=6cm]{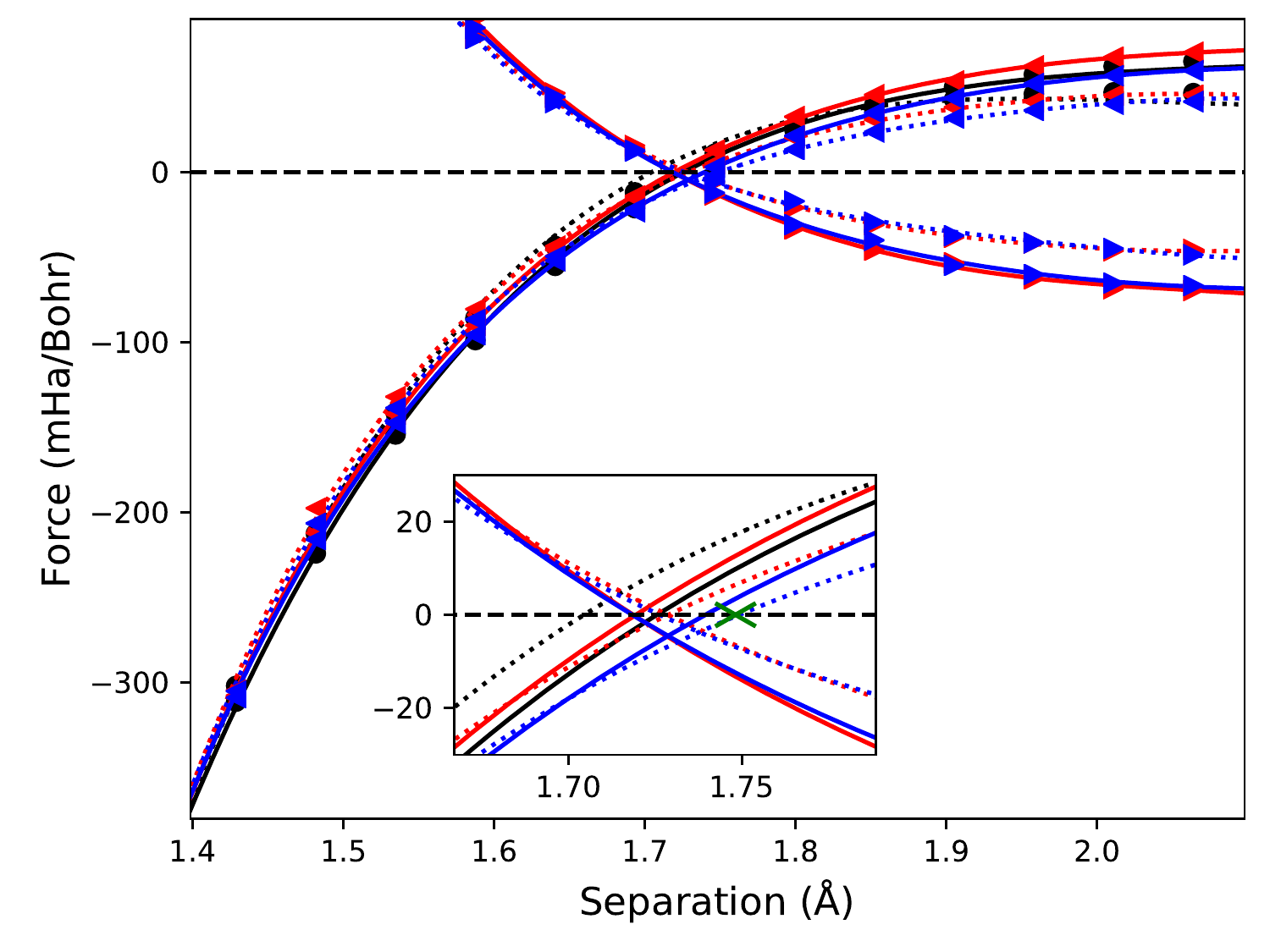}};
    \node at (-3.3,-4.8) {\includegraphics[width=6cm]{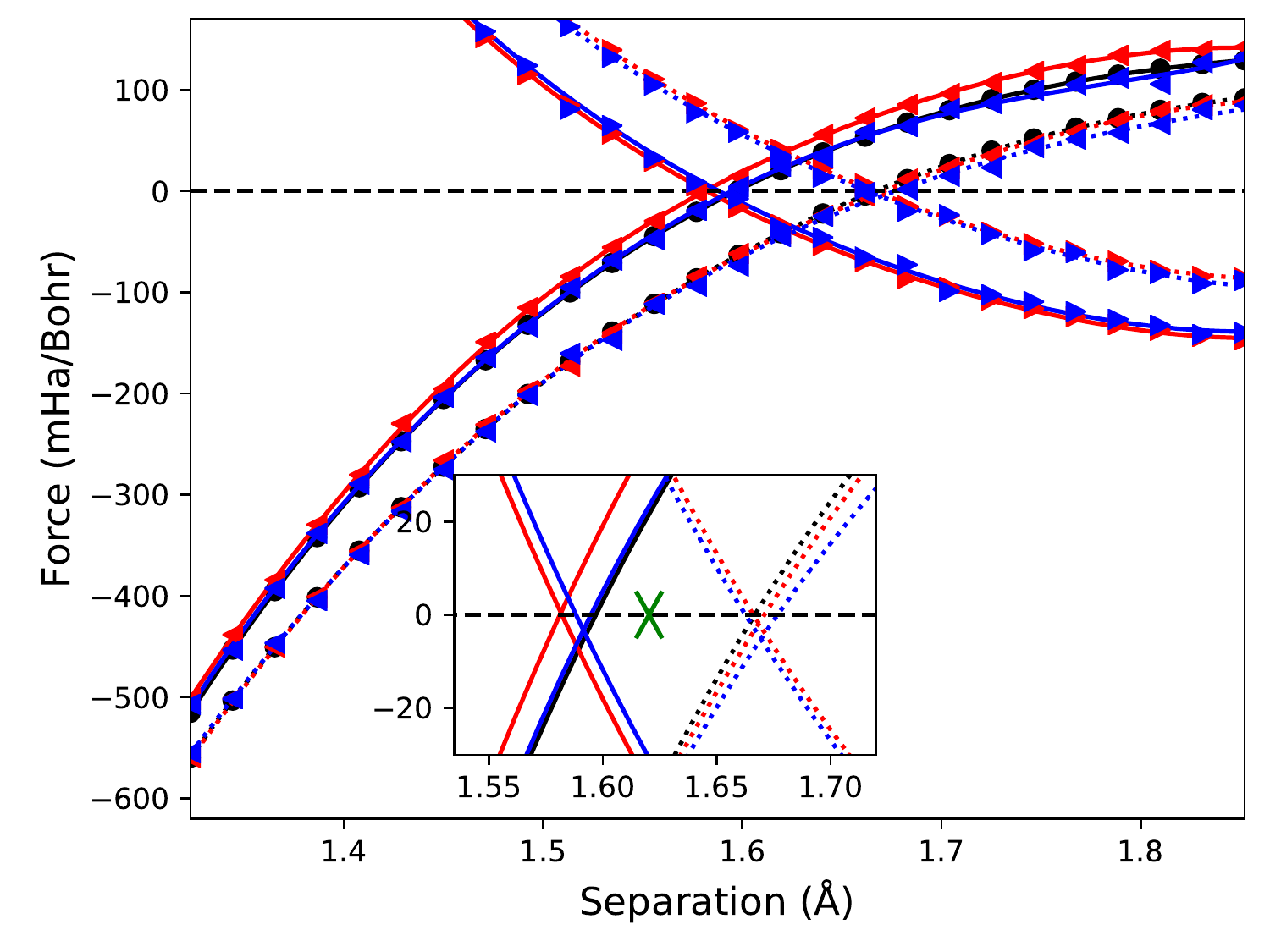}};
    \node at ( 3.3,-4.8) {\includegraphics[width=6cm]{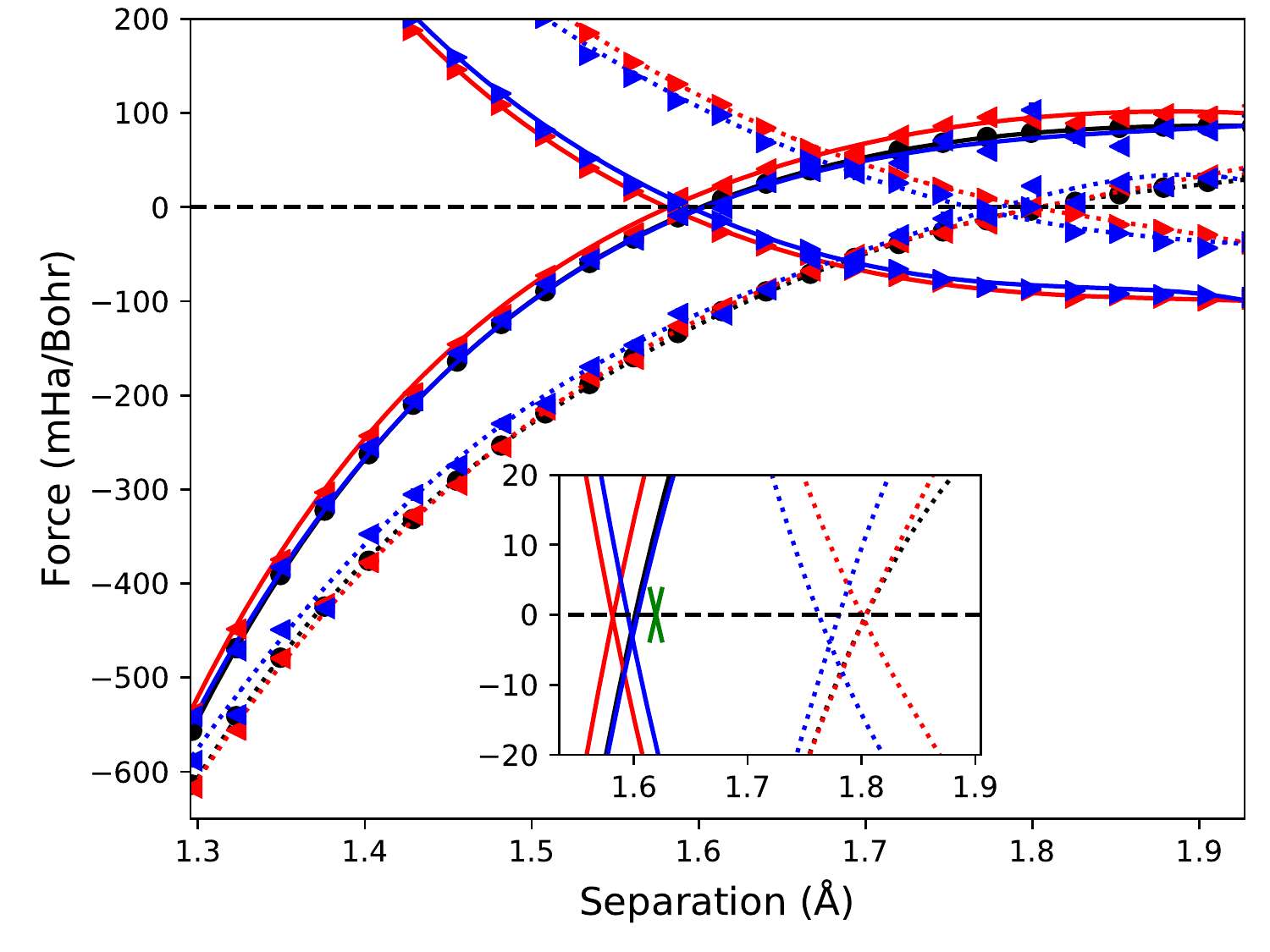}};
    \end{tikzpicture}
    \caption{ZVZB forces for each dimer: LiH (top-left), Li$_2$ (top-right), CO (middle-left), MgO (middle-right), TiO (bottom-left), FeO (bottom-right). The solid (PBE0) and dotted (UHF) lines with black (SCF), red (VMC) and blue (DMC) color are 5-degree polynomial fits and markers are the QMC estimates. The error bars are smaller than the marker size. Green crosses insets are located at the experimental bond lengths with slopes of the experimental force constant.}
    \label{fig:Fcurves}
\end{figure}

\begin{figure}[h]
    \centering
    \begin{tikzpicture}
    \node at (-3.3, 4.8) {\includegraphics[width=6cm]{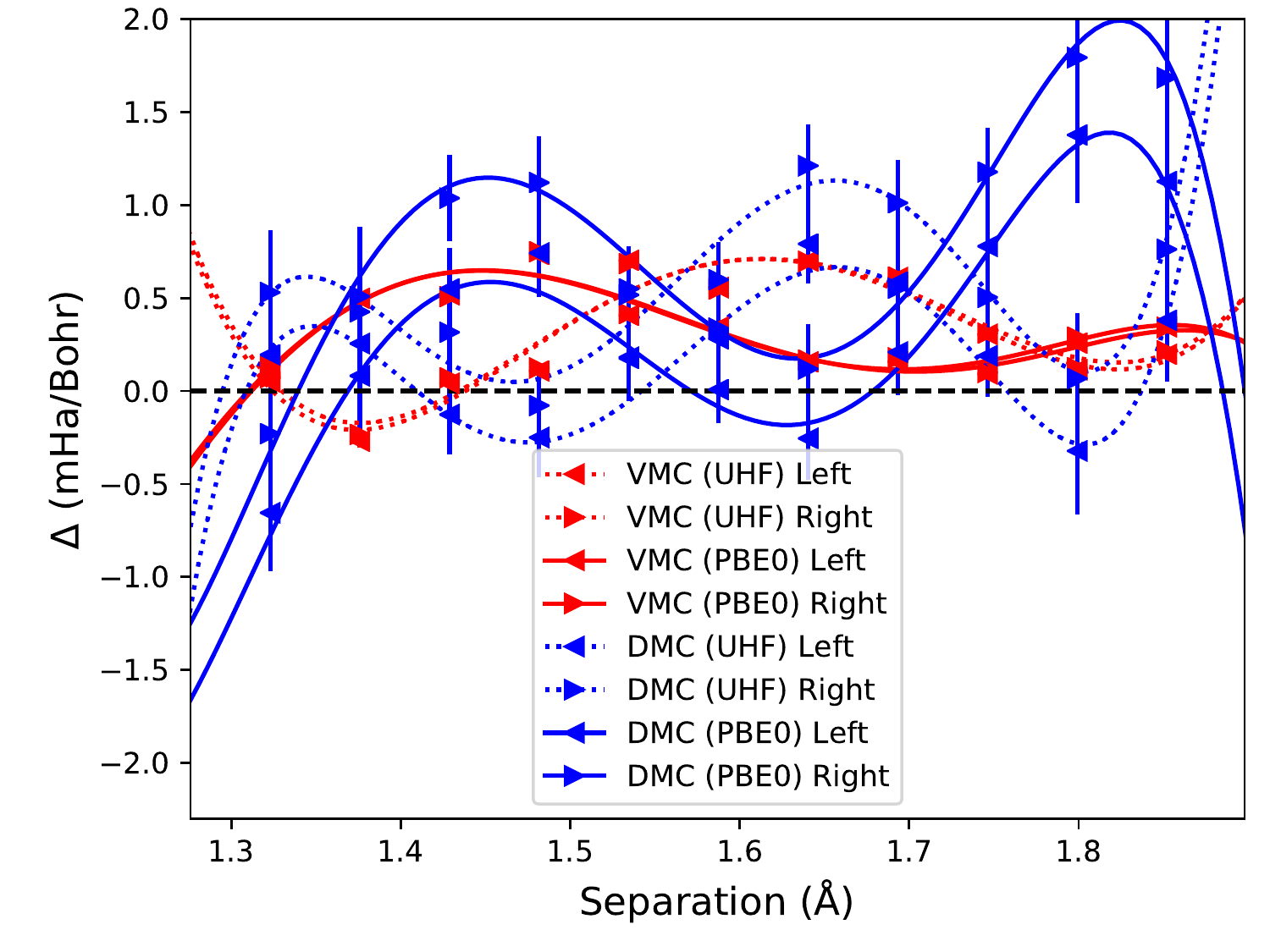}};
    \node at ( 3.3, 4.8) {\includegraphics[width=6cm]{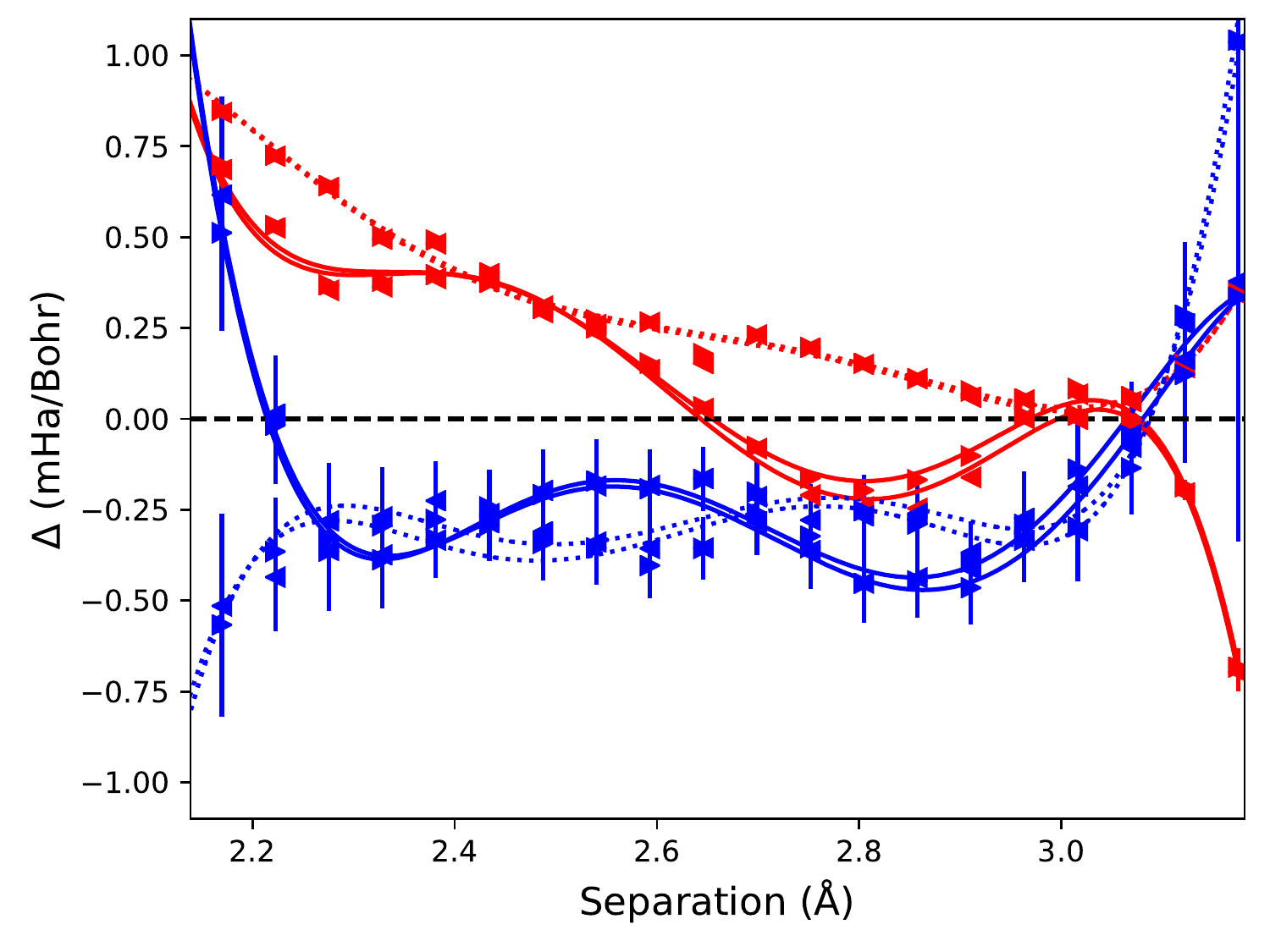}};
    \node at (-3.3, 0) {\includegraphics[width=6cm]{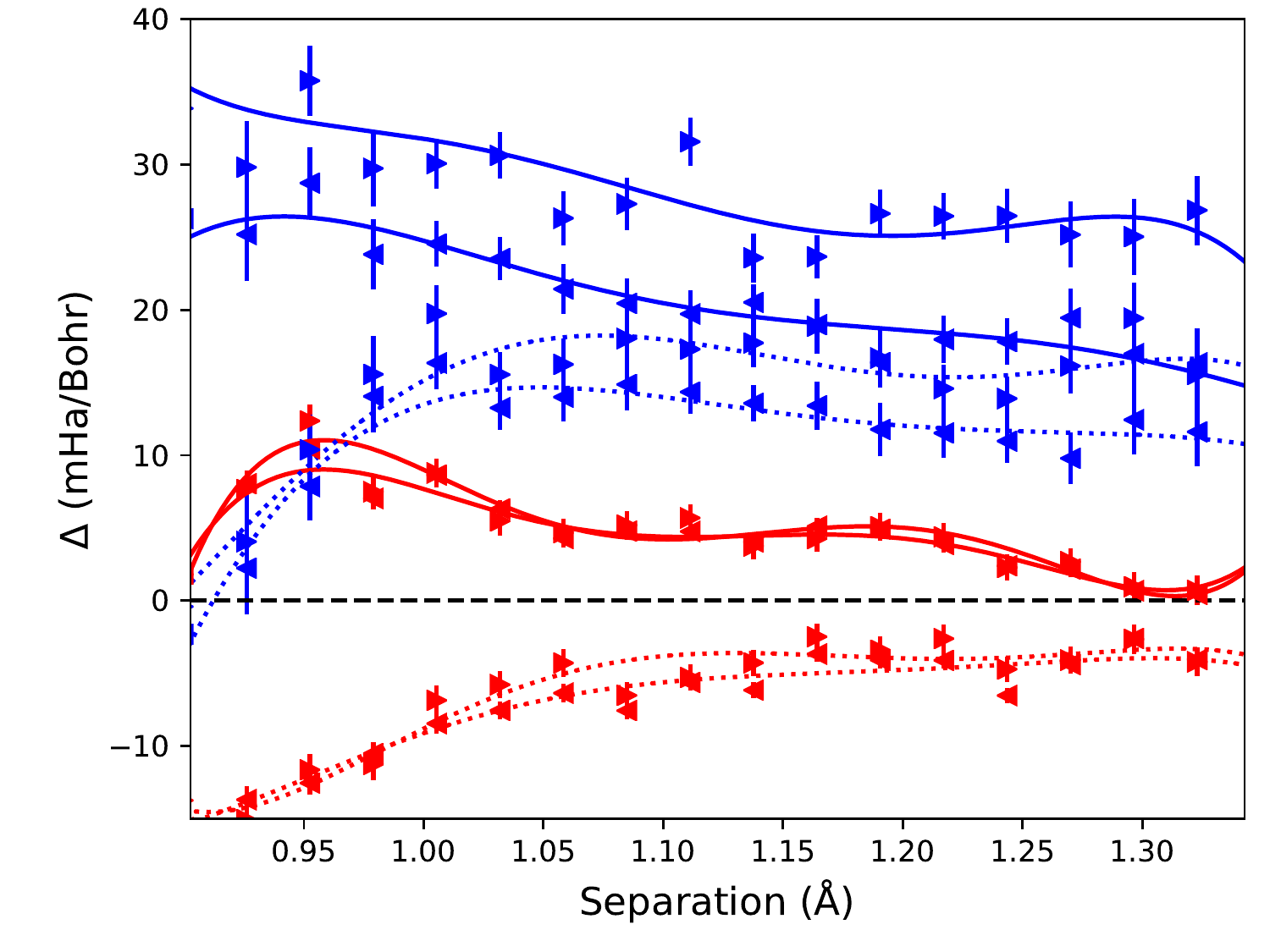}};
    \node at ( 3.3, 0) {\includegraphics[width=6cm]{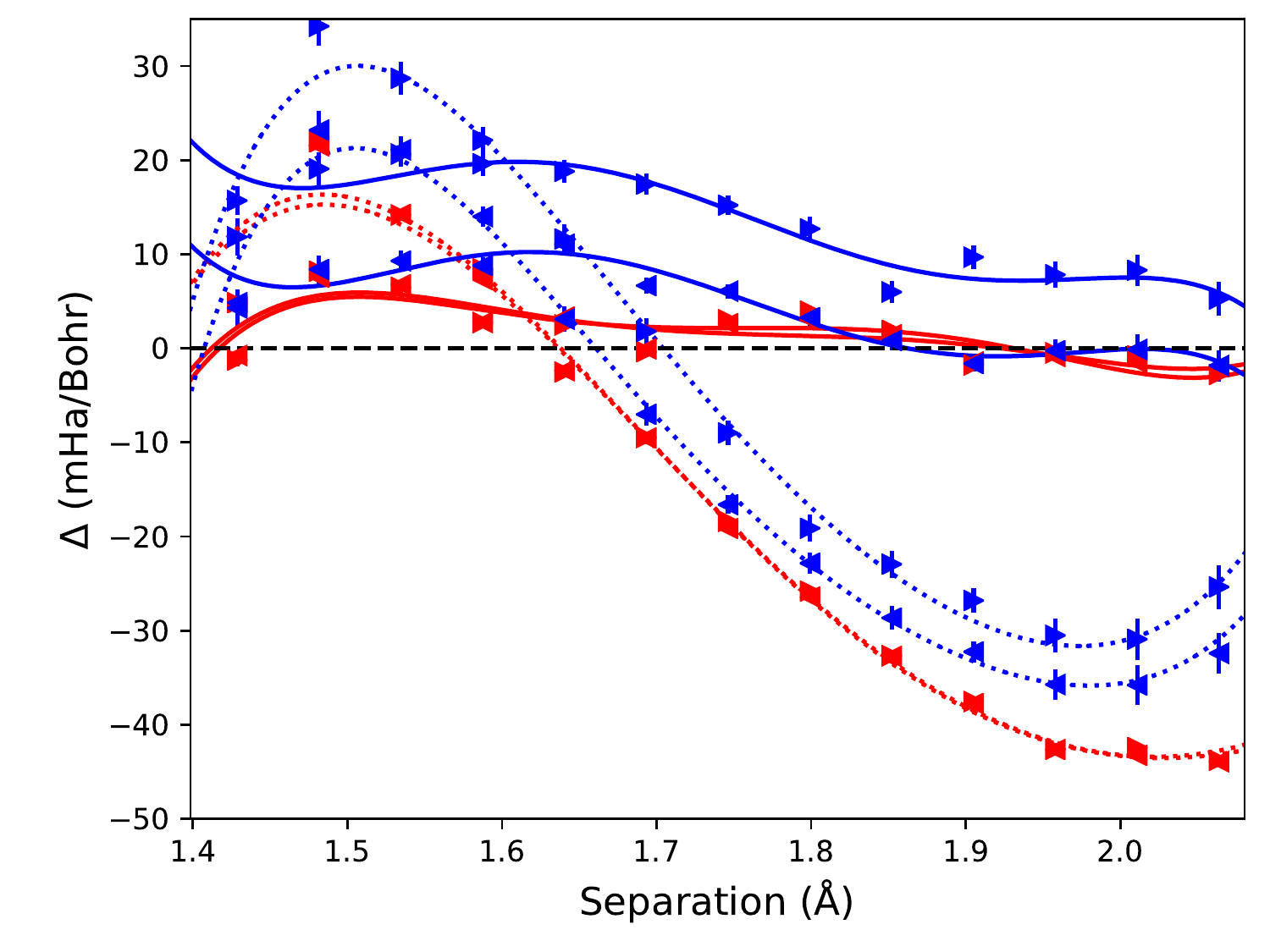}};
    \node at (-3.3,-4.8) {\includegraphics[width=6cm]{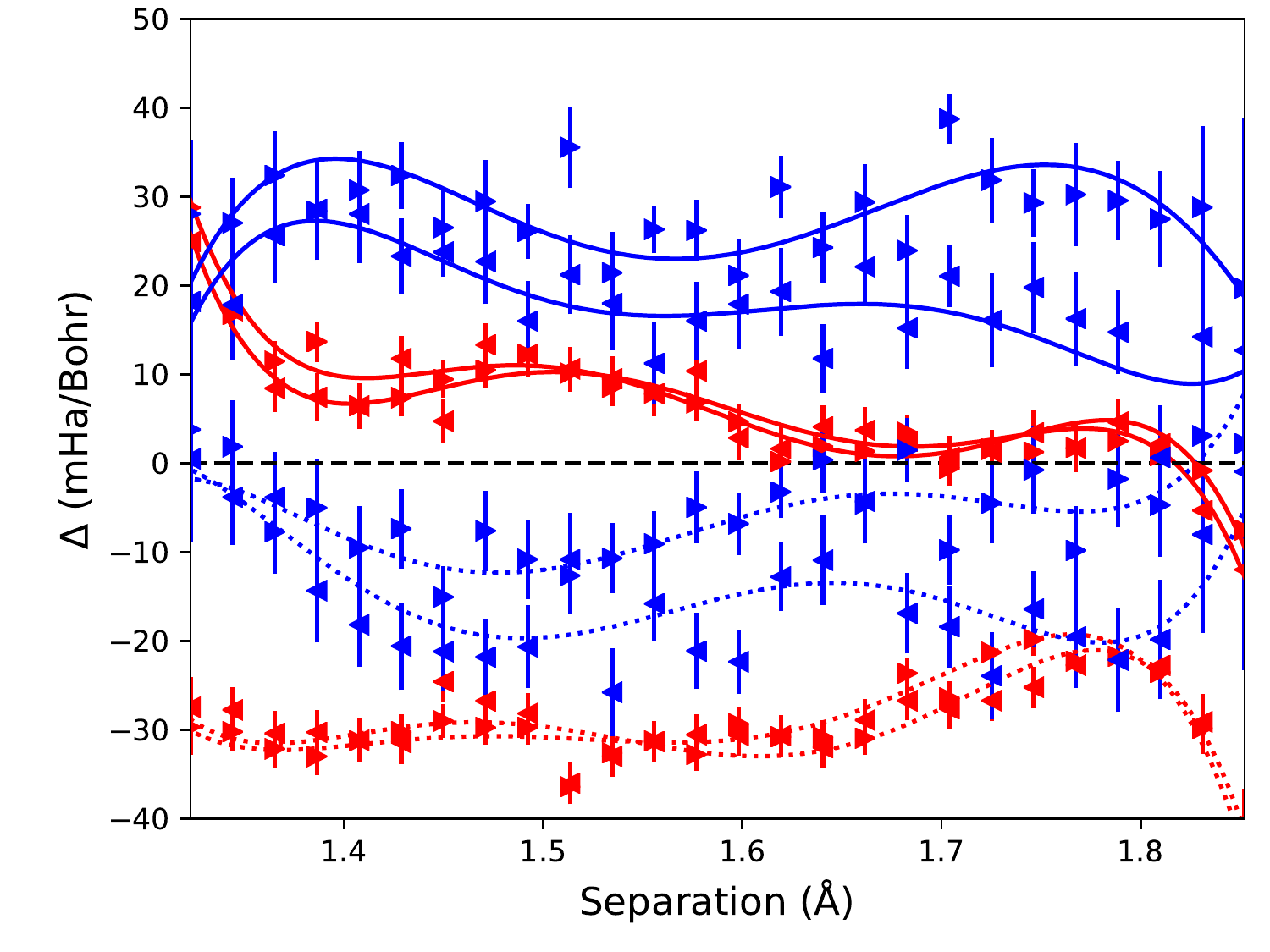}};
    \node at ( 3.3,-4.8) {\includegraphics[width=6cm]{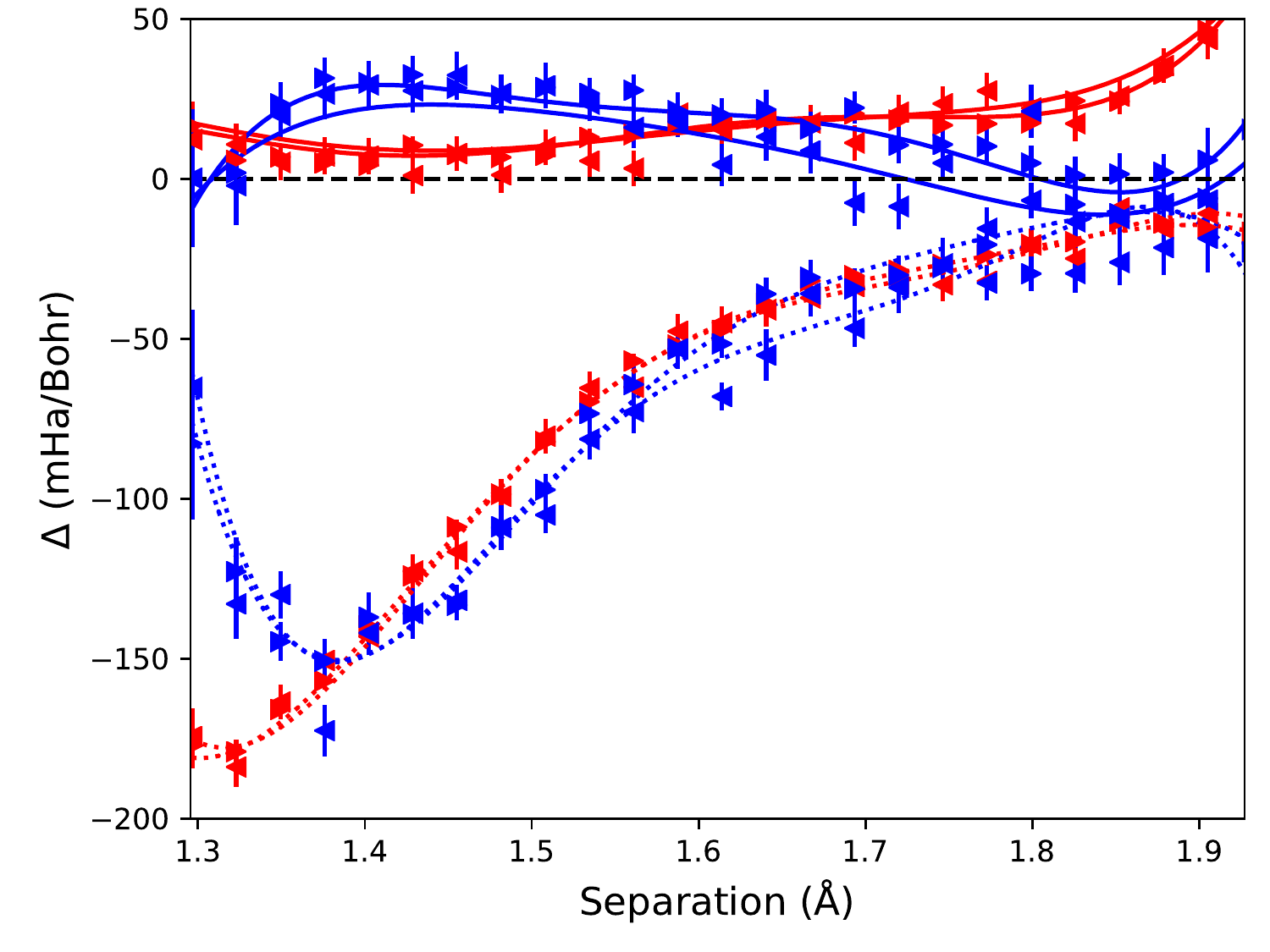}};
    \end{tikzpicture}
    \caption{Self-consistency error $\Delta$ for each dimer: LiH (top-left), Li$_2$ (top-right), CO (middle-left), MgO (middle-right), TiO (bottom-left), FeO (bottom-right). The solid (PBE0) and dotted (UHF) lines with red (VMC) and blue (DMC) color are differences between 5-degree polynomial fit to forces and the derivative of 6-degree polynomial fit to energies. The markers are original QMC forces adjusted by the energy derivatives.}
    \label{fig:SCcurves}
\end{figure}

Binding curves (PES) for each dimer are presented in Figs.~\ref{fig:Ecurves} with SCF, VMC and DMC, and trial wavefunctions based on UHF and PBE0. The respective ZVZB force curves are presented in \ref{fig:Fcurves}, and self-consistency curves (force plus the PES gradient) in Fig.~\ref{fig:SCcurves}. Coarse fluctuations in the self-consistency curves are mostly due to statistical and residual errors in fitting to the PES. The basis sets are the following:  LiH (VTZ), Li$_2$ (VTZ), CO (VQZ), MgO (VQZ), TiO (VTZ) and FeO (VTZ). 

\FloatBarrier

\section{UHF bond lengths and vibrational frequencies}

Data for estimated bond lengths and vibrational frequencies from SCF, VMC and DMC calculations based on UHF orbitals are presented in Table~\ref{tab:rwdata_uhf}.

\begin{table*}[h]
    \centering
    \begin{tabular}{llllllll}
    & Method          & $r_E$           & $r$ (left)      & $r$ (right)     & $\omega_E$      & $\omega$ (left) &  $\omega$ (right)  \\ \hline\hline
LiH & UHF             & 1.60609         &                 &                 & 1415.4          &                 &                   \\
    & VMC             & 1.60333(14)     & 1.59723(5)      & 1.59728(7)      & 1383(3)         & 1406.6(0.2)     & 1406.5(0.3)       \\
    & DMC             & 1.603(2)        & 1.6022(3)       & 1.5995(5)       & 1330(30)        & 1385(2)         & 1399(3)           \\
Li2 & UHF             & 2.77294         &                 &                 & 337.9           &                 &                   \\
    & VMC             & 2.7257(3)       & 2.71943(7)      & 2.71920(8)      & 343.6(0.4)      & 342.83(6)       & 342.93(6)         \\
    & DMC             & 2.704(3)        & 2.7161(7)       & 2.7171(7)       & 333(5)          & 341.7(0.5)      & 341.5(0.5)        \\
MgO & UHF             & 1.7125          &                 &                 & 800.9           &                 &                   \\
    & VMC             & 1.7008(3)       & 1.72821(13)     & 1.7273(8)       & 929(2)          & 711.5(0.5)      & 715(3)            \\
    & DMC             & 1.722(2)        & 1.7484(5)       & 1.7292(13)      & 894(14)         & 673(2)          & 699(5)            \\
CO  & UHF             & 1.09744         &                 &                 & 2442.8          &                 &                   \\
    & VMC             & 1.11157(11)     & 1.11356(8)      & 1.1133(2)       & 2318(3)         & 2301.4(1.3)     & 2304(3)           \\
    & DMC             & 1.1195(5)       & 1.1145(2)       & 1.1137(2)       & 2244(14)        & 2269(3)         & 2281(3)           \\
TiO & UHF             & 1.66688         &                 &                 & 956.6           &                 &                   \\
    & VMC             & 1.6326(8)       & 1.6701(10)      & 1.6669(8)       & 1000(7)         & 942(7)          & 955(5)            \\
    & DMC             & 1.661(4)        & 1.677(2)        & 1.6609(13)      & 920(30)         & 891(13)         & 952(10)           \\
FeO & UHF             & 1.80768         &                 &                 & 598.9           &                 &                   \\
    & VMC             & 1.737(3)        & 1.802(6)        & 1.799(2)        & 640(20)         & 670(20)         & 646(8)            \\
    & DMC             & 1.719(13)       & 1.781(5)        & 1.767(3)        & 660(40)         & 690(30)         & 700(14)           \\
                        \hline\hline
    \end{tabular}
    \caption{Equilibrium bond lengths $r$ (\AA) and vibrational frequencies $\omega$ (cm$^{-1}$) calculated in this work using UHF, and VMC and DMC with UHF orbitals. Subscript $E$ indicates that the property was obtained from the PES; otherwise the values are fitted to ZVZB forces on left or right ions.}
    \label{tab:rwdata_uhf}
\end{table*}

\FloatBarrier

\section{Comparison to earlier works}

We present comparisons to earlier works on QMC forces \cite{Assaraf2000, Filippi2000, Assaraf2003,Lee2005, Chiesa2005, Lee2007, Casalegno2003,Badinski2007}, where applicable, and selected works on QMC energy surfaces \cite{Cleland2016,Krogel2016}. Comparison of equilibrium force values is done in Table~\ref{tab:Feqms_references}, and binding curves and vibrational frequencies in Table~\ref{tab:rwdata_references}. 

Our observations about the quality of VMC forces are consistent with results from prior works \cite{Filippi2000,Assaraf2003,Casalegno2003,Lee2005,Badinski2007,Cleland2016} that use equivalent single reference trial wavefunctions. Comparable prior force results are available for the LiH, Li$_2$ and CO dimers.  For LiH, we find a force derived bond length of $1.5975(5)$ \AA{} versus $1.5987(6)$\cite{Badinski2007}-$1.610(4)$\cite{Casalegno2003} \AA{} from prior works and $1.5948994$ \AA{} from experiment. Our LiH force derived vibrational frequency is $1408.3(4)$, which compares well with prior works (1372(9)\cite{Badinski2007}-1407(4)\cite{Lee2005} cm$^{-1}$) and experiment (1405.5 cm$^{-1}$).   Li$_2$ is a challenging system from the point of view of obtaining an accurate bond length, largely due to the very soft nature of the bond ($351.41$ cm$^{-1}$ experimentally).  Small errors in energy lead to large errors in the bond length, as seen both in our results and those from prior works.  While the VMC vibrational frequency is well described here ($340.9(5)$ cm$^{-1}$) and elsewhere ($366(9)$ cm$^{-1}$)\cite{Assaraf2003}, the bond length is not.  Our force derived VMC bond length overestimates the experimental value by $0.04246(6)$ \AA{}, while prior works overestimate by $0.0540(11)\cite{Filippi2000}-0.156(15)\cite{Assaraf2003}$ \AA{}.  Based on the experimental force constant for Li$_2$, errors in the PES of about $0.02$ eV in the vicinity of the minimum are all that is required to produce errors in the bond length both we and others observe. 
For CO, one prior force based VMC estimate for the bond length and vibrational frequency is available\cite{Lee2005}.  That work finds a both length $-0.0197(6)$ \AA{} smaller than experiment, similar to our underestimation by $-0.0170(1)$ \AA{}, while overestimating $\omega$ by $369(16)$ cm$^{-1}$ compared to $125(4)$ cm$^{-1}$ currently.  Other more recent studies, such as Refs.~\cite{Badinski2010,Moroni2014,Rios2019}, feature improved wave functions and methods, although results from those works are not represented here due to lack of overlapping data. 

Our results for LiH and Li$_2$ with mixed DMC forces are more accurate than the mixed estimates from Ref.~\cite{Assaraf2003}, but not the improved estimates obtained by hybrid extrapolation. Advanced DMC estimators, such as Ref.~\cite{Badinski2010}, show impressive accuracy with LiH but also other dimers outside of this study.

Similar to the VMC case, the DMC bond length of Li$_2$ is significantly overestimated, despite its apparent simplicity.  In this case the DMC fixed node approximation is exact since the wavefunction is nodeless (a two electron singlet following the removal of the He core via pseudization).  The only source of uncontrolled approximation in the PES enters through approximate pseudopotential localization via the T-move scheme. We have confirmed that the quality of the pseudopotential itself is not at issue, since the CCSD(T) binding curve yields an essentially exact bond length.  A prior pseudopotential study shows a similar overestimation of the PES bond length, while also demonstrating superior performance with multi-reference wavefunctions\cite{Cleland2016}.  Mixed DMC forces result in a slight increase in the overestimation, showing combined effects of self-consistency error and pseudopotential localization. A pioneering all-electron DMC force study from Ref.~\cite{Filippi2000} reproduces Li$_2$ binding properties with great accuracy using multi-reference trial wavefunction and correlated sampling instead of Pulay corrections. 

For the CO dimer, one previous DMC force study exists\cite{Lee2005}.  There, using Hartree-Fock based trial wavefunctions and mixed DMC forces, the authors find good bond length (accurate to $-0.0086(11)$ \AA{}) and moderate quality frequency of vibration (within $81(26)$ cm$^{-1}$ of experiment).  Our mixed force results are somewhat worse, with the bond length and vibration frequency deviating from experiment by $-0.0161(2)$ \AA{} and $100(2)$ cm$^{-1}$, but still within the range of variability seen for other systems. 

No earlier DMC force data is available for the metal oxide dimers. Compared to earlier benchmark studies \cite{Cleland2016,Krogel2016} of the DMC PES, this present work obtains a similar accuracy for the derived bond length and vibration frequencies, but generally with larger error bars due the the sensitivities of fitting to polynomials instead of the Morse potential.

\begin{table}[]
    \centering
    \begin{tabular}{llllll}
Dimer & Reference               & Method      & Orbitals  & $F$ (left)      & $F$ (right)     \\ \hline\hline
LiH & This work                 & VMC         & SD-PBE0   & -0.617(14)       & 0.64(2)       \\
    &                           & DMC         &           & -0.98(8)         & 0.47(10)    \\
    & Ref.~\cite{Assaraf2000}   & VMC         & SD-HF     & -26.3(2)         &                 \\
    &                           & DMC         &           & -12.5(9)         &                 \\
    &                           & hybrid      &           & -1.3(1.1)        &                 \\
    & Ref.~\cite{Casalegno2003} & VMC         & SD-HF     & -1.9(4)          &                 \\
    &                           & DMC         &           & -0.2(2)          &                 \\
Li$_2$ & This work              & VMC         & SD-PBE0   & -1.460(7)        & 1.472(7)     \\
    &                           & DMC         &           & -1.39(5)         & 1.36(5)     \\
    & Ref.~\cite{Assaraf2000}   & VMC         & SD-HF     & -196(2)          &                 \\
    &                           & DMC         &           & -96(3)           &                 \\
    &                           & hybrid      &           & -4(4)            &                 \\
    & Ref.~\cite{Assaraf2003}   & VMC         & MD-VB     &  6.92(9)         &                 \\
    &                           & DMC         &           &  3.58(19)        &                 \\
    &                           & hybrid      &           &  0.24(39)        &                 \\
CO  & This work                 & VMC         & SD-PBE0   &  40.0(0.4)       &-40.5(0.9)       \\
    &                           & DMC         &           &  29.7(0.6)       &-38.7(1.4)       \\
    & Ref.~\cite{Lee2007}       & VMC         & SD-HF     &  28(5)           &                 \\
    &                           & DMC         &           &  16(8)           &                 \\
                        \hline\hline
    \end{tabular}
    \caption{Comparison of this work (PBE0 with V5Z) and earlier QMC studies on left and right forces (mHa/Bohr), when applicable, at experimental equilibrium geometry. \label{tab:Feqms_references}}
\end{table}

\begin{table}[]
    \centering \scriptsize{
    \begin{tabular}{llllllllll}
Dimer& Reference                & Method      & Orbitals  & $r$ (PES)       & $r$ (left)      & $r$ (right)     & $\omega$ (PES)  & $\omega$ (left) &  $\omega$ (right) \\ \hline\hline
LiH & Expt                      &             &           & 1.5948994       &                 &                 & 1405.5          &                 &                   \\
    & This work                 & VMC         & SD-PBE0   & 1.6002(2)       & 1.59750(5)      & 1.59746(8)      & 1417(3)         & 1408.3(0.2)     & 1408.3(0.4)       \\
    &                           & DMC         &           & 1.602(2)        & 1.6026(3)       & 1.6003(5)       & 1400(30)        & 1389(2)         & 1392(3)           \\
    &                           & hybrid      &           &                 & 1.6073(6)       & 1.6036(10)      &                 & 1366(4)         & 1375(5)           \\
    & Ref.~\cite{Assaraf2003}   & VMC         & MD-VB     &                 & 1.646(9)        &                 &                 & 1559(40)        &                   \\
    &                           & DMC         &           &                 & 1.617(4)        &                 &                 & 1549(6)         &                   \\
    &                           & hybrid      &           &                 & 1.588(8)        &                 &                 & 1519(31)        &                   \\
    & Ref.~\cite{Casalegno2003} & VMC         & SD-HF     & 1.595(14)       & 1.610(4)        &                 &                 &                 &                   \\ 
    &                           & DMC         &           & 1.609(3)        & 1.601(3)        &                 &                 &                 &                   \\
    & Ref.~\cite{Chiesa2005}    & DMC         & SD-HF     &                 & 1.592(4)        &                 &                 & 1402(4)         &                   \\
    & Ref.~\cite{Lee2005}       & VMC         & SD-HF     &                 & 1.6076(6)       &                 &                 & 1407(4)         &                   \\ 
    &                           & DMC         &           &                 & 1.5981(6)       &                 &                 & 1445(20)        &                   \\ 
    & Ref.~\cite{Badinski2007}  & VMC         & SD-HF     & 1.5952(4)       & 1.6063(8)       & 1.5911(8)       & 1416(15)        & 1392(13)        & 1352(12)          \\
    &                           & DMC         &           & 1.6000(3)       & 1.5995(8)       & 1.5970(12)      & 1395(20)        & 1424(24)        & 1398(22)          \\
    & Ref.~\cite{Cleland2016}   & VMC         & SD-B3LYP  & 1.6007(10)      &                 &                 &                 &                 &                   \\
    &                           & VMC         & MD-MCSCF  & 1.6116(9)       &                 &                 &                 &                 &                   \\
    &                           & DMC         & SD-PBE    & 1.6047(10)      &                 &                 &                 &                 &                   \\
    &                           & DMC         & MD-MCSCF  & 1.6077(10)      &                 &                 &                 &                 &                   \\
Li$_2$ & Expt                   &             &           & 2.6732974       &                 &                 & 351.41          &                 &                   \\
    & This work                 & VMC         & SD-PBE0   & 2.7113(3)       & 2.71630(8)      & 2.71522(8)      & 350.5(0.4)      & 340.56(6)       & 341.32(7)        \\
    &                           & DMC         &           & 2.708(3)        & 2.7145(7)       & 2.7141(7)       & 350(6)          & 340.4(0.5)      & 338.8(0.5)          \\
    &                           & hybrid      &           &                 & 2.7123(14)      & 2.7123(14)      &                 & 338.3(1.0)      & 336.8(1.1)          \\
    & Ref.~\cite{Filippi2000}   & VMC         & MD-MCSCF  &                 & 2.7273(11)      &                 &                 &                 &                   \\
    &                           & DMC         &           &                 & 2.663(2)        &                 &                 &                 &                   \\
    & Ref.~\cite{Assaraf2003}   & VMC         & MD-VB     &                 & 2.829(15)       &                 &                 & 366(9)          &                   \\
    &                           & DMC         &           &                 & 2.752(9)        &                 &                 & 373(5)          &                   \\
    &                           & hybrid      &           &                 & 2.682(15)       &                 &                 & 387(8)          &                   \\
    & Ref.~\cite{Cleland2016}   & VMC         & SD-B3LYP  & 2.6849(40)      &                 &                 &                 &                 &                   \\
    &                           & VMC         & MD-MCSCF  & 2.6930(8)       &                 &                 &                 &                 &                   \\
    &                           & DMC         & SD-PBE    & 2.722(6)        &                 &                 &                 &                 &                   \\
    &                           & DMC         & MD-MCSCF  & 2.6819(20)      &                 &                 &                 &                 &                   \\
MgO & Expt                      &             &           & 1.7481687       &                 &                 & 785.21          &                 &                   \\
    & This work                 & VMC         & SD-PBE0   & 1.7219(5)       & 1.71864(11)     & 1.7185(6)       & 829(3)          & 821.7(0.4)      & 825(2)         \\
    &                           & DMC         &           & 1.751(2)        & 1.7395(3)       & 1.7185(6)       & 784(13)         & 765.1(0.9)      & 805(4)          \\
    &                           & hybrid      &           &                 & 1.7635(7)       & 1.720(2)        &                 & 705(2)          & 786(7)          \\
    & Ref.~\cite{Cleland2016}   & VMC         & SD-B3LYP  & 1.705(3)        &                 &                 &                 &                 &                   \\
    &                           & VMC         & MD-MCSCF  & 1.753(2)        &                 &                 &                 &                 &                   \\
    &                           & DMC         & SD-HF     & 1.708(2)        &                 &                 &                 &                 &                   \\
    &                           & DMC         & MD-MCSCF  & 1.749(2)        &                 &                 &                 &                 &                   \\
CO  & Expt                      &             &           & 1.1283199       &                 &                 & 2169.81         &                 &                   \\
    & This work                 & VMC         & SD-PBE0   & 1.11343(11)     & 1.11122(9)      & 1.1115(2)       & 2266(3)         & 2296.9(1.3)     & 2293(3)          \\
    &                           & DMC         &           & 1.1225(5)       & 1.1140(2)       & 1.1105(2)       & 2190(20)        & 2252(3)         & 2288(3)          \\
    &                           & hybrid      &           &                 & 1.1164(4)       & 1.1097(4)       &                 & 2215(5)         & 2282(6)          \\
    & Ref.~\cite{Lee2005}       & VMC         & SD-HF     &                 & 1.1086(6)       &                 &                 & 2539(16)        &                   \\ 
    &                           & DMC         &           &                 & 1.1197(11)      &                 &                 & 2251(26)        &                   \\ 
    & Ref.~\cite{Cleland2016}   & VMC         & SD-B3LYP  & 1.11642(50)     &                 &                 &                 &                 &                   \\
    &                           & VMC         & MD-MCSCF  & 1.12672(40)     &                 &                 &                 &                 &                   \\
    &                           & DMC         & SD-HF     & 1.12122(40)     &                 &                 &                 &                 &                   \\
    &                           & DMC         & MD-MCSCF  & 1.12572(30)     &                 &                 &                 &                 &                   \\
TiO & Expt                      &             &           & 1.6203          &                 &                 & 1009.18         &                 &                   \\
    & This work                 & VMC         & SD-PBE0   & 1.5874(6)       & 1.5809(8)       & 1.5821(7)       & 1126(7)         & 1117(5)         & 1113(4)         \\
    &                           & DMC         &           & 1.611(3)        & 1.598(2)        & 1.5857(12)      & 1040(30)        & 1063(8)         & 1115(6)          \\
    &                           & hybrid      &           &                 & 1.619(4)        & 1.589(3)        &                 & 980(20)         & 1089(13)          \\
    & Ref.~\cite{Krogel2016}    & DMC         & SD-LDA    & 1.614(1)        &                 &                 & 1019(6)         &                 &                   \\
    &                           & DMC         & SD-LDA    &                 &                 &                 & 1010(6)         &                 &                   \\
FeO & Expt                      &             &           & 1.6193997       &                 &                 & 882.0           &                 &                   \\
    & This work                 & VMC         & SD-PBE0   & 1.6007(11)      & 1.583(3)        & 1.5792(12)      & 882(10)         & 962(13)         & 950(6)           \\
    &                           & DMC         &           & 1.622(3)        & 1.603(4)        & 1.589(3)        & 900(40)         & 860(20)         & 940(12)          \\
    &                           & hybrid      &           &                 & 1.622(10)       & 1.610(6)        &                 & 800(40)         & 900(30)          \\
    & Ref.~\cite{Krogel2016}    & DMC         & SD-LDA    & 1.618(1)        &                 &                 & 894(5)          &                 &                   \\
    &                           & DMC         & SD-LDA    &                 &                 &                 & 881(6)          &                 &                   \\
                        \hline\hline
    \end{tabular}
    \caption{Comparison of this work (PBE0) and selected values from other QMC studies on bond lengths $r$ (\AA) and vibrational frequencies $\omega$ (cm$^{-1}$) based on left or right force, or the PES.}
    \label{tab:rwdata_references}}
\end{table}

\FloatBarrier

\section{Intrinsic variances and their scaling relations with $\zeff$}

The estimated intrinsic variances of energies and ZVZB-TRE forces are presented in Table~\ref{tab:EF_variances_atoms} for VMC and DMC simulations of isolated atoms. The intrinsic variances of ZVZB-TRE and ZVZB-SW-TRE forces are given in Tables~\ref{tab:F_variances_homonuclear_vmc} and \ref{tab:F_variances_homonuclear_dmc} based on VMC and DMC simulations of homonuclear dimers, respectively.

\begin{figure}[h!]
    \centering
    \begin{tikzpicture}
    \node at (0, 0) {\includegraphics[width=6.8cm]{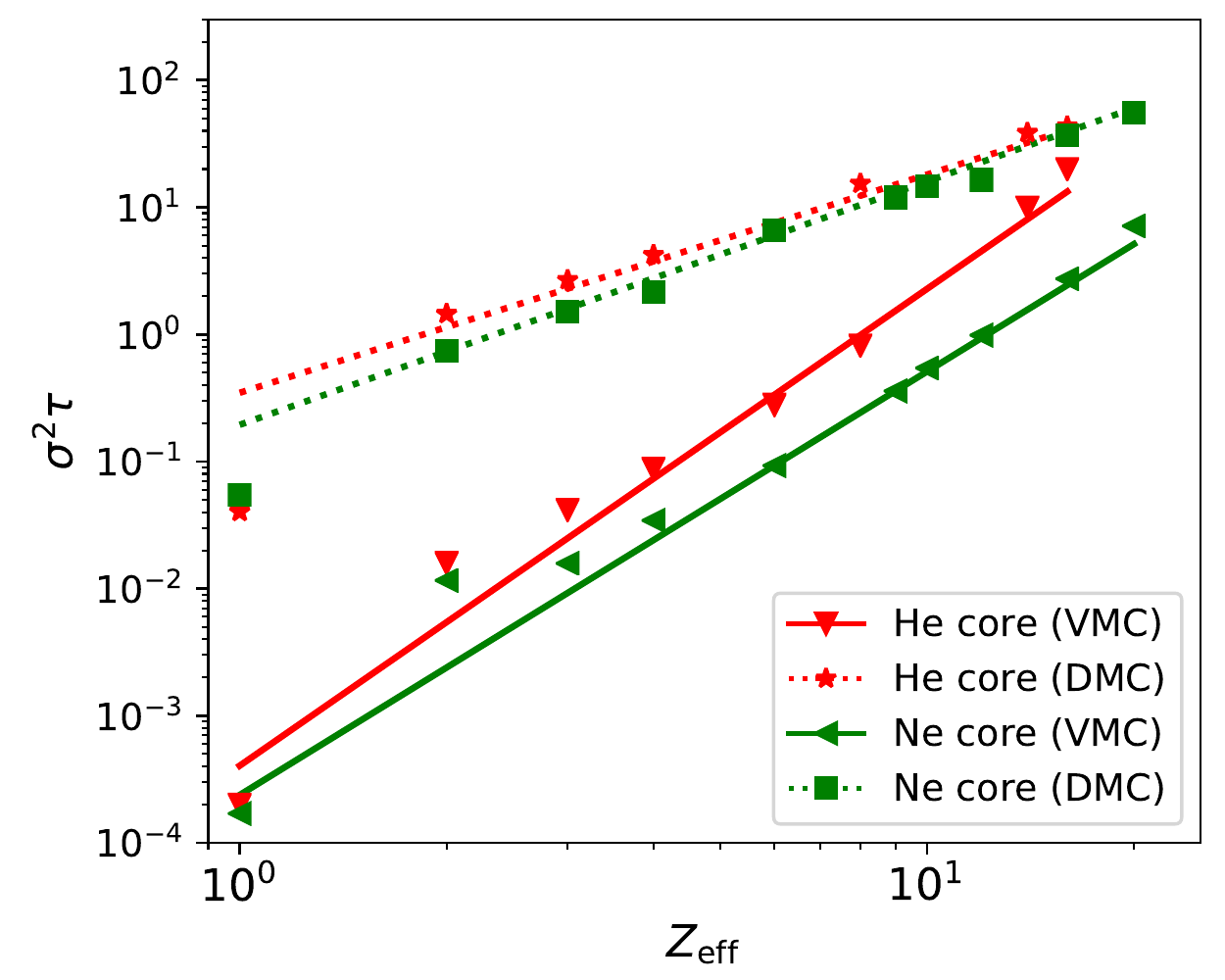}};
    \node at (6.8, 0) {\includegraphics[width=6.8cm]{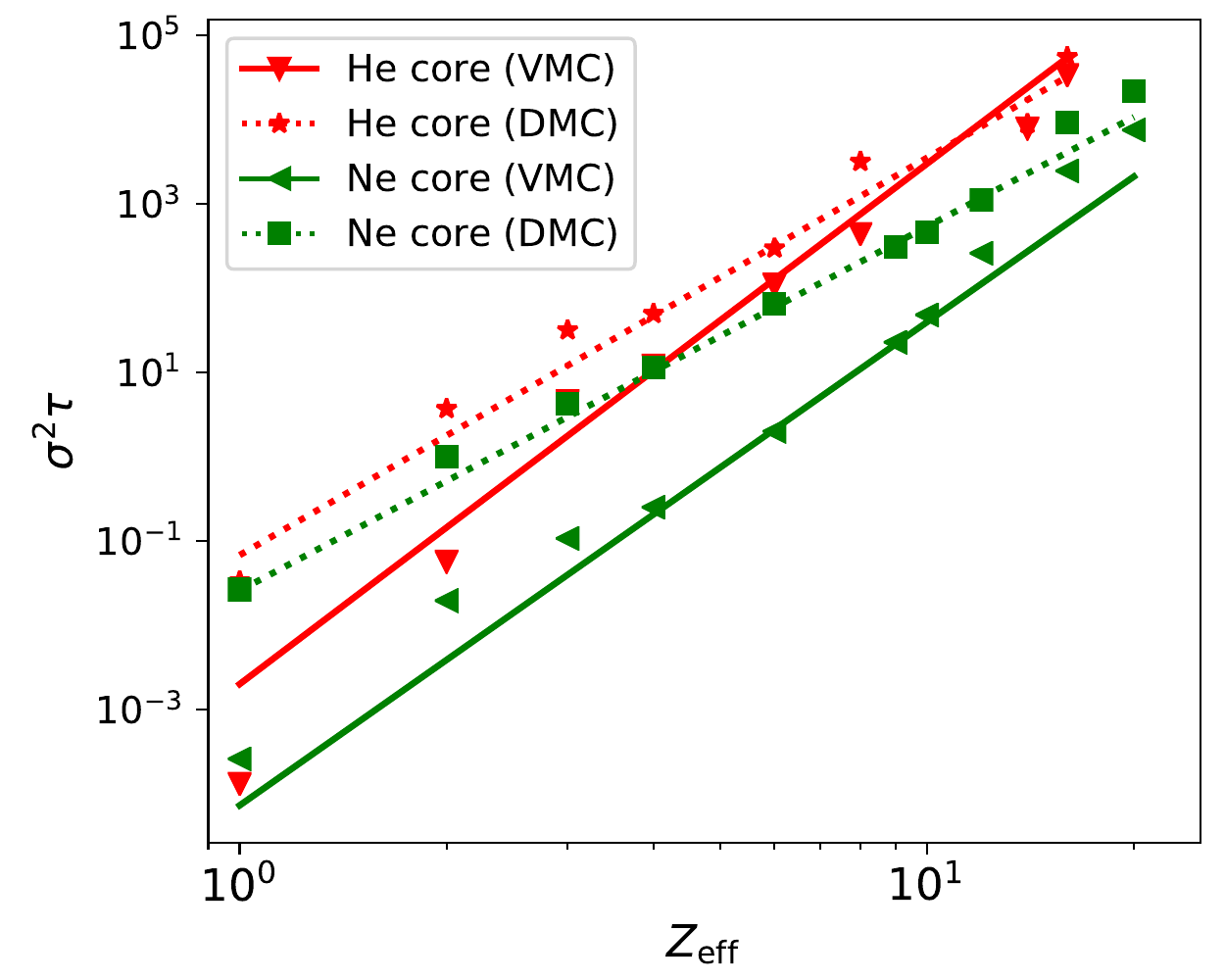}};
    \end{tikzpicture}
    \caption{Autocorrection normalized variances $\sigma^2 \tau$ of DMC and VMC energies and ZVZB-TRE forces.}
    \label{fig:E_F_cost}
\end{figure}

\begin{figure}[h!]
    \centering
    \begin{tikzpicture}
    \node at (0, 0) {\includegraphics[width=6.8cm]{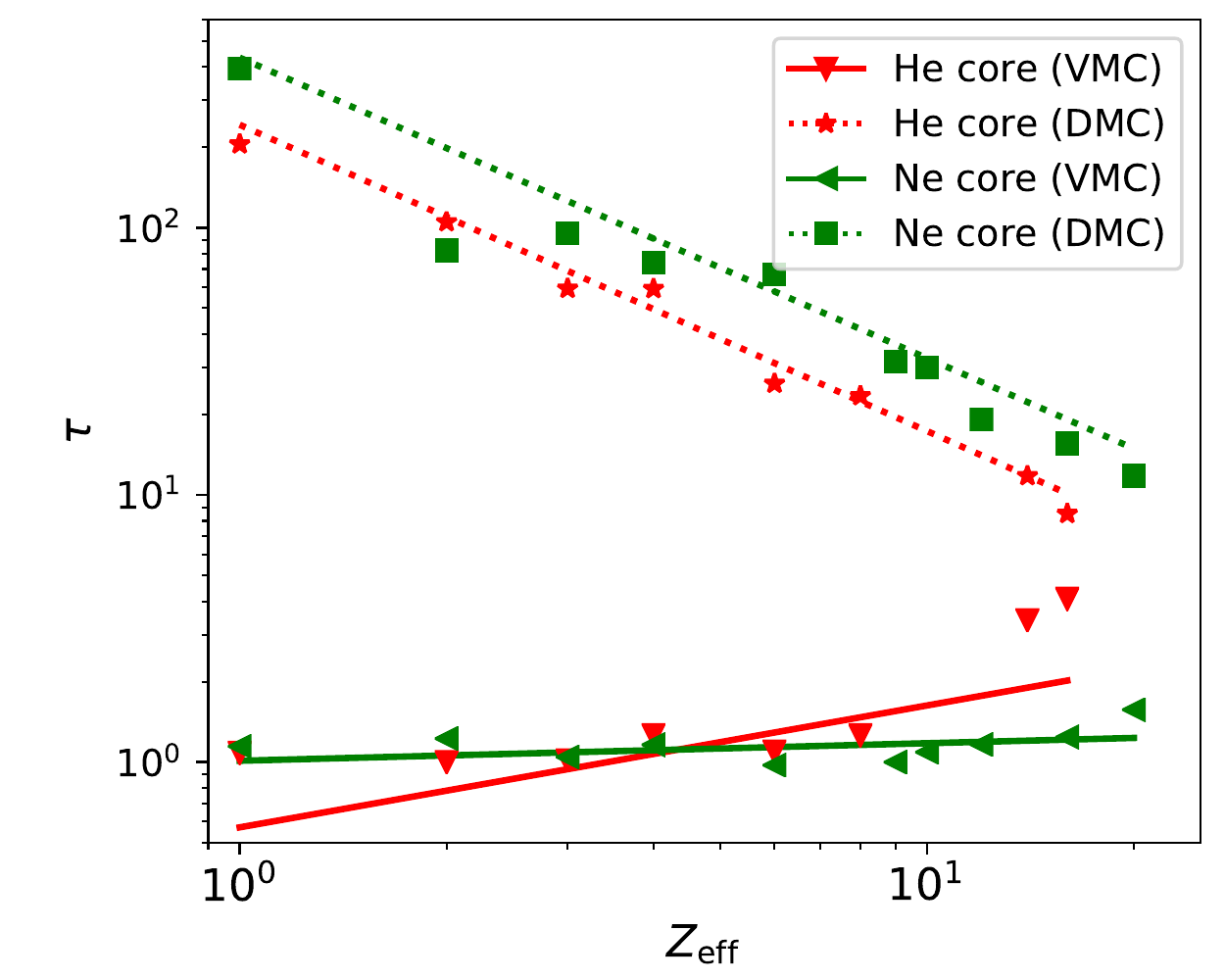}};
    \node at (6.8, 0) {\includegraphics[width=6.8cm]{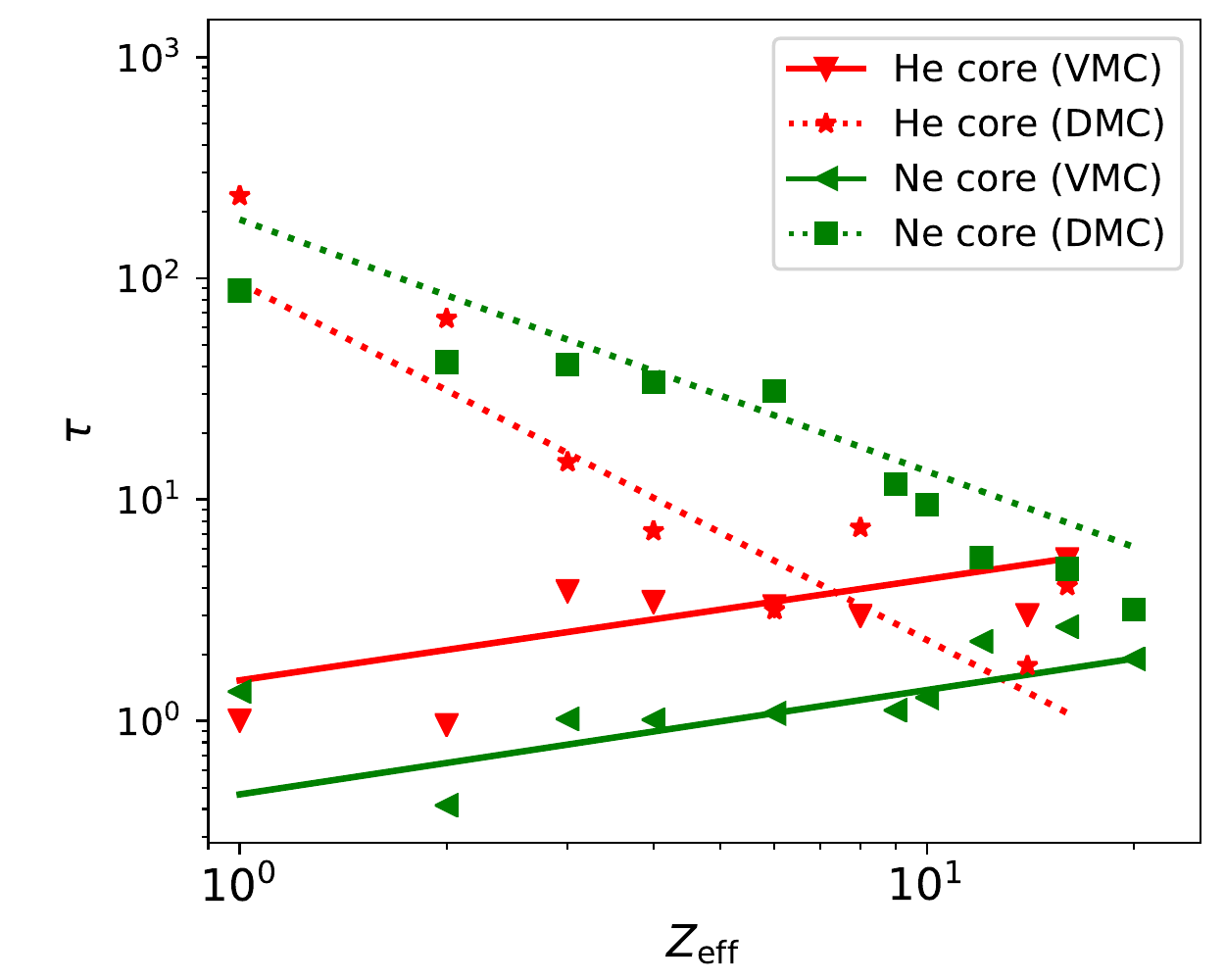}};
    \end{tikzpicture}
    \caption{Sample autocorrelation times $\tau$ of DMC energies and ZVZB-TRE forces.}
    \label{fig:E_F_tau}
\end{figure}

\begin{figure}[h!]
    \centering
    \begin{tikzpicture}
    \node at (0, 0) {\includegraphics[width=7cm]{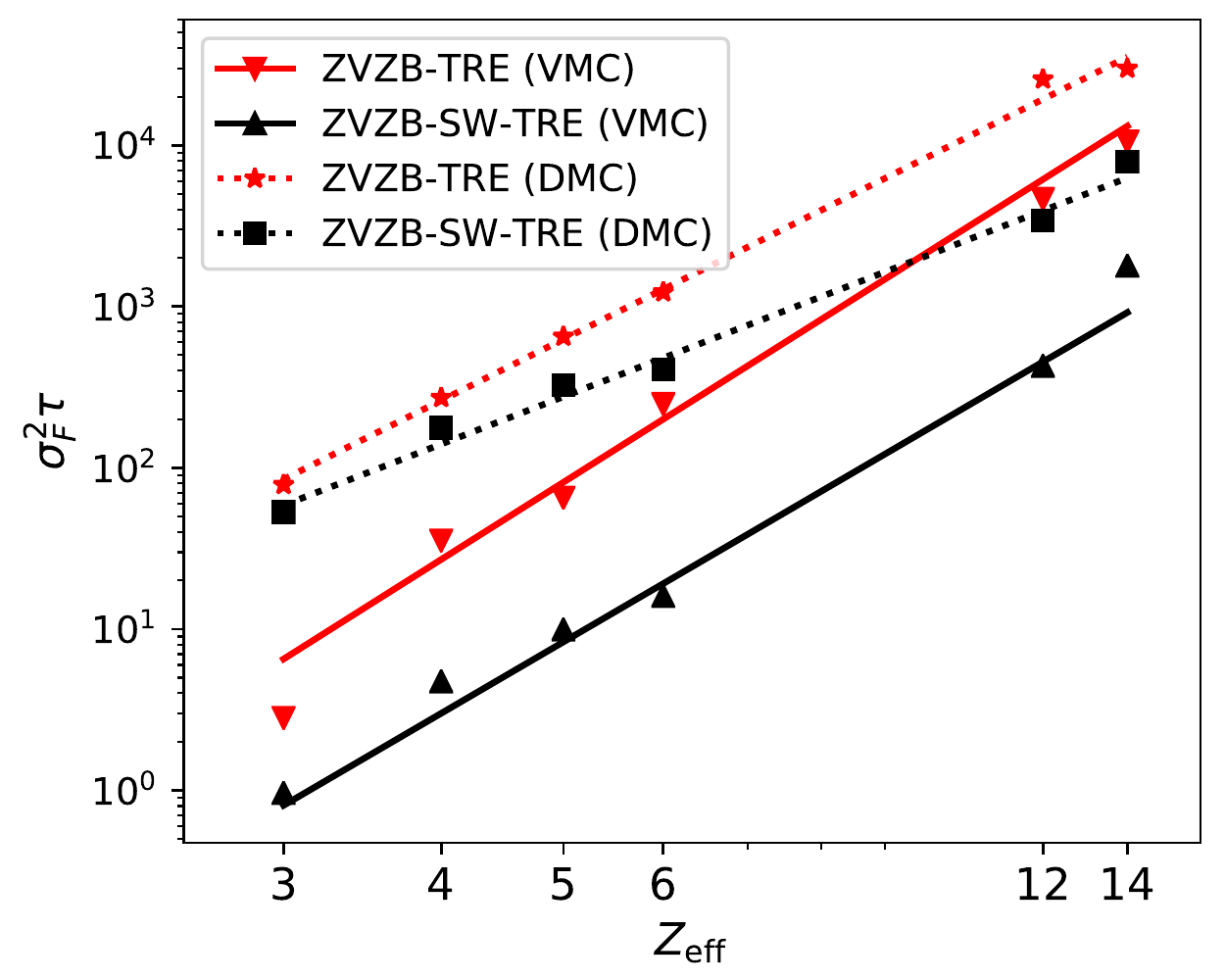}};
    \end{tikzpicture}
    \caption{Autocorrection normalized variances $\sigma^2 \tau$ of DMC and VMC ZVZB-SW-TRE and ZVZB-TRE forces.}
    \label{fig:E_F_cost_homonuclear}
\end{figure}

Scaling relations of form $a \zeff^b$ have studied using Theil-Sen fits of the energies (E) and ZVZB-TRE forces (F) in VMC and DMC simulation: intrinsic variances $\sigma^2$, autocorrelation times $\tau$, and the autocorrelation renormalized variance $\sigma^2\tau$. The fitted coefficients are presented in Table~\ref{tab:fitdata_si}, and Figs.~\ref{fig:E_F_cost}, \ref{fig:E_F_tau}, and \ref{fig:E_F_cost_homonuclear}.

\begin{table}[h]
    \centering
    \begin{tabular}{lllllll}
Atom   & Core    & $\zeff$& E (VMC)  & E (DMC)  & F (VMC)  & F (DMC)  \\ \hline \hline
Li     & He      &      1 &  0.00018 &   0.0002 &  0.00013 &  0.00014 \\
Be     &         &      2 &    0.016 &    0.014 &     0.06 &    0.056 \\
B      &         &      3 &    0.041 &    0.045 &      1.1 &      2.3 \\
C      &         &      4 &    0.069 &    0.072 &      3.5 &      7.2 \\
O      &         &      6 &     0.25 &     0.26 &       35 &       94 \\
Ne     &         &      8 &     0.65 &     0.66 &      150 &      410 \\
S      &         &     14 &      2.9 &      3.3 &     2600 &     5000 \\
Ar     &         &     16 &      4.9 &      5.1 &     6300 &    14000 \\
Na     & Ne      &      1 &  0.00015 &  0.00014 &   0.0002 &   0.0003 \\
Mg     &         &      2 &   0.0095 &    0.009 &    0.052 &    0.024 \\
Al     &         &      3 &    0.015 &    0.016 &     0.11 &     0.11 \\
Si     &         &      4 &     0.03 &    0.029 &     0.25 &     0.41 \\
S      &         &      6 &    0.095 &    0.098 &      1.9 &      2.1 \\
K      &         &      9 &     0.36 &     0.38 &       20 &       26 \\
Ca     &         &     10 &      0.5 &     0.49 &       38 &       49 \\
Ti     &         &     12 &     0.85 &     0.86 &      110 &      200 \\
Fe     &         &     16 &      2.2 &      2.4 &      930 &     1900 \\
Zn     &         &     20 &      4.5 &      4.7 &     3900 &     6800 \\ \hline \hline
    \end{tabular}
    \caption{Intrinsic variances of energies (E) and ZVZB-TRE forces (F) of isolated atoms with VMC and DMC}
    \label{tab:EF_variances_atoms}
\end{table}

\begin{table}[h]
    \centering
    \begin{tabular}{lllllll}
Atom   & Core    & $\zeff$ & ZVZB-TRE (z) & ZVZB-SW-TRE (x,y) & ZVZB-TRE (x,y) & ZVZB-SW-TRE (x,y) \\ \hline \hline
B     & He      &      3 &      1.9 &     0.61 &      1.9 &    0.056 \\
C     &         &      4 &       13 &      4.5 &        8 &     0.55 \\
N     &         &      5 &       33 &        9 &       20 &      0.9 \\
O     &         &      6 &       60 &       12 &       60 &      1.1 \\
Si    &         &     12 &     2000 &      270 &     2000 &     0.53 \\
S     &         &     14 &     5100 &      700 &     4400 &      1.9 \\
\hline \hline
    \end{tabular}
    \caption{Intrinsic variances of ZVZB-TRE and ZVZB-SW-TRE forces of homonuclear dimers with Helium core from VMC.}
    \label{tab:F_variances_homonuclear_vmc}
\end{table}

\begin{table}[h]
    \centering
    \begin{tabular}{lllllll}
Atom   & Core    & $\zeff$ & ZVZB-TRE (z) & ZVZB-SW-TRE (x,y) & ZVZB-TRE (x,y) & ZVZB-SW-TRE (x,y) \\ \hline \hline
B     & He      &      3 &      3.8 &     0.77 &      3.5 &    0.095 \\
C     &         &      4 &       24 &      5.3 &       23 &     0.85 \\
N     &         &      5 &       66 &       12 &       47 &      1.4 \\
O     &         &      6 &      130 &       52 &      140 &      1.9 \\
Si    &         &     12 &     3400 &      320 &     3200 &     0.79 \\
S     &         &     14 &     8900 &      870 &     7700 &     1800 \\ \hline \hline
    \end{tabular}
    \caption{Intrinsic variances of ZVZB-TRE and ZVZB-SW-TRE forces of homonuclear dimers with Helium core from DMC.}
    \label{tab:F_variances_homonuclear_dmc}
\end{table}

\begin{table}[h]
    \centering
    \begin{tabular}{lllll}
Quantity                & Core          & Method      & $a$           & $b$ \\ \hline \hline
$\sigma^2_E$            & He            & VMC         & 0.0012        & 3.0(5)     \\
                        & He            & DMC         & 0.0011        & 3.0(4)     \\
                        & Ne            & VMC         & 0.00036       & 3.1(2)     \\
                        & Ne            & DMC         & 0.00038       & 3.1(2)     \\
$\tau_E$                & He            & VMC         & 0.57          & 0.5(4)  \\
                        & He            & DMC         & 240           &-1.1(0.2)     \\
                        & Ne            & VMC         & 1.0           & 0.1(1)     \\
                        & Ne            & DMC         & 430           &-1.1(2)     \\
$\sigma^2_E \tau_E$     & He            & VMC         & 0.00040       & 3.8(7)     \\
                        & He            & DMC         & 0.35          & 1.7(5)     \\
                        & Ne            & VMC         & 0.00024       & 3.3(3)     \\
                        & Ne            & DMC         & 0.19          & 1.9(2)     \\
$\sigma^2_F$            & He            & VMC         & 0.0017        & 5.5(6)     \\
                        & He            & DMC         & 0.0022        & 5.9(9)     \\
                        & Ne            & VMC         & 0.00010       & 5.5(4)     \\
                        & Ne            & DMC         & 0.000096      & 5.6(4)     \\
$\tau_F$                & He            & VMC         & 1.5           & 0.5(5)     \\
                        & He            & DMC         & 95            &-1.6(8)     \\
                        & Ne            & VMC         & 0.47          & 0.5(3)     \\
                        & Ne            & DMC         & 180           &-1.1(2)     \\
$\sigma^2_F \tau_F$     & He            & VMC         & 0.0020        & 6.2(1.0)     \\
                        & He            & DMC         & 0.068         & 4.7(6)     \\
                        & Ne            & VMC         & 0.000073      & 5.7(5)     \\
                        & Ne            & DMC         & 0.026         & 4.3(3)     \\ \hline \hline
    \end{tabular}
    \caption{Fitting parameters to $aZ_{\mathrm{eff}}^b$ of intrinsic variances of local energy $\sigma^2_E$ and ZVZB-TRE forces from VMC and DMC simulations of isolated atoms with variable pseudized cores.  90\% confidence intervals of the exponents $b$ from Theil-Sen fits are given in parentheses. The prefactors $a$ are in the units of Ha$^2$ for energies and (Ha/Bohr)$^2$ for forces. }
    \label{tab:fitdata_si}
\end{table}

\FloatBarrier

\section{System sizes with equivalent statistical cost}

\begin{figure}[h]
    \centering
    \begin{tikzpicture}
    \node at (-3.5, 0) {\includegraphics[width=7cm]{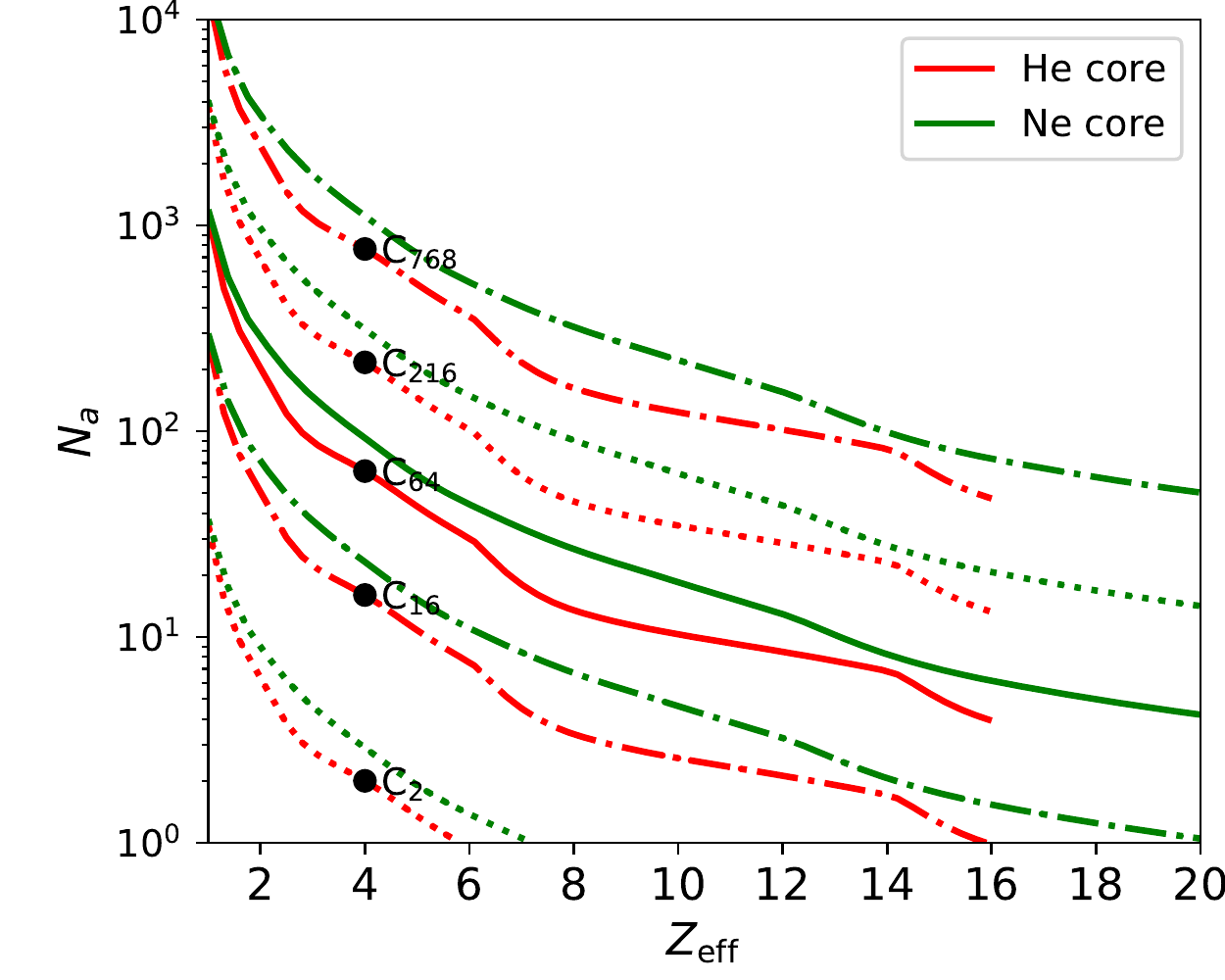}};
    \node at ( 3.5, 0) {\includegraphics[width=7cm]{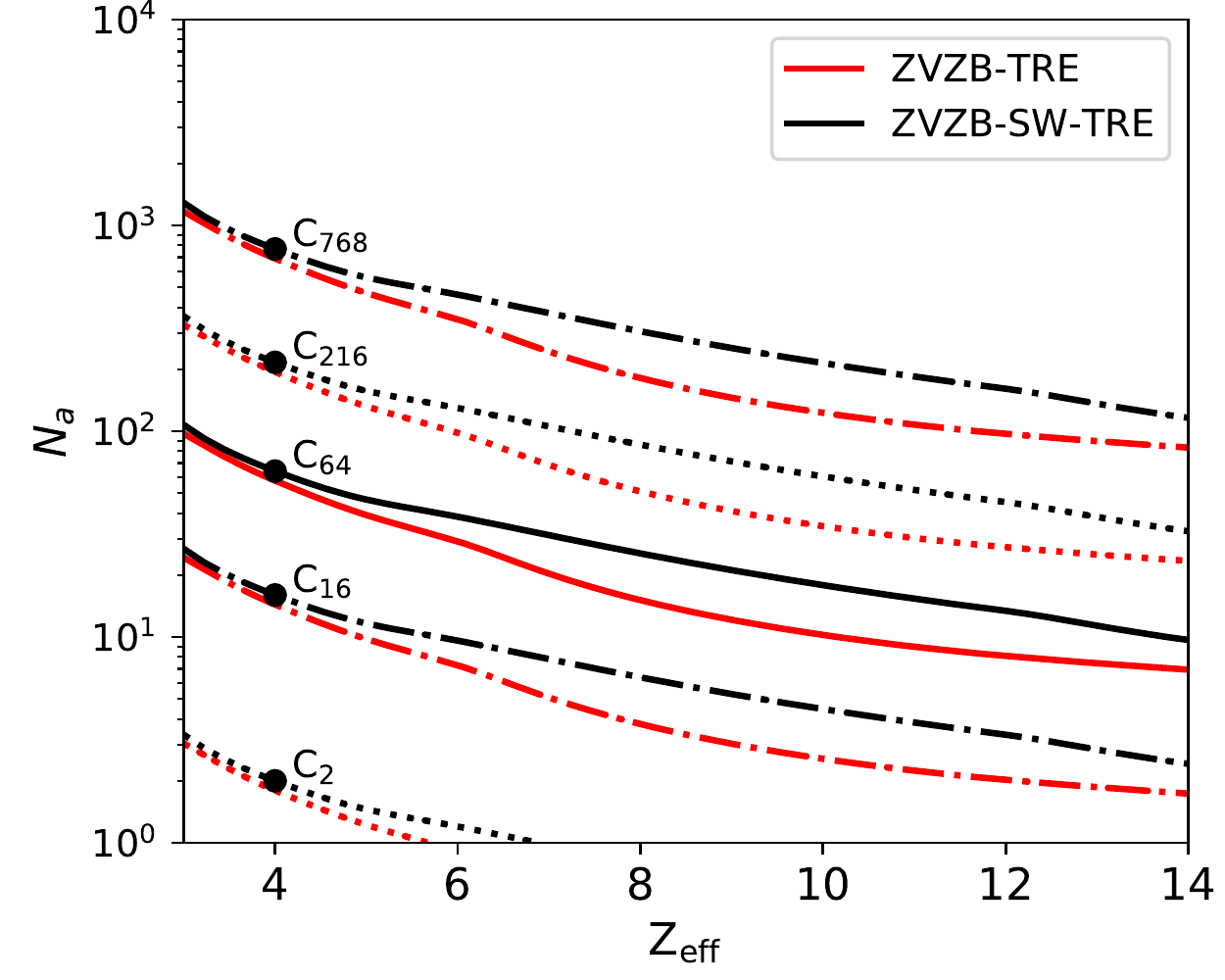} };
    \end{tikzpicture}
    \caption{Cost isocontours for DMC simulations with variable system sizes $N$ and $\zeff$ based on ZVZB-TRE forces of isolated atoms (left) and ZVZB-TRE and ZVZB-SW-TRE forces on homonuclear dimers (right)}
    \label{fig:Fdmc_cost_isocontours}
\end{figure}

System sizes with equivalent statistical costs to 64, 126 and 768 Carbon atoms are presented based on ZVZB-TRE forces on isolated atoms from VMC (Table~\ref{tab:equivalent_cost_vmc}) or DMC (Table~\ref{tab:equivalent_cost_dmc}) simulations. Similar data is presented based on comparison of ZVZB-TRE forces and ZVZB-SW-TRE forces from VMC (Table~\ref{tab:equivalent_cost_vmcsw}) and DMC (Table~\ref{tab:equivalent_cost_dmcsw}). The cost isocontours for VMC are presented and discussed in Sec.~V of the main body, and similar presentation for DMC is given in Fig.~\ref{fig:Fdmc_cost_isocontours}.

\begin{table}[h]
    \centering
    \begin{tabular}{l|ll|ll|ll}
 & C$_{64}$ & & C$_{216}$ & & C$_{768}$ & \\
$\zeff$ & He & Ne & He & Ne & He & Ne \\ \hline \hline
1         & 3150    & 2656   & 10634    & 8967     & 37810    & 31883    \\
2         & 410     & 536    & 1385     & 1810     & 4926     & 6436     \\
3         & 101     & 258    & 342      & 872      & 1216     & 3103     \\
4         & 63      & 168    & 215      & 568      & 767      & 2022     \\
5         & 38      & 106    & 131      & 358      & 466      & 1275     \\
6         & 27      & 73     & 91       & 249      & 326      & 888      \\
7         & 20      & 51     & 68       & 173      & 241      & 618      \\
8         & 15      & 37     & 52       & 127      & 186      & 453      \\
9         & 12      & 29     & 41       & 100      & 146      & 356      \\
10        & 9       & 22     & 32       & 76       & 117      & 273      \\
11        & 7       & 16     & 26       & 56       & 95       & 201      \\
12        & 6       & 13     & 22       & 44       & 80       & 156      \\
13        & 5       & 10     & 19       & 34       & 68       & 121      \\
14        & 4       & 8      & 16       & 27       & 59       & 98       \\
15        & 3       & 6      & 12       & 23       & 46       & 82       \\
16        & 3       & 5      & 10       & 20       & 37       & 71       \\
17        & -       & 5      & -        & 17       & -        & 63       \\
18        & -       & 4      & -        & 15       & -        & 56       \\
19        & -       & 4      & -        & 14       & -        & 50       \\
20        & -       & 3      & -        & 12       & -        & 46       \\ \hline \hline 
    \end{tabular}
    \caption{Numbers of Helium (He) and Neon (Ne) core atoms with variable $\zeff$ that can be computed with computational cost equivalent of 64, 216 or 768 Carbon (He) atoms, using ZVZB-TRE force estimator for VMC, according to data of isolated atoms.}
    \label{tab:equivalent_cost_vmc}
\end{table}

\begin{table}[h]
    \centering
    \begin{tabular}{l|ll|ll|ll}
 & C$_{64}$ & & C$_{216}$ & & C$_{768}$ & \\
$\zeff$ & He & Ne & He & Ne & He & Ne \\ \hline \hline
1         & 1124    & 1191   & 3796     & 4020     & 13499    & 14295    \\
2         & 206     & 285    & 695      & 963      & 2472     & 3425     \\
3         & 88      & 146    & 299      & 495      & 1065     & 1761     \\
4         & 63      & 92     & 215      & 312      & 767      & 1111     \\
5         & 43      & 60     & 145      & 205      & 517      & 730      \\
6         & 30      & 44     & 101      & 148      & 362      & 529      \\
7         & 17      & 33     & 60       & 113      & 215      & 404      \\
8         & 13      & 26     & 45       & 90       & 161      & 320      \\
9         & 11      & 22     & 38       & 74       & 138      & 265      \\
10        & 10      & 18     & 34       & 62       & 123      & 221      \\
11        & 9       & 15     & 31       & 52       & 111      & 185      \\
12        & 8       & 12     & 28       & 43       & 101      & 154      \\
13        & 7       & 10     & 25       & 34       & 91       & 122      \\
14        & 6       & 8      & 23       & 27       & 81       & 98       \\
15        & 4       & 6      & 16       & 23       & 59       & 83       \\
16        & 3       & 6      & 13       & 20       & 46       & 73       \\
17        & -       & 5      & -        & 18       & -        & 65       \\
18        & -       & 4      & -        & 16       & -        & 59       \\
19        & -       & 4      & -        & 15       & -        & 54       \\
20        & -       & 4      & -        & 14       & -        & 50       \\ 
\hline \hline 
    \end{tabular}
    \caption{Numbers of Helium (He) and Neon (Ne) core atoms with variable $\zeff$ that can be computed with computational cost equivalent of 64, 216 or 768 Carbon (He) atoms, using ZVZB-TRE force estimator for DMC, according to data of isolated atoms.}
    \label{tab:equivalent_cost_dmc}
\end{table}

\begin{table}[h]
    \centering
    \begin{tabular}{l|ll|ll|ll}
 & C$_{64}$& & C$_{216}$ & & C$_{768}$ & \\ 
$\zeff$ & ZVZB-TRE & ZVZB-SW-TRE & ZVZB-TRE & ZVZB-SW-TRE & ZVZB-TRE & ZVZB-SW-TRE \\ \hline \hline
3         & 90      & 118    & 305      & 399      & 1084     & 1418     \\
4         & 38      & 63     & 130      & 215      & 464      & 767      \\
5         & 28      & 44     & 94       & 151      & 336      & 539      \\
6         & 19      & 34     & 64       & 116      & 229      & 413      \\
7         & 13      & 25     & 45       & 86       & 162      & 307      \\
8         & 10      & 19     & 34       & 65       & 123      & 234      \\
9         & 8       & 15     & 27       & 52       & 98       & 185      \\
10        & 6       & 12     & 22       & 42       & 81       & 151      \\
11        & 5       & 10     & 19       & 35       & 69       & 127      \\
12        & 5       & 9      & 16       & 30       & 60       & 109      \\
13        & 4       & 7      & 14       & 23       & 50       & 84       \\
14        & 3       & 5      & 12       & 19       & 43       & 68       \\ 
\hline \hline 
    \end{tabular}
    \caption{Numbers of Helium-core atoms with variable $\zeff$ that can be computed with computational cost equivalent of 64, 216 or 768 Carbon atoms, using ZVZB-TRE or ZVZB-TRE-SW force estimators for VMC, according to data of homonuclear dimers.}
    \label{tab:equivalent_cost_vmcsw}
\end{table}

\begin{table}[h]
    \centering
    \begin{tabular}{l|ll|ll|ll}
 & C$_{64}$& & C$_{216}$ & & C$_{768}$ & \\ 
$\zeff$ & ZVZB-TRE & ZVZB-SW-TRE & ZVZB-TRE & ZVZB-SW-TRE & ZVZB-TRE & ZVZB-SW-TRE \\ \hline \hline
3         & 97      & 107    & 328      & 362      & 1167     & 1287     \\
4         & 57      & 63     & 194      & 215      & 689      & 767      \\
5         & 39      & 46     & 131      & 156      & 469      & 557      \\
6         & 29      & 38     & 98       & 129      & 349      & 460      \\
7         & 20      & 31     & 68       & 105      & 243      & 374      \\
8         & 15      & 25     & 50       & 85       & 181      & 305      \\
9         & 12      & 21     & 40       & 71       & 145      & 253      \\
10        & 10      & 17     & 34       & 60       & 122      & 214      \\
11        & 8       & 15     & 30       & 51       & 107      & 184      \\
12        & 8       & 13     & 27       & 45       & 97       & 160      \\
13        & 7       & 11     & 25       & 38       & 89       & 136      \\
14        & 6       & 9      & 23       & 32       & 83       & 116      \\ 
\hline \hline
    \end{tabular}
    \caption{Numbers of Helium-core atoms with variable $\zeff$ that can be computed with computational cost equivalent of 64, 216 or 768 Carbon atoms, using ZVZB-TRE or ZVZB-TRE-SW force estimators for DMC, according to data of homonuclear dimers.}
    \label{tab:equivalent_cost_dmcsw}
\end{table}

\FloatBarrier

\section{Mapping between bond stiffness and force uncertainty}

\begin{figure}[h]
    \centering
    \begin{tikzpicture}
    \node at (0, 0) {\includegraphics[width=7cm]{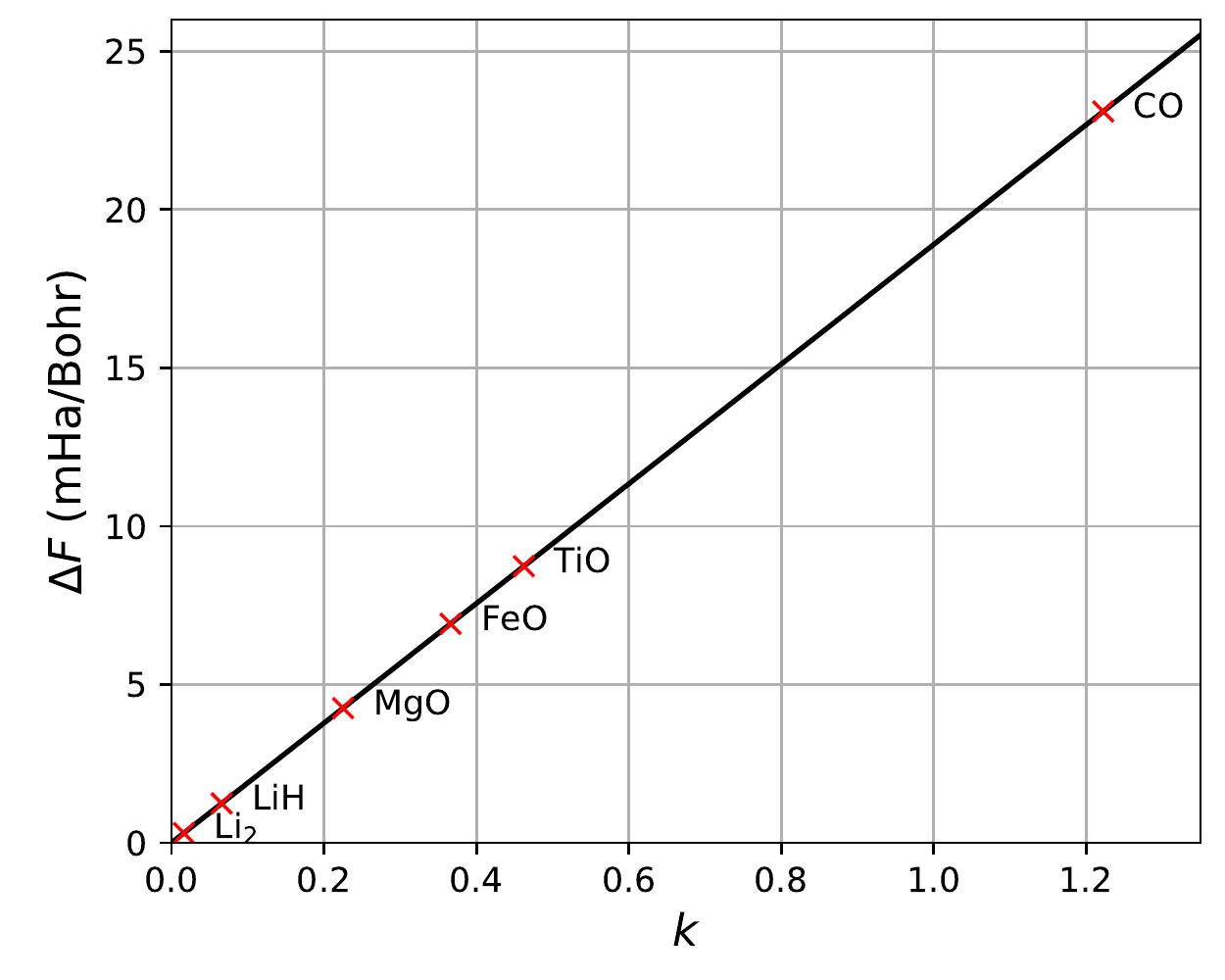}};
    \end{tikzpicture}
    \caption{Mapping between stiffness $k$ and the maximum force uncertainty required for spatial accuracy of $\sim 0.01$ \AA.}
    \label{fig:dF_vs_k}
\end{figure}

A mapping between the bond stiffness (force constant) of the dimers in this study and force uncertainty to reach the precision of $\sim 0.01$ \AA{} is presented in Fig.~\ref{fig:dF_vs_k}. 

\section{Implementation Details of Forces for Non-Local Pseudopotentials}

QMCPACK uses the following form of the non-local pseudopotential, following the normalization and quadrature rules of Mitas and Ceperley\cite{Mitas1991}:
\begin{eqnarray}
\frac{\hat{V}_{NL}\Psi_T}{\Psi_T} & = & \sum_I \sum_i \sum_\ell \left(
(2\ell+1) v_l(r_{iI})
\sum_{\alpha}\left[ w_\alpha P_\ell \left(\cos \theta^\alpha_{iI} \right) \frac{\Psi_T(\ldots \mathbf{q}^\alpha_{iI} \ldots)}{\Psi_T(\ldots \mathbf{r}_i \ldots)}   
\right]\right) \\
\mathbf{r}_{iI} &=& \mathbf{r}_i - \mathbf{R}_I \\
r_{iI} &=& |\mathbf{r}_i - \mathbf{R}_I| \\
\mathbf{q}^\alpha_{iI} &=& \mathbf{R}_I + r_{iI} \mathbf{u}_\alpha \\ 
\cos \theta^\alpha_{iI} &=& (\mathbf{r}_{iI}\cdot \mathbf{u}_\alpha) r_{iI}^{-1}
\end{eqnarray}

Lower case vectors and indices denote electron coordinates, upper case vectors and indices denote ionic coordinates.  $P_\ell(x)$ is a Legendre polynomial, $w_\alpha$ is a weight given by whichever quadrature rule is being used, and $v_\ell(r)$ the radial potential for the $\ell$ angular momentum channel.

We now consider the derivative of this expression w.r.t the ion displacement $R_{J,\gamma}$.  Formally, 

\begin{eqnarray}
\scriptstyle \frac{\partial}{\partial R_{J,\gamma}}\frac{\hat{V}_{NL}\Psi_T}{\Psi_T} & = & \scriptstyle \sum_I \sum_i \sum_\ell 
(2\ell+1) \frac{\partial}{\partial R_{J,\gamma}} v_l(r_{iI})
\sum_{\alpha}\left[ w_\alpha P_\ell \left(\cos \theta^\alpha_{iI} \right) \frac{\Psi_T(\ldots \mathbf{q}^\alpha_{iI} \ldots)}{\Psi_T(\ldots \mathbf{r}_i \ldots)}\right] \\
&+& \scriptstyle (2\ell+1) v_l(r_{iI}) \sum_{\alpha}\left[ w_\alpha  \frac{\partial}{\partial R_{J,\gamma}}P_\ell \left(\cos \theta^\alpha_{iI} \right) \frac{\Psi_T(\ldots \mathbf{q}^\alpha_{iI} \ldots)}{\Psi_T(\ldots \mathbf{r}_i \ldots)}\right] \\
&+& \scriptstyle (2\ell+1) v_l(r_{iI}) \sum_{\alpha}\left[w_\alpha P_\ell \left(\cos \theta^\alpha_{iI} \right) \frac{\partial}{\partial R_{J,\gamma}}\frac{\Psi_T(\ldots \mathbf{q}^\alpha_{iI} \ldots)}{\Psi_T(\ldots \mathbf{r}_i \ldots)}\right]\\ 
\end{eqnarray}
First, note that the first two terms vanish unless $I=J$.  The last term can have Pulay like contributions coming from explicit $\Psi_T$ dependence on $i_{J,\gamma}$.  Before crunching on evaluating this whole expression, lets take some intermediate derivatives.  Using Einstein notation, with the big caveat that capital indices are not summed over (e.g. $I$ and $J$): 
\begin{equation}
\frac{\partial}{\partial R_{J,\gamma}} r_{{iI},\mu} = -\delta_{IJ}\delta_{\gamma\mu}
\end{equation}
\begin{equation}
\frac{\partial}{\partial R_{J,\gamma}} r_{iI} = -\delta_{IJ}\frac{r_{iI,\gamma}}{r_{iI}}
\end{equation}
\begin{equation}
\frac{\partial}{\partial R_{J,\gamma}} r_{iI,\mu}u^{\alpha}_\mu = -\delta_{IJ}\delta_{\gamma\mu}u^\alpha_\mu=-\delta_{IJ}u^\alpha_\gamma
\end{equation}

These allow us to evaluate the derivatives of the quadrature points and angles:
\begin{equation}
\frac{\partial}{\partial R_{J,\gamma}} q^\alpha_{iI,\mu} = \delta_{IJ} \left[ \delta_{\gamma\mu} - \frac{r_{iI,\gamma} u^{\alpha}_\mu }{r_{iI}}\right]
\end{equation}
\begin{equation}
\frac{\partial}{\partial R_{J,\gamma}} \cos \theta^\alpha_{iI} = \frac{\partial}{\partial R_{J,\gamma}}\left( r_{iI,\mu}u^{\alpha}_\mu r_{iI}^{-1}\right) = -\delta_{IJ}r^{-1}_{iI}\left[ u^{\alpha}_\gamma - \cos \theta^\alpha_{iI} \frac{r_{iI,\gamma}}{r_{iI}}\right] 
\end{equation}

Now let's consider what happens to the wavefunction derivative:
\begin{equation}
\frac{\partial}{\partial R_{J,\gamma}}\frac{\Psi_T(\ldots \mathbf{q}^\alpha_{iI} \ldots)}{\Psi_T(\ldots \mathbf{r}_i \ldots)}  =  \frac{ \frac{\partial}{\partial R_{J,\gamma}} \Psi_T(\ldots \mathbf{q}^\alpha_{iI} \ldots)}{\Psi_T(\ldots \mathbf{r}_i \ldots)}-\frac{ \Psi_T(\ldots \mathbf{q}^\alpha_{iI} \ldots)}{\Psi_T(\ldots \mathbf{r}_i \ldots)}\frac{ \frac{\partial}{\partial R_{J,\gamma}} \Psi_T(\ldots \mathbf{r}_i \ldots)}{\Psi_T(\ldots \mathbf{r}_i \ldots)}
\end{equation}

The ion derivative in the second term is a standard Pulay term and is evaluated as such.  The first term is complicated by the potential explicit dependence of $\Psi_T$ on the ion coordinates (as in the second term), but also by the ion dependence of the quadrature points.  The total derivative of the numerator is:

\begin{eqnarray}
\scriptstyle \frac{d}{d R_{J,\gamma}} \Psi_T(\ldots \mathbf{q}^\alpha_{iI} \ldots)  & = & \scriptstyle \frac{\partial}{\partial R_{J,\gamma}} \Psi_T(\ldots \mathbf{q}^\alpha_{iI} \ldots) + \partial_{i,\mu}\Psi_T(\ldots q^\alpha_{iI,\mu} \ldots) \frac{\partial q^\alpha_{iI,\mu}}{\partial R_{J,\gamma}} \\
& = &  \scriptstyle \frac{\partial}{\partial R_{J,\gamma}} \Psi_T(\ldots \mathbf{q}^\alpha_{iI} \ldots) + \delta_{IJ}\partial_{i,\mu}\Psi_T(\ldots q^\alpha_{iI,\mu} \ldots)\left[ \delta_{\gamma\mu} - \frac{r_{iI,\gamma} u^{\alpha}_\mu }{r_{iI}}\right]
\end{eqnarray}

Now for the radial potential derivative:
\begin{equation}
\frac{\partial}{\partial R_{J,\gamma}} v_l(r_{iI}) = -\delta_{IJ} v'_l(r_{iI})\frac{r_{iI,\gamma}}{r_{iI}}
\end{equation}

The legendre polynomial...
\begin{eqnarray}
  \frac{\partial}{\partial R_{J,\gamma}} P_\ell( \cos \theta^\alpha_{iI} ) &=& P'_\ell( \cos \theta^\alpha_{iI} ) \frac{\partial}{\partial R_{J,\gamma}} \cos \theta^\alpha_{iI} \\
  &=& -\delta_{IJ} P'_\ell( \cos \theta^\alpha_{iI} ) r^{-1}_{iI}\left[ u^{\alpha}_\gamma - \cos \theta^\alpha_{iI} \frac{r_{iI,\gamma}}{r_{iI}}\right]  
\end{eqnarray}

The Legendre Polynomials in QMCPACK are computed via the recursion relation:
\begin{equation}
(n+1)P_{n+1}(x) = (2n+1)x P_n(x) - n P_{n-1}(x)
\end{equation}
$P_0(x)=1$, and $P_1(x)=x$.

We can take the derivative of the above equation to obtain a recursion relation for the derivatives of the Legendre polynomials:
\begin{equation}
(n+1)P'_{n+1}(x) = (2n+1)\left[x P'_n(x) + P_n(x) \right] - n P'_{n-1}(x)
\end{equation}
With of course, $P'_0(x)=0$, and $P'_1(x)=1$.

We now take all these results and plug them in:
\begin{eqnarray}
\scriptstyle
\frac{\partial}{\partial R_{J,\gamma}}\frac{\hat{V}_{NL}\Psi_T}{\Psi_T} & = & \scriptstyle -\sum_i \sum_\ell (2\ell+1) \underbrace{ \scriptstyle v'_l(r_{iJ})\frac{r_{iJ,\gamma}}{r_{iJ}}}
\sum_{\alpha}\left[ w_\alpha P_\ell \left(\cos \theta^\alpha_{iJ} \right) \frac{\Psi_T(\ldots \mathbf{q}^\alpha_{iJ} \ldots)}{\Psi_T(\ldots \mathbf{r}_i \ldots)}\right] \\
\scriptstyle & - & \scriptstyle \sum_i \sum_\ell (2\ell+1) v_l(r_{iJ}) \sum_{\alpha}\left[ w_\alpha \underbrace{ \scriptstyle P'_\ell( \cos \theta^\alpha_{iJ} ) r^{-1}_{iJ}\left( u^{\alpha}_\gamma - \cos \theta^\alpha_{iJ} \frac{r_{iJ,\gamma}}{r_{iJ}}\right)} \frac{\Psi_T(\ldots \mathbf{q}^\alpha_{iJ} \ldots)}{\Psi_T(\ldots \mathbf{r}_i \ldots)} \right] \\
 & + &  \scriptstyle \sum_i \sum_\ell (2\ell+1) v_l(r_{iJ}) \sum_\alpha \left[ w_\alpha P_\ell \left(\cos \theta^\alpha_{iJ} \right) \underbrace{\scriptstyle \frac{\partial_{i,\mu}\Psi_T(\ldots \mathbf{q}^\alpha_{iJ} \ldots)}{\Psi_T(\ldots \mathbf{r}_{i} \ldots)}\left( \delta_{\gamma\mu} - \frac{r_{iJ,\gamma} u^{\alpha}_\mu }{r_{iJ}} \right)}\right] \\
 & + & \scriptstyle \frac{\hat{V}_{NL}\frac{\partial \Psi_T}{\partial R_{J,\gamma}}}{\Psi_T} - \left(\frac{\hat{V}_{NL}\Psi_T}{\Psi_T}\right)\frac{\frac{\partial \Psi_T}{\partial R_{J,\gamma}}}{\Psi_T}
\end{eqnarray}

I've put braces under the terms that need to be computed \textit{beyond} what's already computed for the nonlocal pseudopotential energy evaluation.  
In the last line, we have first what's essentially a Pulay term, and the last is a ZVZB like term.

\FloatBarrier

\section{Acknowledgements}

This research has been provided by the US Department of Energy, Office of Science, Basic Energy Sciences, Materials Sciences and Engineering Division, as part of the Computational Materials Sciences Program and Center for Predictive Simulation of Functional Materials. This research used resources of the Compute and Data Environment for Science (CADES) at the Oak Ridge National Laboratory, which is supported by the Office of Science of the U.S. Department of Energy under Contract No. DE-AC05-00OR22725.

This manuscript has been authored in part by UT-Battelle, LLC, under contract DE-AC05-00OR22725 with the US Department of Energy (DOE). The US government retains and the publisher, by accepting the article for publication, acknowledges that the US government retains a nonexclusive, paid-up, irrevocable, worldwide license to publish or reproduce the published form of this manuscript, or allow others to do so, for US government purposes. DOE will provide public access to these results of federally sponsored research in accordance with the DOE Public Access Plan (http://energy.gov/downloads/doe-public-access-plan).

Sandia National Laboratories is a multimission laboratory managed and operated by National Technology \& Engineering Solutions of Sandia, LLC, a wholly owned subsidiary of Honeywell International Inc., for the U.S. Department of Energy’s National Nuclear Security Administration under contract DE-NA0003525. This paper describes objective technical results and analysis. Any subjective views or opinions that might be expressed in the paper do not necessarily represent the views of the U.S. Department of Energy or the United States Government.

The data that support the findings of this study are openly available in The Materials Data Facility at http://doi.org/[doi], reference number [reference number].

\section{References}
\bibliography{references}